\documentclass[floats,aps,epsf,amsfonts]{revtex4}
\usepackage{graphicx}

\begin{document}
\draft

\title{Gravitational self force on a particle in circular orbit
around a Schwarzschild black hole}

\author{Leor Barack and Norichika Sago}
\address
{School of Mathematics, University of Southampton, Southampton,
SO17 1BJ, United Kingdom}

\date{\today}

\begin{abstract}
We calculate the gravitational self force acting on a pointlike particle of
mass $\mu$, set in a circular geodesic orbit around a Schwarzschild black hole.
Our calculation is done in the Lorenz gauge: For given orbital radius,
we first solve directly for the Lorenz-gauge metric perturbation using numerical
evolution in the time domain; We then compute the (finite) back-reaction force
from each of the multipole modes of the perturbation; Finally, we apply
the ``mode sum'' method to obtain the total, physical self force.
The {\em temporal} component of the self force (which is gauge invariant) describes
the dissipation of orbital energy through gravitational radiation.
Our results for this component are consistent, to within the computational accuracy,
with the total flux of gravitational-wave energy radiated to infinity and through the
event horizon. The {\em radial} component of the self force
(which is gauge dependent) is calculated here for the first time. It describes
a conservative shift in the orbital parameters away from their geodesic values.
We thus obtain the $O(\mu)$ correction to the specific energy and angular momentum parameters
(in the Lorenz gauge), as well as the $O(\mu)$ shift in the orbital frequency
(which is gauge invariant).
\end{abstract}

\maketitle

%%%%%%%%%%%%%%%%%%%%%%%%%%%%%%%%%%%%%%%%%%%%%%%%%%%
\section{Introduction and summary} \label{Sec:intro}
%%%%%%%%%%%%%%%%%%%%%%%%%%%%%%%%%%%%%%%%%%%%%%%%%%%

The problem of calculating the back-reaction force, or {\it self force} (SF),
experienced by a point particle as it moves in curved spacetime is now understood
well enough to allow actual computations of this effect in systems comprising of a
small object orbiting a large black hole. The fundamental formulation of the
problem and its solution was set in works by
DeWitt and Brehme \cite{DeWitt:1960fc} (electromagnetic SF),
Mino, Sasaki and Tanaka \cite{Mino:1996nk} and Quinn and Wald \cite{Quinn:1996am}
(gravitational SF), and Quinn \cite{Quinn:2000wa} (scalar field SF).
An alternative formulation was introduced by
Detweiler and Whiting \cite{Detweiler:2002mi}, also clarifying the relation between the SF
picture (``forced motion on a background geometry'') and the standard description
based on the principle of equivalence (``geodesic motion in a perturbed geometry'').
A number of authors later devised a practical calculation method for the SF in black
hole spacetimes---the ``mode sum scheme''---which is based on multipole decomposition
of the retarded field, and relies on standard methods of black hole perturbation theory
\cite{Barack:1999wf,Barack:2001bw,Barack:2001gx,Barack:2002bt,Barack:2002mh}.
This method has since been implemented by various
authors on a case-by case basis, so far mostly for calculations of the {\em scalar field}
SF. Work so far included the cases of a static particle in Schwarzschild \cite{Burko:1999zy}
or along the rotation axis of a Kerr black hole \cite{Burko:2001kr};
radial plunge trajectories \cite{Barack:2000zq} and circular orbits around a Schwarzschild
Black hole \cite{Burko:2000xx,Detweiler:2002gi,Diaz-Rivera:2004ik,Hikida:2004jw}; and ongoing
work on eccentric orbits in Schwarzschild \cite{Hikida,Hass}.
The {\em gravitational} SF has been calculated so far only for radial trajectories in
Schwarzschild \cite{Barack:2002ku} and for static (non-geodesic) particles in Schwarzschild
\cite{Keidl:2006wk}.  The case of an orbiting particle has been tackled only under
the post-Newtonian (PN) approximation \cite{Hikida:2004hs}. A comprehensive review of the
subject, including a self-contained account of SF fundamentals, is
provided by Poisson \cite{Poisson:2003nc}. For a snapshot of the current activity
in the field, the reader may refer to \cite{Lousto:2005an}.

One of the main motivations for the work on self forces draws from the need to
devise accurate theoretical waveforms for the gravitational radiation from extreme
mass-ratio inspirals (EMRIs)---of the prime targets for LISA, the planned space-based gravitational
wave detector \cite{Barack:2003fp}. This requires solving the SF problem in the {\em gravitational}
case, for generic inspiral orbits around a Kerr black hole. The main challenge in extending
the analysis from the scalar-field toy model to the gravitational case
has to do with the gauge freedom in the latter case. The problem can be summarized as follows.
The gravitational perturbation in the vicinity of the point particle is best
described using the {\em Lorenz} gauge (see Appendix \ref{AppA}), which preserves the
local isotropic nature of the point singularity. On the other hand, the field equations
that govern the global evolution of the metric perturbation are more tractable in
gauges which comply well with the global symmetry of the black hole background---best
known examples of which are the ``radiation'' gauges \cite{Chrzanowski:1975wv} or the
Regge-Wheeler gauge \cite{Regge:1957td}. Now, in calculating the local SF we need,
essentially, to subtract a suitable local, divergent piece of the perturbation
from the full (retarded) perturbation field.  In doing so, both fields (local and
global) must be given in the same gauge; the ``gauge problem'' arises since
the two fields are normally calculated in different gauges.
Indeed, the only fully-worked-out example of the gravitational SF so far is the
case of radial orbits in Schwarzschild \cite{Barack:2002ku}, where the gauge problem is
avoided simply because, in this particular setup, the singular piece of the Regge-Wheeler
perturbation happens to coincide with that of the Lorenz-gauge perturbation.

One approach to the problem has been to try and calculate the local divergent
piece in one of the ``global'' gauges---specifically the Regge Wheeler gauge in
the Schwarzschild spacetime \cite{Mino:1998qg,Mino:1998gp,Nakano:2000ne}.
This has been implemented so far only within a post-Newtonian approximation
\cite{Hikida:2004hs}. In the current work we take a complementary approach:
We solve the perturbation equations, and obtain the ``global'' retarded field,
directly in the Lorenz gauge. The calculation is therefore done entirely within
the Lorenz gauge, the ``subtraction'' procedure necessary for constructing the
SF is implemented in a straightforward way, and the gauge problem is avoided
altogether. Other advantages of working in the Lorenz gauge include the fact
that the field equations then take a fully hyperbolic form (making them
especially suitable for time-domain integration); and the fact that the
Lorenz-gauge metric perturbation is better behaved near the particle
compared with the perturbation in other gauges \cite{Barack:2001ph} (which, again,
makes it more suitable for numerical implementation). The better regularity
of the Lorenz-gauge perturbation is manifested in the behavior of individual
multipole modes of the field\footnote{By ``multipole modes'' we mean
here, and throughout this paper, the components of a spacetime function in a
spherical-harmonic basis defined on 2-spheres {\em centered at the black hole}
(and not at the particle), as usual in black hole perturbation theory.
}:
It is well known, for example, that the multipole modes of the Regge--Wheeler
perturbation from a point particle in Schwarzschild generally show a
discontinuity across the particle. In contrast, the modes of the Lorenz-gauge
perturbation are always continuous at the particle.

Our ``all-Lorenz-gauge'' approach is made possible (at least is the Schwarzschild
case) following a recent work by Barack and Lousto \cite{Barack:2005nr} (hereafter BL), which
provided a practical formulation of the Lorenz-gauge perturbation equations in the
Schwarzschild geometry. Our calculation is based on the BL formulation, and our
numerical code incorporates the code developed in BL (with a few improvements).
In the current work we focus on circular geodesic orbits, for simplicity. However, since
our treatment is based on a time-domain evolution, our code could be amended to
deal with generic orbits (in Schwarzschild) in an almost straightforward
manner. We shall discuss this extension of the analysis in the concluding section,
and also comment there on the important extension to the Kerr case.

Our calculation of the SF proceeds as follows. We first write down the 10
(coupled) evolution equations for the 10 tensorial-harmonic components of the
Lorenz-gauge metric perturbation, in the form given in BL (with a slight modification).
The energy-momentum of the orbiting particle is represented by a suitable
delta-function distribution, whose tensor-harmonic components serve as sources for
the evolution equations. For given orbital radius and given multipole numbers $l$ and
$m$ we solve the equations numerically through time-domain evolution in 1+1 dimensions
(time+radius), using a 2nd-order-convergent finite-difference scheme on a staggered
grid based on characteristic coordinates. We integrate long enough to allow any
spurious initial radiation to dissipate efficiently (this takes $\sim 3$ orbital
periods for strong-field orbits). We then record the values of the metric
perturbation and it temporal and radial derivatives at the location of the particle.
[Recall that individual multipole modes of the perturbation are continuous at the
particle. Their first derivatives have a finite jump discontinuity across the
particle (in the radial direction) and so we record both values of the derivatives.]
We repeat this calculation for all multipole modes with $2\leq l\leq l_{\rm max}$, where
$l_{\rm max}$ is determined experimentally so that our standard of accuracy (of
$< 10^{-3}$ fractional error in the final SF) is met. In practice we found
it sufficient to take $l_{\rm max}=15$ for the radial component and $l_{\rm max}=5$--$9$
for the $t$ component (depending on the orbital radius). The modes $l=0,1$ are calculated
separately, using the method of Detweiler and Poisson \cite{Detweiler:2003ci}.
The values of the metric perturbation and its derivatives at the particle are then
used as input for the mode-sum scheme. Within this scheme, each of the modes is
``regularized'' using functions known analytically
\cite{Barack:2001gx,Barack:2002bt,Barack:2002mh}, and the sum over modes
yields the desired SF.

Our main results are summarized in Tables \ref{table:compare-Edot} and
\ref{table:result-Fr} (along with Fig.\ \ref{fig:r0vsFr}).
The tables display the values of both radial and temporal
components of the SF as a function of the orbital radius.
The temporal component of the SF is simply related, in our stationary
circular-orbit setup, to the rate of change of the orbital energy parameter,
and hence to the flux of energy carried in gravitational waves
to null infinity and down the event horizon. Our results demonstrate
this energy balance, which provides a reassuring validity test for our code.
The radial component of the SF (which is itself gauge dependent) describes the
conservative back-reaction effect on the orbital parameters. Based on our
results we calculate the conservative shift in the
energy and angular momentum of the circular geodesic, as well as the shift
in the orbital frequency---the latter being gauge invariant. Our results for
the shifts are plotted in Figs.\ \ref{fig:ELvsr0} and \ref{fig:r0vsOmg2}.
%We find that, in the Lorenz gauge, the radial component of the SF is {\em repulsive}
%(pointing outward, away from the black hole) and that the values of the orbital
%energy and angular momentum both {\em decrease} by the effect of the conservative
%force.

To check the validity and robustness of our code, and assess the accuracy of our
results, we performed the following tests.
(i) {\it Numerical convergence:} For each of the modes calculated, we repeated our
computation with a handful of different numerical resolutions, checking that the
answer converges quadratically to a limiting value with decreasing step size.
To determine the limiting value we used a Richardson extrapolation (over step size),
and recorded the estimated error from this extrapolation.
(ii) {\it Effect of spurious initial waves:} We compared the values of the SF at two
different (late) evolution times (recall that in our stationary setup the physical SF
does not depend on time), in order to asses the effect of residual spurious waves.
(iii) {\it Large $l$ behavior:} The behavior of the SF modes at large multipole
numbers $l$ is known analytically with high precision [see Eqs.\
(\ref{ModeSum})--(\ref{B}) below].
We verified that our numerically calculated modes have the right behavior at large
$l$, through all three known leading terms in the $1/l$ expansion. This agreement
is necessary, in fact, for a successful implementation of the mode-sum scheme.
(iv) {\em Error from large-$l$ tail:} The mode-sum scheme involves summation over
all modes $l$. In practice we computed all modes up to $l=l_{\rm max}$,
and used an extrapolation to estimate the contribution from the remaining
$l>l_{\rm max}$ tail.
%(the regularized SF modes that enter the mode sum fall off
%at large $l$ as $\propto l^{-2}$.
We assessed and recorded the error from this extrapolation.
(v) {\it Comparison of one-sided forces:} The mode sum scheme can be implemented in
two essentially independent ways, by using either the ``external'' or the ``internal''
values of the SF modes (i.e., values calculated by taking one-sided derivatives of
the metric perturbation from outside or inside the orbit, respectively). Of course,
the final value of the SF should not depend on our choice. We used our code to
work out both values, and checked that they are the same to a very good accuracy.
We recorded the (tiny) difference between the two values. Our total computation
error was taken to be the combined error from the extrapolation over step size
[item (i) above], the deviation from stationarity [item (ii) above], the
extrapolation over $l$ [item (iv)], and the small discrepancy between external
and internal values. We made sure that this combined error is kept under 0.1\%.
(vi) {\it Comparison with energy flux:} We checked that the computed temporal
component of the SF balances the flux of energy to null infinity and down the horizon.
We found an excellent agreement.
%(vii) {\it Comparison with post-Newtonian calculations:}
%[\P see if we can say anything here]

The structure of this paper is as follows.
In Sec.\ \ref{Sec:theory} we review the formalism for constructing the metric
perturbation in the Lorenz gauge and for calculating the SF via the mode sum
scheme (focusing on the case of circular orbits in Schwarzschild).
We also discuss the effect of the SF on the geodesic parameters (energy,
angular momentum, angular velocity), and how these depend on the choice of
gauge. Sec.\ \ref{Sec:numerics} describes our numerical method, including a detailed
description of the finite-difference scheme. In Sec.\ \ref{Sec:validation} we
present a few validation tests for our code, and explain how we estimated
the computation error. Sec.\ \ref{Sec:results} gives the results: We tabulate
both temporal and radial components of the SF as a function of
the orbital radius, and calculate the shift in the orbital
parameters due to the conservative piece of the force.
%In Sec.\ \ref{Sec:PN} we use our results for the SF to infer analytic
%post-Newtonian expressions for large orbital radii. We compare these to
%results from the post-Newtonian literature.
Finally, in Sec.\ \ref{Sec:summary} we discuss the extension of this work
to more general orbits in Schwarzschild, and to orbits in Kerr.

Throughout this work we use standard geometrized units (with $c=G=1$)
and metric signature $({-}{+}{+}{+})$. The Riemann tensor is defined as in
Ref.\ \cite{MTW}.

%%%%%%%%%%%%%%%%%%%%%%%%%%%%%%%%%%%%%%%%%%%%%%%%%%%%%%%
\section{Review of theory:
         Self force in Lorenz gauge} \label{Sec:theory}
%%%%%%%%%%%%%%%%%%%%%%%%%%%%%%%%%%%%%%%%%%%%%%%%%%%%%%%

%We review here the calculating the Lorenz-gauge SF in Lorenz gauge focus on a
%circular orbit. To obtain the self force, we adopt the
%mode sum scheme
%\cite{Barack:1999wf,Barack:2001bw,Barack:2001gx,Barack:2002bt,Barack:2002mh},
%in which the singular (S) part is subtracted from the
%full force by modes.
%The analytic formula of the S part is given in
%\cite{Barack:2001gx} based on the scalar harmonic expansion.
%On the other hand, we solve the field equations under the
%tensor harmonic expansion in order to obtain the full force.
%To perform the subtraction of the S part, therefore,
%we need to rearrange the tensor harmonic modes to
%the spherical harmonic modes. We also present a scheme
%to do that.

\subsection{Orbital setup and equation of motion}

Consider a pointlike particle of mass $\mu$, in a circular orbit around
a Schwarzschild black hole with mass $M\gg\mu$. Let the worldline of
the particle be represented by $x^{\alpha}=x_{\rm p}^{\alpha}(\tau)$,
with tangent four velocity $u^{\alpha}= dx_{\rm p}^{\alpha}/d\tau$.
At the limit $\mu\to 0$ (i.e., neglecting SF effects) the particle traces a
geodesic $x_{\rm p}^{\alpha}=x_{0}^{\alpha}(\tau)$, with associated four
velocity $u^{\alpha}_0= dx_{\rm 0}^{\alpha}/d\tau$. Without limiting the
generality, we adopt a Schwarzschild coordinate system $t,r,\theta,\varphi$
in which the orbit is confined to the equatorial plane. Then
%~~~~~~~~~~~~~~~~~~~~~~~~~~~~~~~~~~~~~~~~~~~~~~~~~~~~~~~~~~~~~~~~
\begin{equation} \label{z}
x_{0}^{\alpha}(\tau)=\left[t_0(\tau),r_0={\rm const},\theta_0=\pi/2,\varphi_0(\tau)\right].
\end{equation}
%~~~~~~~~~~~~~~~~~~~~~~~~~~~~~~~~~~~~~~~~~~~~~~~~~~~~~~~~~~~~~~~~
This circular geodesic can be parametrized by the radius $r_0$, or,
alternatively, by the angular velocity (with respect to time $t$)
%~~~~~~~~~~~~~~~~~~~~~~~~~~~~~~~~~~~~~~~~~~~~~~~~~~~~~~~~~~~~~~~~
\begin{equation} \label{Omega0}
\Omega_0\equiv d\varphi_0/dt=\sqrt{M/r_0^3},
\end{equation}
%~~~~~~~~~~~~~~~~~~~~~~~~~~~~~~~~~~~~~~~~~~~~~~~~~~~~~~~~~~~~~~~~
by the specific energy parameter,
%~~~~~~~~~~~~~~~~~~~~~~~~~~~~~~~~~~~~~~~~~~~~~~~~~~~~~~~~~~~~~~~~
\begin{equation} \label{E0}
{\cal E}_0\equiv -u_{0t}=(1-2M/r_0)(1-3M/r_0)^{-1/2},
\end{equation}
%~~~~~~~~~~~~~~~~~~~~~~~~~~~~~~~~~~~~~~~~~~~~~~~~~~~~~~~~~~~~~~~~
or by the specific angular momentum parameter,
%~~~~~~~~~~~~~~~~~~~~~~~~~~~~~~~~~~~~~~~~~~~~~~~~~~~~~~~~~~~~~~~~
\begin{equation} \label{L0}
{\cal L}_0\equiv u_{0\varphi}=(M r_0)^{1/2}(1-3M/r_0)^{-1/2}.
\end{equation}
%~~~~~~~~~~~~~~~~~~~~~~~~~~~~~~~~~~~~~~~~~~~~~~~~~~~~~~~~~~~~~~~~
The subscripts `0' here indicate that the above are values associated with the
geodesic $x_{\rm 0}^{\alpha}$ (below we will consider the correction to
these values due the SF effect).  As always in our perturbative
treatment, tensorial indices are ``raised'' and ``lowered'' using the
background metric.

%The four velocity of the particle is
%then given (in Schwarzschild coordinates) by
%%~~~~~~~~~~~~~~~~~~~~~~~~~~~~~~~~~~~~~~~~~~~~~~~~~~~~~~~~~~~~~~~~
%\begin{equation} \label{u}
%u^{\alpha}=({\cal E}/f_0)[1,0,0,\omega].
%\end{equation}
%%~~~~~~~~~~~~~~~~~~~~~~~~~~~~~~~~~~~~~~~~~~~~~~~~~~~~~~~~~~~~~~~~
%we have
%\begin{equation}
%{\cal E}^2=f_0(1+{\cal L}^2), \quad\quad
%{\cal E}=f_0(1-3M/r_0)^{-1/2}, \quad\quad
%{\cal L}=(M/r_0)^{1/2}(1-3M/r_0)^{-1/2}.
%\end{equation}

Now assume that $\mu$ is finite (but still very small compared to $M$).
The equation of motion can be written as
%~~~~~~~~~~~~~~~~~~~~~~~~~~~~~~~~~~~~~~~~~~~~~~~~~~~~~~~~~~~~~~~~
\begin{equation} \label{EOM}
\mu \frac{D^2x_{\rm p}^{\alpha}}{D\tau^2}=
\mu \frac{Du^{\alpha}}{D\tau}=
F^{\alpha},
\end{equation}
%~~~~~~~~~~~~~~~~~~~~~~~~~~~~~~~~~~~~~~~~~~~~~~~~~~~~~~~~~~~~~~~~
where the covariant derivatives are taken with respect to the {\it background}
(Schwarzschild) geometry, and $F^{\alpha}[\sim O(\mu^2)]$ is the gravitational SF.
Clearly, the symmetry of the problem implies $F^{\theta}=0$.
Also, assuming the four-velocity is kept normalized along the worldline,
i.e., $u_{\alpha}u^{\alpha}=-1$ , we have $D(u_{\alpha}u^{\alpha})/D\tau=0$,
leading to $u_{\alpha}F^{\alpha}=0$, and the four components of the SF are not
independent. In the circular orbit case we have the relation
$u_tF^t+u_{\varphi}F^{\varphi}=0$, which we may write, through leading order in $\mu$,
as
%~~~~~~~~~~~~~~~~~~~~~~~~~~~~~~~~~~~~~~~~~~~~~~~~~~~~~~~~~~~~~~~~
\begin{equation} \label{Fvarphi}
F^{\varphi}=({\cal E}_0/{\cal L}_0) F^t.
\end{equation}
%~~~~~~~~~~~~~~~~~~~~~~~~~~~~~~~~~~~~~~~~~~~~~~~~~~~~~~~~~~~~~~~~
Hence, we need only calculate two of the components of the SF: the $r$
component and (say) the $t$ component. For simplicity we shall refer to
these as the ``radial'' and ``temporal'' components. Our goal would be to
calculate both components, as a function of the orbital radius $r_0$.
Note that it is sufficient, for the sake of obtaining the leading-order
$[O(\mu^2)]$ SF, to assume that the motion is momentarily geodesic.
%Deviations from geodesic motion would affect the local SF
%only through higher order in $\mu$.

The SF affects the motion of the particle in two ways: Firstly, $-u_t$ and
$u_{\varphi}$ are no longer conserved over time, so we can speak of the
``rate of change'' of the energy and angular momentum of the orbit.
Secondly, at each given time, the values of $-u_t$ and $u_{\varphi}$ are
shifted with respect to their corresponding geodesic values $-u_{0t}$ and
$u_{0\varphi}$. In the case of a circular orbit, the first, ``dissipative''
effect is due entirely to $F^t$ (and $F^{\varphi}$), while the second,
``conservative'' effect is due entirely to $F^{r}$.
To see this, start by defining
${\cal E}\equiv -u_t$ and ${\cal L}\equiv u_{\varphi}$.
The $t$ and $\varphi$ components of Eq.\ (\ref{EOM}) immediately give
%~~~~~~~~~~~~~~~~~~~~~~~~~~~~~~~~~~~~~~~~~~~~~~~~~~~~~~~~~~~~~~~~
\begin{equation} \label{EdotLdot}
\frac{d{\cal E}}{d\tau}=-\mu^{-1}F_t \quad \text{and} \quad
\frac{d{\cal L}}{d\tau}=\mu^{-1}F_\varphi,
\end{equation}
%~~~~~~~~~~~~~~~~~~~~~~~~~~~~~~~~~~~~~~~~~~~~~~~~~~~~~~~~~~~~~~~~
respectively, describing the dissipative effect of the SF. The change of
energy and angular momentum is precisely balanced by the flux of energy
and angular momentum carried away by gravitational waves.
For the conservative effect, consider the $r$ component of Eq.\ (\ref{EOM}),
along with the normalization condition $u_{\alpha}u^{\alpha}=-1$.
Solving these two equations simultaneously for $u^t$ and $u^{\varphi}$
(Recalling $u^r=du^r/d\tau=0$), one obtains
$(u^t)^2=r_0\left(1-r_0F_r/\mu\right)/(r_0-3M)$ and
$(u^\varphi)^2=\left(M/r_0^2-F^r/\mu\right)/(r_0-3M)$, or, through $O(\mu)$,
%~~~~~~~~~~~~~~~~~~~~~~~~~~~~~~~~~~~~~~~~~~~~~~~~~~~~~~~~~~~~~~~~
\begin{equation} \label{EandL}
{\cal E}={\cal E}_0\left[1-\left(\frac{r_0}{2\mu}\right)F_r\right],
\quad\quad
{\cal L}={\cal L}_0\left[1-\left(\frac{r_0^2}{2M\mu}\right) F^r\right].
\end{equation}
%~~~~~~~~~~~~~~~~~~~~~~~~~~~~~~~~~~~~~~~~~~~~~~~~~~~~~~~~~~~~~~~~
Given the radial component of the SF, the last two equations give the
``conservative'' shift in the energy and angular momentum parameters.
It is also useful to look at the shift in the orbital frequency
$\Omega\equiv d\varphi_p/dt=u^{\varphi}/u^t$. Based on the above
expressions for $u^t$ and $u^{\varphi}$ we obtain, through $O(\mu)$,
%~~~~~~~~~~~~~~~~~~~~~~~~~~~~~~~~~~~~~~~~~~~~~~~~~~~~~~~~~~~~~~~~
\begin{equation} \label{Omega}
{\Omega}={\Omega}_0\left[1-\left(\frac{r_0(r_0-3M)}{2M\mu}\right)F_r\right],
\end{equation}
%~~~~~~~~~~~~~~~~~~~~~~~~~~~~~~~~~~~~~~~~~~~~~~~~~~~~~~~~~~~~~~~~
which describes the shift in the ``frequency at infinity'' due to the
conservative piece of the SF.

\subsection{Gauge dependence} \label{subsec:gauge}

It is important to understand how the above quantities depend on the choice
of gauge. Let $h_{\alpha\beta}[\sim O(\mu)]$ be the metric perturbation due
to the particle, given in a specific gauge. Consider an infinitesimal gauge
transformation
%~~~~~~~~~~~~~~~~~~~~~~~~~~~~~~~~~~~~~~~~~~~~~~~~~~~~~~~~~~~~~~~~
\begin{equation} \label{xi}
x^{\mu}\to x'^{\mu}=x^{\mu}+\xi^{\mu}.
\end{equation}
%~~~~~~~~~~~~~~~~~~~~~~~~~~~~~~~~~~~~~~~~~~~~~~~~~~~~~~~~~~~~~~~~
This will change the metric perturbation by an amount
%~~~~~~~~~~~~~~~~~~~~~~~~~~~~~~~~~~~~~~~~~~~~~~~~~~~~~~~~~~~~~~~~
\begin{equation} \label{h-gauge}
\delta_{\xi}h_{\alpha\beta}\equiv h'_{\alpha\beta}-h_{\alpha\beta}=
-(\xi_{\alpha;\beta}+\xi_{\beta;\alpha}).
\end{equation}
%~~~~~~~~~~~~~~~~~~~~~~~~~~~~~~~~~~~~~~~~~~~~~~~~~~~~~~~~~~~~~~~~
It will also induce a change in the SF, given by \cite{Barack:2001ph}
%~~~~~~~~~~~~~~~~~~~~~~~~~~~~~~~~~~~~~~~~~~~~~~~~~~~~~~~~~~~~~~~~
\begin{equation}\label{deltaF}
\delta_{\xi} F_{\alpha}\equiv F'_{\alpha}-F_{\alpha}=
\mu\left[\left[g_{\alpha\lambda}(x_0)+u_{0\alpha}u_{0\lambda}\right]
\frac{D^2\xi^{\lambda}}{D\tau^2}
+R_{\alpha\mu\lambda\nu}(x_0)u_{0}^{\mu}\xi^{\lambda}u_{0}^{\nu}\right].
\end{equation}
%~~~~~~~~~~~~~~~~~~~~~~~~~~~~~~~~~~~~~~~~~~~~~~~~~~~~~~~~~~~~~~~~
Here $g_{\alpha\lambda}(x_0)$ and $R_{\alpha\mu\lambda\nu}(x_0)$ are the
background (Schwarzschild) metric and Riemann tensors, respectively, evaluated
at the particle.

We now assume that the original gauge is ``physically reasonable''. In our case
(a circular equatorial orbit in Schwarzschild), this would mean that the
perturbation $h_{\alpha\beta}$ in that gauge reflects the stationarity of the
problem, and also retains the symmetry of reflection through the equatorial plane.
The Lorenz gauge (see Appendix \ref{AppA} for definition) is an example of such a gauge.
Restricting our discussion to gauge transformations within the family of
``physically reasonable'' gauges, we should clearly require, in our case,
$d\xi_{\alpha}/d\tau=0$, as well as $\xi^{\theta}=0$.
From Eq.\ (\ref{deltaF}) we then get
$\delta_{\xi} F_{t}=\delta_{\xi} F_{\theta}=\delta_{\xi} F_{\varphi}=0$, along with
%~~~~~~~~~~~~~~~~~~~~~~~~~~~~~~~~~~~~~~~~~~~~~~~~~~~~~~~~~~~~~~~~
\begin{equation}\label{deltaFr}
\delta_{\xi} F_{r}=-3\mu({\cal L}^2_0/r_0^4)\xi^{r}.%,\quad\quad
%\delta F_{\theta}=\mu({\cal L}_0/r_0)^2\xi^{\theta}.
\end{equation}
%~~~~~~~~~~~~~~~~~~~~~~~~~~~~~~~~~~~~~~~~~~~~~~~~~~~~~~~~~~~~~~~~
Hence, ``physically reasonable'' gauge transformations may affect the radial
component of the SF (they do so if they have $\xi^r\ne 0$), but not the other
components.

One immediate consequence of the above is that the quantities $d{\cal E}/d\tau$ and
$d{\cal L}/d\tau$ in Eq.\ (\ref{EdotLdot}) are invariant under a gauge transformation
(as should be expected on physical grounds). However, the quantities $\cal E$
and $\cal L$ themselves are {\it not} gauge invariant. To see this, use Eq.\
(\ref{deltaFr}) in Eq.\ (\ref{EandL}), along with
${\cal E}_0\to {\cal E}_0+(d{\cal E}_0/dr_0)\xi^r$, and
${\cal L}_0\to {\cal L}_0+(d{\cal L}_0/dr_0)\xi^r$.
[Note here that the coordinate location $r_0$ (and hence also ${\cal E}_0$ and
${\cal L}_0$) is obviously {\em not} gauge invariant: Under the gauge transformation
(\ref{xi}) we have simply $r_0\to r_0+\xi^r$.]
Denoting $\delta_{\xi} {\cal E}\equiv {\cal E}'-{\cal E}$ and
$\delta_{\xi} {\cal L}\equiv {\cal L}'-{\cal L}$, and keeping only terms linear in $\xi^r$,
one then obtains the following gauge transformation formulas for $\cal E$ and
$\cal L$:
%~~~~~~~~~~~~~~~~~~~~~~~~~~~~~~~~~~~~~~~~~~~~~~~~~~~~~~~~~~~~~~~~
\begin{equation}\label{deltaE}
\delta_{\xi} {\cal E}=\frac{2M/r_0^2}{\sqrt{1-3M/r_0}}\,\xi^r,
\quad\quad
\delta_{\xi} {\cal L}=\frac{2}{\sqrt{1-3M/r_0}}\, \xi^r.
\end{equation}
%~~~~~~~~~~~~~~~~~~~~~~~~~~~~~~~~~~~~~~~~~~~~~~~~~~~~~~~~~~~~~~~~

We can construct orbital parameters that {\it are} invariant under the transformation
(\ref{xi}). One example is the frequency $\Omega$, given in Eq.\ (\ref{Omega}).
Detweiler pointed out recently \cite{Detweiler:2005kq} that the combination
${\cal E}-\Omega {\cal L}\equiv S$ is also gauge invariant (for circular orbits).
For both $\Omega$ and $S$, it is a straightforward exercise to show that, under
the gauge transformation (\ref{xi}),
%~~~~~~~~~~~~~~~~~~~~~~~~~~~~~~~~~~~~~~~~~~~~~~~~~~~~~~~~~~~~~~~~
\begin{equation}\label{deltaOmega}
\delta_{\xi} \Omega=0,  \quad\quad
\delta_{\xi} S= \delta_{\xi}({\cal E}-\Omega {\cal L})=0.
\end{equation}
%~~~~~~~~~~~~~~~~~~~~~~~~~~~~~~~~~~~~~~~~~~~~~~~~~~~~~~~~~~~~~~~~

\subsection{Metric Perturbation in Lorenz gauge} \label{subsec:MP}

In this work we calculate the SF in the Lorenz gauge. This will require knowledge
of the full (retarded) metric perturbation $h_{\alpha\beta}$ in the Lorenz gauge.
We briefly review here the construction of the Lorenz-gauge perturbation, referring
the reader to BL \cite{Barack:2005nr} for further details.

In the formulation by BL, the Lorenz-gauge metric perturbation in Schwarzschild
is constructed through\footnote{We use here the same notation and definitions as in BL,
except for a redefinition $\bar h^{(3)}[{\rm here}]=(1-2M/r)^{-1}\bar h^{(3)}[{\rm
BL}]$. This simplifies slightly the form of the perturbation equations (\ref{FE}),
and is motivated by the fact that $\bar h^{(3)}[{\rm BL paper}]$ vanishes
at the horizon. Also, we fix here an error in Eq.\ (20) of BL: The factor
``$i$'' in the expressions for $h^{lm}_{r\varphi}$ and $h^{lm}_{\theta\varphi}$
is erroneous and we omit it here.}
%~~~~~~~~~~~~~~~~~~~~~~~~~~~~~~~~~~~~~~~~~~~~~~~~~~~~~~~~~~~~~~~~~~~~~~~
\begin{equation}\label{h construction1}
h_{\alpha\beta}=\frac{\mu}{2r}\sum_{l=0}^{\infty}
\sum_{m=-l}^{l} h^{lm}_{\alpha\beta},
\end{equation}
%~~~~~~~~~~~~~~~~~~~~~~~~~~~~~~~~~~~~~~~~~~~~~~~~~~~~~~~~~~~~~~~~~~~~~~~
with
%~~~~~~~~~~~~~~~~~~~~~~~~~~~~~~~~~~~~~~~~~~~~~~~~~~~~~~~~~~~~~~~~~~~~~~~
\begin{eqnarray}\label{h construction2}
h^{lm}_{tt}&=& \left(\bar h^{(1)}+f\bar h^{(6)}\right)Y^{lm}, \nonumber\\
h^{lm}_{tr}&=& f^{-1}\bar h^{(2)}Y^{lm}, \nonumber\\
h^{lm}_{rr}&=& f^{-2}\left(\bar h^{(1)}-f\bar h^{(6)}\right)Y^{lm}, \nonumber\\
h^{lm}_{t\theta}&=& r\left(\bar h^{(4)}Y^{lm}_{\rm V1}
                   +\bar h^{(8)}Y^{lm}_{\rm V2}\right),\nonumber\\
h^{lm}_{t\varphi}&=& r\sin\theta\left(\bar h^{(4)}Y^{lm}_{\rm V2}
                   -\bar h^{(8)}Y^{lm}_{\rm V1}\right),\nonumber\\
h^{lm}_{r\theta}&=& rf^{-1}\left(\bar h^{(5)}Y^{lm}_{\rm V1}
                   +\bar h^{(9)}Y^{lm}_{\rm V2}\right), \nonumber\\
h^{lm}_{r\varphi}&=& rf^{-1}\sin\theta\left(\bar h^{(5)}Y^{lm}_{\rm V2}
                   -\bar h^{(9)}Y^{lm}_{\rm V1}\right), \nonumber\\
h^{lm}_{\theta\theta}&=& r^2\left(\bar h^{(3)}Y^{lm}
 +\bar h^{(7)}Y^{lm}_{\rm T1}+\bar h^{(10)}Y^{lm}_{\rm T2}\right), \nonumber\\
h^{lm}_{\theta\varphi}&=& r^2\sin\theta\left(\bar h^{(7)}Y^{lm}_{\rm T2}
                   -\bar h^{(10)}Y^{lm}_{\rm T1}\right), \nonumber\\
h^{lm}_{\varphi\varphi}&=&
             r^2\sin^2\theta\,\left(\bar h^{(3)}Y^{lm}
             -\bar h^{(7)}Y^{lm}_{\rm T1}-\bar h^{(10)}Y^{lm}_{\rm T2}\right).
\end{eqnarray}
%~~~~~~~~~~~~~~~~~~~~~~~~~~~~~~~~~~~~~~~~~~~~~~~~~~~~~~~~~~~~~~~~~~~~~~~
Here $f=1-2M/r$, and
%~~~~~~~~~~~~~~~~~~~~~~~~~~~~~~~~~~~~~~~~~~~~~~~~~~~~~~~~~~~~~~~~
$Y^{lm}_{\rm V1}$, $Y^{lm}_{\rm V2}$, $Y^{lm}_{\rm T1}$, and
$Y^{lm}_{\rm T2}$ are angular functions constructed from the standard
spherical harmonics $Y^{lm}(\theta,\varphi)$ through
%~~~~~~~~~~~~~~~~~~~~~~~~~~~~~~~~~~~~~~~~~~~~~~~~~~~~~~~~~~~~~~~~~~~~~~~
\begin{eqnarray} \label{eqIII30}
Y^{lm}_{\rm V1} &\equiv & \frac{1}{l(l+1)}\, Y^{lm}_{,\theta}
    \quad \text{(for $l>0$)}, \nonumber\\
Y^{lm}_{\rm V2} &\equiv & \frac{1}{l(l+1)}\,\sin^{-1}\theta\,Y^{lm}_{,\varphi}
     \quad \text{(for $l>0$)},\nonumber\\
Y^{lm}_{\rm T1} &\equiv & \frac{(l-2)!}{(l+2)!} \left[
\sin\theta \left(\sin^{-1}\theta\, Y^{lm}_{,\theta}\right)_{,\theta}
-\sin^{-2}\theta\, Y^{lm}_{,\varphi\varphi}\right]
     \quad \text{(for $l>1$)}, \nonumber\\
Y^{lm}_{\rm T2} &\equiv & \frac{2(l-2)!}{(l+2)!}\,
\left(\sin^{-1}\theta\, Y^{lm}_{,\varphi}\right)_{,\theta}
 \quad \text{(for $l>1$)}.
\end{eqnarray}
%~~~~~~~~~~~~~~~~~~~~~~~~~~~~~~~~~~~~~~~~~~~~~~~~~~~~~~~~~~~~~~~~~~~~~~~
The functions $\bar h^{(i)lm}$ [$i=1,\ldots,10$; the indices $l,m$ were
omitted in Eq.\ (\ref{h construction2}) for brevity] depend on $r$ and $t$
only, and form our basic set of perturbation fields. These time-radial
fields are obtained as solutions to the coupled set of hyperbolic
[in a 2-dimensional (2-D) sense], scalar-like equations
%~~~~~~~~~~~~~~~~~~~~~~~~~~~~~~~~~~~~~~~~~~~~~~~~~~~~~~~~~~~~~~~~~~~~~~~
\begin{equation}\label{FE}
\square \bar h^{(i)lm}+
{\cal M}^{(i)l}_{\;(j)}\bar h^{(j)lm}=S^{(i)lm}\quad (i=1,\ldots,10).
\end{equation}
%~~~~~~~~~~~~~~~~~~~~~~~~~~~~~~~~~~~~~~~~~~~~~~~~~~~~~~~~~~~~~~~~~~~~~~~
Here `$\square$' represents the 2-D scalar field operator,
$\square= \partial_{uv}+ V(r)$, where $v$ and $u$ are the standard
Eddington--Finkelstein null coordinates, and the potential is
$V(r)=\frac{1}{4}f\left[2M/r^3+l(l+1)/r^2\right]$.
The ``coupling'' terms ${\cal M}^{(i)l}_{\;(j)}\bar h^{(j)lm}$ involve
first derivatives of the $\bar h^{(j)lm}$'s at most (no second derivatives),
and $S^{(i)lm}$ are source terms. Both ${\cal M}^{(i)l}_{\;(j)}\bar h^{(j)lm}$
and $S^{(i)lm}$ are given explicitly in Appendix \ref{AppA} for our circular
orbit case. In addition to the evolution equations (\ref{FE}), the functions
$\bar h^{(i)lm}$ also satisfy four elliptic equations, which stem from the
gauge conditions. These ``constraint'' equations are also given in
Appendix \ref{AppA}. The set (\ref{FE}) incorporates ``divergence dissipating''
terms, which guarantee that violations of the Lorenz gauge conditions are
efficiently damped during the evolution \cite{Barack:2005nr}.

BL demonstrated in \cite{Barack:2005nr} how Eqs.\ (\ref{FE}) can be evolved numerically
(for $l\geq 2)$, in the time domain, with a delta-function source term representing
an orbiting point particle. They also demonstrated that the solutions preserve
the Lorenz gauge condition throughout the late-time evolution (initial violations of
the gauge conditions are suppressed over a time-scale $\sim M$).

The perturbation modes $l=0$ and $l=1$ require a separate treatment.
For $l=0$, the set (\ref{FE}) reduces to 4 equations (for $\bar h^{(1,2,3,6)}$),
which describe the spherically-symmetric, monopole mass perturbation.
Sec.\ III-D of BL gives the solution for the Lorenz-gauge monopole perturbation in
analytic form (based on analysis by Detweiler and Poisson \cite{Detweiler:2003ci}).
For $l=1,m=\pm 1$, the set (\ref{FE}) reduces to 6 equations
(for $\bar h^{(1\text{---}6)}$).
These describe the rotational dipole piece of the perturbation, which is
(in Newtonian terms) due to the motion of the black hole about the center
of mass of the black hole--particle system. The Lorenz-gauge solution for
this mode was obtained by Detweiler and Poisson \cite{Detweiler:2003ci}, using a procedure
that reduces the problem to the solution of 3 coupled ordinary differential
equations (see also Ori \cite{Ori:2003cm} for a fully analytic weak-field treatment).
Finally, for $l=1,m=0$---the axisymmetric dipole perturbation---the set (\ref{FE})
reduces to a single equation (for $\bar h^{(8)}$), describing the perturbation in the
angular momentum due to the particle. Analytic solution for this mode was
obtained long ago be Zerilli \cite{Zerilli:1971wd}, and identified in \cite{Detweiler:2003ci}
as a Lorenz gauge solution. BL later obtained analytic Lorenz-gauge solutions
for {\it all} axisymmetric ($m=0$) modes with odd $l$. These are given in Sec. III-C
of BL.

In what follows we will prescribe the construction of the gravitational SF
directly in terms of the functions $\bar h^{(i)lm}$.

\subsection{Self force via the mode-sum method}

In the mode sum scheme, the gravitational SF is calculated through
\cite{Barack:2001bw,Barack:2001gx,Barack:2002bt,Barack:2002mh}
%~~~~~~~~~~~~~~~~~~~~~~~~~~~~~~~~~~~~~~~~~~~~~~~~~~~~~~~~~~~~~~~~~~~~~~~
\begin{equation}\label{ModeSum}
F^{\alpha}=\sum_{l=0}^{\infty}
\left[\left[F_{\rm full}^{\alpha l}(x_0)\right]_{\pm}-A^{\alpha}_{\pm}L-B^{\alpha}\right],
\end{equation}
%~~~~~~~~~~~~~~~~~~~~~~~~~~~~~~~~~~~~~~~~~~~~~~~~~~~~~~~~~~~~~~~~~~~~~~~
where $L\equiv l+1/2$, and
$F_{\rm full}^{\alpha l}$ are the modes of the ``full'' force, obtained
from the modes of the full (retarded) metric perturbation in a manner described
below. The subscript $\pm$ refers to the two possible values of $F_{\rm full}^{\alpha l}$
at $x_0$, resulting from taking one-sided radial derivatives of the metric perturbation
from either $r\to r_0^+$ or $r\to r_0^-$ (recall that the modes of the Lorenz-gauge
metric perturbation are continuous, but their gradients generally have a finite jump
discontinuity across the particle).
$A^{\alpha}_{\pm}$ and $B^{\alpha}$ are the ``regularization
parameters'', which are known analytically. For circular equatorial geodesics in
Schwarzschild we have $A^{\alpha}_{\pm}=B^{\alpha}=0$ for $\alpha=t,\theta,\varphi$,
and
%~~~~~~~~~~~~~~~~~~~~~~~~~~~~~~~~~~~~~~~~~~~~~~~~~~~~~~~~~~~~~~~~~~~~~~~
\begin{equation} \label{A}
A^r_{\pm}=\mp \frac{\mu^2}{r_0^2}\left(1-\frac{3M}{r_0}\right)^{1/2},
\end{equation}
\begin{equation}\label{B}
B^r=\frac{\mu^2r_0{\cal E}_0^2}{\pi({\cal L}_0^2+r_0^2)^{3/2}}
\left[\hat E(w)-2\hat K(w)\right],
\end{equation}
%~~~~~~~~~~~~~~~~~~~~~~~~~~~~~~~~~~~~~~~~~~~~~~~~~~~~~~~~~~~~~~~~~~~~~~~
where $\hat K(w)\equiv \int_{0}^{\pi/2}(1-w\sin^2x)^{-1/2}dx$ and
$\hat E(w)\equiv \int_{0}^{\pi/2}(1-w\sin^2x)^{1/2}dx$
are complete elliptic
integrals of the first and second kind, respectively, and $w\equiv (r_0/M-2)^{-1}$.
The final SF $F^{\alpha}$ can be calculated using either ``external'' ($+$) or
``internal'' ($-$) values; The difference
$\left[F_{\rm full}^{\alpha l}(x_0)\right]_{\pm}-A^{\alpha}_{\pm}L$
is guaranteed to be direction-independent. One can also use the average value,
$\bar F_{\rm full}^{\alpha l}\equiv
\left\{[F_{\rm full}^{\alpha l}]_{+}+[F_{\rm full}^{\alpha l}]_{-}\right\}/2$,
in terms of which the mode-sum formula takes the more compact form
%~~~~~~~~~~~~~~~~~~~~~~~~~~~~~~~~~~~~~~~~~~~~~~~~~~~~~~~~~~~~~~~~~~~~~~~
\begin{equation}\label{ModeSum2}
F^{\alpha}=\sum_{l=0}^{\infty}
\left[\bar F_{\rm full}^{\alpha l}(x_0)-B^{\alpha}\right].
\end{equation}
%~~~~~~~~~~~~~~~~~~~~~~~~~~~~~~~~~~~~~~~~~~~~~~~~~~~~~~~~~~~~~~~~~~~~~~~

Recall here we are interested in the $r$ and $t$ components of the SF.
For the $r$ component, the contribution from the individual full modes is
$\propto L$ at large $L$, and the sum over $\left[F_{\rm full}^{\alpha l}(x_0)\right]_{\pm}$
diverges. However, the contribution from the ``regularized'' modes
$\left[F_{\rm full}^{r l}(x_0)\right]_{\pm}-A^{r}_{\pm}L-B^{r}$
falls off as $\propto L^{-2}$ at large $L$, and their sum converges (and gives
the correct physical SF). The regularized modes admit the large-$l$
expansion
%~~~~~~~~~~~~~~~~~~~~~~~~~~~~~~~~~~~~~~~~~~~~~~~~~~~~~~~~~~~~~~~~~~~~~~~
\begin{equation}\label{Freg}
F_{\rm reg}^{rl}\equiv \left[F_{\rm full}^{r l}(x_0)\right]_{\pm}-A^{r}_{\pm}L-B^{r}
=\frac{D_2}{L^2}+\frac{D_4}{L^4}+\cdots,
\end{equation}
%~~~~~~~~~~~~~~~~~~~~~~~~~~~~~~~~~~~~~~~~~~~~~~~~~~~~~~~~~~~~~~~~~~~~~~~
where $D_2,D_4,\ldots$ are coefficients that may depend on the orbital
parameters, but not on $L$.
As for the $t$ component: In the special case of a circular orbit
the temporal full-force modes require no regularization (recall $A^t=B^t=0$), and
their sum converges. In fact, in this case the
mode sum can be shown to converge {\rm exponentially} at large $l$ (this
will be demonstrated experimentally in Sec.\ \ref{Sec:validation} below).
For later convenience we write
%~~~~~~~~~~~~~~~~~~~~~~~~~~~~~~~~~~~~~~~~~~~~~~~~~~~~~~~~~~~~~~~~~~~~~~~
\begin{equation}\label{Ftreg}
F_{\rm reg}^{tl}\equiv F_{\rm full}^{tl},
\end{equation}
%~~~~~~~~~~~~~~~~~~~~~~~~~~~~~~~~~~~~~~~~~~~~~~~~~~~~~~~~~~~~~~~~~~~~~~~
and the mode-sum formula becomes, for either the $r$ or the $t$ component,
%~~~~~~~~~~~~~~~~~~~~~~~~~~~~~~~~~~~~~~~~~~~~~~~~~~~~~~~~~~~~~~~~~~~~~~~
\begin{equation}\label{ModeSum3}
F^{\alpha}=\sum_{l=0}^{\infty} F_{\rm reg}^{\alpha l}.
\end{equation}
%~~~~~~~~~~~~~~~~~~~~~~~~~~~~~~~~~~~~~~~~~~~~~~~~~~~~~~~~~~~~~~~~~~~~~~~

\subsection{Construction of the full-force modes}

We now describe the construction of the full modes
$\left[F_{\rm full}^{\alpha l}(x_0)\right]_{\pm}$ out of the Lorenz-gauge
perturbation fields $\bar h^{(i)lm}$.

First, following \cite{Barack:2001bw}, define the ``full force field'' as
a tensor field at arbitrary spacetime point $x$, for a given (fixed) worldline point
$x_0$ (where the SF is to be calculated):
%~~~~~~~~~~~~~~~~~~~~~~~~~~~~~~~~~~~~~~~~~~~~~~~~~~~~~~~~~~~~~~~~~~~~~~~
\begin{equation}\label{Ffull}
F_{\rm full}^{\alpha}(x;x_0)=\mu k^{\alpha\beta\gamma\delta} \bar h_{\beta\gamma;\delta}.
\end{equation}
%~~~~~~~~~~~~~~~~~~~~~~~~~~~~~~~~~~~~~~~~~~~~~~~~~~~~~~~~~~~~~~~~~~~~~~~
Here $\bar h_{\alpha\beta}=h_{\alpha\beta}
-\frac{1}{2}g_{\alpha\beta}g^{\mu\nu}h_{\mu\nu}$ is
the ``trace-reversed'' Lorenz-gauge metric perturbation at $x$, and
%~~~~~~~~~~~~~~~~~~~~~~~~~~~~~~~~~~~~~~~~~~~~~~~~~~~~~~~~~~~~~~~~~~~~~~~
\begin{equation}\label{k}
k^{\alpha\beta\gamma\delta}(x;x_0)=
         g^{\alpha\delta}u^{\beta}u^{\gamma}/2
        -g^{\alpha\beta}u^{\gamma}u^{\delta}
        -u^{\alpha}u^{\beta}u^{\gamma}u^{\delta}/2
        +u^{\alpha}g^{\beta\gamma}u^{\delta}/4
        +g^{\alpha\delta}g^{\beta\gamma}/4,
\end{equation}
%~~~~~~~~~~~~~~~~~~~~~~~~~~~~~~~~~~~~~~~~~~~~~~~~~~~~~~~~~~~~~~~~~~~~~~~
where $g^{\alpha\delta}$ is the background metric at $x$, and $u^{\alpha}$
are the values of the contravariant components of the four-velocity at $x_0$
(treated as fixed coefficients). Obviously, the full force $F_{\rm full}^{\alpha}$
diverges for $x\to x_0$ (like $\sim$distance$^{-2}$).

Next, expand the metric perturbation in tensor harmonics as in Eq.\
(\ref{h construction1}), and substitute in Eq.\ (\ref{Ffull}).
%%~~~~~~~~~~~~~~~~~~~~~~~~~~~~~~~~~~~~~~~~~~~~~~~~~~~~~~~~~~~~~~~~~~~~~~
%\begin{equation} \label{IV-20}
%F^{\alpha}_{\rm full}=
%\mu \sum_{i\ell m}k^{\alpha\beta\gamma\delta}
%\left[\bar{h}^{(i)\ell m}(r,t)\,Y^{(i)\ell m}_{\beta\gamma}(\theta,\varphi)
%\right]_{;\delta}
%\equiv \mu \sum_{i\ell m} F^{\alpha(i)\ell m}(t,r,\theta,\varphi),
%\end{equation}
%%~~~~~~~~~~~~~~~~~~~~~~~~~~~~~~~~~~~~~~~~~~~~~~~~~~~~~~~~~~~~~~~~~~~~~~
%Start with
%\begin{equation}
%F^{\alpha}=k^{\alpha\beta\gamma\delta} \bar h_{\beta\gamma;\delta}=
%\frac{1}{r_0^2}\sum_{i\ell m}k^{\alpha\beta\gamma\delta}
%\left[\bar{h}^{(i)\ell m}(r,t)\,Y^{(i)\ell m}_{\beta\gamma}(\theta,\varphi)
%\right]_{;\delta},
%\end{equation}
%where I assume the ``constant'' extension of $u^{\alpha}$ and natural
%extension of $g^{\alpha\beta}$. [Recall that can allow for $O(\delta x^2)$ deviations
%without changing values of RP \cite{BO2003}.]
Taking the limits $r\to r_0$ and $t\to t_0$ (but maintaining the $\theta,\varphi$
dependence), the full force takes the form
%%~~~~~~~~~~~~~~~~~~~~~~~~~~~~~~~~~~~~~~~~~~~~~~~~~~~~~~~~~~~~~~~~~~~~~~
\begin{eqnarray} \label{Ffull2}
\left[F_{\rm full}^{\alpha}(\theta,\varphi;r_0,t_0)\right]_{\pm}
&=&\frac{\mu^2}{r_0^2}\sum_{l=0}^{\infty}\sum_{m=-l}^{l}
\left\{
f_{0\pm}^{\alpha lm} Y^{lm}+
f_{1\pm}^{\alpha lm} \sin^2\theta\, Y^{lm}+
f_{2\pm}^{\alpha lm} \cos\theta\sin\theta\, Y^{lm}_{,\theta}\right.\nonumber\\
& &
+\left.f_{3\pm}^{\alpha lm} \sin^2\theta\, Y^{lm}_{,\theta\theta}
+f_{4\pm}^{\alpha} (\cos\theta Y^{lm}-\sin\theta Y^{lm}_{,\theta})\right.\nonumber\\
& &
+\left.f_{5\pm}^{\alpha lm} \sin\theta\, Y^{lm}_{,\theta}+
f_{6\pm}^{\alpha lm} \sin^3\theta\, Y^{lm}_{,\theta}+
f_{7\pm}^{\alpha lm} \cos\theta\sin^2\theta\, Y^{lm}_{,\theta\theta}
\right\},
\end{eqnarray}
%%~~~~~~~~~~~~~~~~~~~~~~~~~~~~~~~~~~~~~~~~~~~~~~~~~~~~~~~~~~~~~~~~~~~~~~
where $Y^{lm}(\theta,\varphi)$ are the spherical harmonics, and the
coefficients $f_{n\pm}^{\alpha lm}$ are constructed from the perturbation fields
$\bar h^{(i)lm}$ and their first $r$ and $t$ derivatives, all evaluated at $x_0$.
The labels $+/-$ correspond to taking external/internal $r$ derivatives, respectively.
The explicit expressions for the $f_{n\pm}^{\alpha lm}$'s are quite lengthy, and we give
them separately, in Appendix \ref{AppB}.

The individual $l$ modes in Eq.\ (\ref{Ffull2}) are {\em not} quite yet the full force
modes needed in the mode-sum formula (\ref{ModeSum}): A little complication arises
because the mode-sum formula requires the decomposition of the full force in
{\em scalar} harmonics. That is, to obtain the modes
$\left[F_{\rm full}^{\alpha l}(x_0)\right]_{\pm}$
for use in Eq.\ (\ref{ModeSum}) we are required to ignore the vectorial nature of the
full force, and expand each of its components in {\em scalar} harmonics.
To obtain this, we need only re-expand all angular functions in Eq.\ (\ref{Ffull2})
in spherical harmonics, and rearrange the terms in the sum. The necessary
expansion formulas are given in Appendix \ref{AppC}. We find that each of the
tensor-harmonic $l$ modes in Eq.\ (\ref{Ffull2}) couples to a finite number of
scalar harmonics (there is no coupling between different $m$ modes). For the $r$
component, each tensor-harmonic $l$ generally couples to 5 scalar harmonic modes
$l'$ with $l-2\leq l'\leq l+2$; for the $t$ component, it generally couples to the
7 modes $l-3\leq l'\leq l+3$. It should be commented that this finite coupling
is characteristic of {\it all} (equatorial) orbits in Schwarzschild, not necessarily
circular (although in general both $t$ and $r$ components would involve coupling
to 7 scalar harmonics). We also comment that, in our circular orbit case, it is
not quite necessary to expand the $t$ component in scalar harmonics, since this
component requires no regularization and so the mode-sum can be evaluated directly
using the tensor harmonic expansion. In this work we nevertheless choose to expand
the $t$ component too in spherical harmonics, for two reasons: Firstly, this
would allow us to test our general treatment of the coupling between modes,
since the computed value of the $t$ component of the SF could readily be verified
by comparing with the flux of radiated energy. Secondly, our code could then be
more easily adapted to deal with eccentric orbits, where scalar-harmonic
decomposition of both $r$ and $t$ components is necessary.

Hence, we now re-expand Eq.\ (\ref{Ffull2}) in spherical harmonics (using the relations
given in Appendix \ref{AppC}), then rearrange the terms in the sum by collecting all
terms with the same scalar-harmonic multipole number $l$, and, finally, set
$\theta\to \theta_0$ and $\varphi\to \varphi_0$.
%This yields a finite value for each $l$, although for the $r$ component this
%value depends of whether the $r$ derivatives of the perturbation [Eq.\ (\ref{fr})]
%where taken from $r_0^{+}$ or from $r_0^{-}$.
The resulting ``scalar harmonic'' $l$ modes of the full force
%, which are the ones to be incorporated in the mode
%sum formula (\ref{ModeSum}),
take the final form
%%~~~~~~~~~~~~~~~~~~~~~~~~~~~~~~~~~~~~~~~~~~~~~~~~~~~~~~~~~~~~~~~~~~~~~~~
%\begin{equation}\label{Ffull3}
%F_{\rm full}^{\alpha}=\sum_{l=0}^{\infty} (F_{\rm full}^{\alpha l})_{\pm}
%\end{equation}
%%~~~~~~~~~~~~~~~~~~~~~~~~~~~~~~~~~~~~~~~~~~~~~~~~~~~~~~~~~~~~~~~~~~~~~~~
%~~~~~~~~~~~~~~~~~~~~~~~~~~~~~~~~~~~~~~~~~~~~~~~~~~~~~~~~~~~~~~~~~~~~~~~
\begin{equation}\label{Flfull}
\left[F_{\rm full}^{\alpha l}(x_0)\right]_{\pm}= \frac{\mu^2}{r_0^2}\sum_{m=-l}^{l} Y^{lm}(\theta_0,\varphi_0)\times
\left\{{\cal F}^{\alpha l-3,m}_{(-3)}+{\cal F}^{\alpha l-2,m}_{(-2)}+{\cal F}^{\alpha l-1,m}_{(-1)}
+{\cal F}^{\alpha lm}_{(0)}+ {\cal F}^{\alpha l+1,m}_{(+1)}+{\cal F}^{\alpha l+2,m}_{(+2)}
+{\cal F}^{\alpha l+3,m}_{(+3)}\right\}_{\pm},
\end{equation}
%~~~~~~~~~~~~~~~~~~~~~~~~~~~~~~~~~~~~~~~~~~~~~~~~~~~~~~~~~~~~~~~~~~~~~~~
%(the subscript $\pm$ refers to one-sided derivatives taken from $r\to r_0^{\pm})$,
where
%~~~~~~~~~~~~~~~~~~~~~~~~~~~~~~~~~~~~~~~~~~~~~~~~~~~~~~~~~~~~~~~~~~~~~~~
\begin{eqnarray} \label{calF}
{\cal F}^{\alpha lm}_{(-3)} &=&
\zeta_{(+3)}^{lm} f_{6\pm}^{\alpha lm}
+ \xi_{(+3)}^{lm}f_{7\pm}^{\alpha lm},
\nonumber \\
{\cal F}^{\alpha lm}_{(-2)} &=&
\alpha_{(+2)}^{lm} f_{1\pm}^{\alpha lm}
+ \beta_{(+2)}^{lm} f_{2\pm}^{\alpha lm}
+ \gamma_{(+2)}^{lm} f_{3\pm}^{\alpha lm},
\nonumber \\
{\cal F}^{\alpha lm}_{(-1)} &=&
\epsilon_{(+1)}^{lm} f_{4\pm}^{\alpha lm}
+ \delta_{(+1)}^{lm} f_{5\pm}^{\alpha lm}
+ \zeta_{(+1)}^{lm} f_{6\pm}^{\alpha lm}
+ \xi_{(+1)}^{lm} f_{7\pm}^{\alpha lm},
\nonumber\\
{\cal F}^{\alpha lm}_{(0)} &=&
f_{0\pm}^{\alpha lm}
+ \alpha_{(0)}^{\alpha lm} f_{1\pm}^{\alpha lm}
+ \beta_{(0)}^{lm} f_{2\pm}^{\alpha lm}
+ \gamma_{(0)}^{lm} f_{3\pm}^{\alpha lm},
\nonumber \\
{\cal F}^{\alpha lm}_{(+1)} &=&
\epsilon_{(-1)}^{lm} f_{4\pm}^{\alpha lm}
+ \delta_{(-1)}^{lm} f_{5\pm}^{\alpha lm}
+ \zeta_{(-1)}^{lm} f_{6\pm}^{\alpha lm}
+ \xi_{(-1)}^{lm} f_{7\pm}^{\alpha lm},
\nonumber \\
{\cal F}^{\alpha lm}_{(+2)} &=&
\alpha_{(-2)}^{lm} f_{1\pm}^{\alpha lm}
+ \beta_{(-2)}^{lm} f_{2\pm}^{\alpha lm}
+ \gamma_{(-2)}^{lm} f_{3\pm}^{\alpha lm},
\nonumber \\
{\cal F}^{\alpha lm}_{(+3)} &=&
\zeta_{(-3)}^{lm} f_{6\pm}^{\alpha lm}
+ \xi_{(-3)}^{lm}f_{7\pm}^{\alpha lm},
\end{eqnarray}
%~~~~~~~~~~~~~~~~~~~~~~~~~~~~~~~~~~~~~~~~~~~~~~~~~~~~~~~~~~~~~~~~~~~~~~~
%Here all functions $f^{lm}_n$ are evaluated at the location of the particle ($r\to r_0$
%and $t\to t_0$), and
with the various coefficients $\alpha,\beta,\gamma,\delta,\epsilon,\zeta$, and $\xi$ given in
Appendix \ref{AppC}. Note $f_{6\pm}^r=f_{7\pm}^r=0$, so that
Eq.\ (\ref{Flfull}) simplifies considerably for the $r$ component. The spherical harmonic
$Y^{lm}(\theta_0,\varphi_0)$ is given explicitly, for $\theta_0=\pi/2$, by
%~~~~~~~~~~~~~~~~~~~~~~~~~~~~~~~~~~~~~~~~~~~~~~~~~~~~~~~~~~~~~~~~~~~~~~~
\begin{equation} \label{Ylm}
Y^{lm}(\pi/2,\varphi_0)=
e^{im\varphi_0}\times\left\{\begin{array}{ll}
(-1)^{(l+m)/2} \left[\frac{(2l+1)(l+m-1)!!(l-m-1)!!}{4\pi(l+m)!!(l-m)!!}\right]^{1/2},
& l-m \  \text{even} \\
0, & l-m \ \text{odd}.
\end{array}\right.
\end{equation}
%~~~~~~~~~~~~~~~~~~~~~~~~~~~~~~~~~~~~~~~~~~~~~~~~~~~~~~~~~~~~~~~~~~~~~~~
Hence, for given $l$, only $m$ modes with even $l-m$ contribute to the sum in
Eq.\ (\ref{Flfull}). A further simplification arises since the individual $m$ modes
in the sum in Eq.\ (\ref{Flfull}) are invariant under $m\to -m$, which allows us
to fold the $m<0$ part of the sum over to $m>0$. In practice, therefore, one
is required to compute only $l/2+1$ $m$-modes for each even $l$-mode, and $(l+1)/2$
$m$-modes for each odd $l$-mode.
Finally, note
$\alpha_{(+2)}^{lm} = \beta_{(+2)}^{lm} = \gamma_{(+2)}^{lm}
= \epsilon_{(+1)}^{lm} = \delta_{(+1)}^{lm} = \zeta_{(+1)}^{lm}
= \xi_{(+1)}^{lm} = \zeta_{(+3)}^{lm} = \xi_{(+3)}^{lm} = 0$
for $l<0$, such that no functions $f_n^{\alpha lm}$ with $l<0$ occur in Eq.\ (\ref{Flfull}).

%$\zeta_{(+3)}^{l-3,m}=\xi_{(+3)}^{l-3,m}=0$ for $m=l,l-2$, as well as
%$\alpha_{(+2)}^{l-2,m}=\beta_{(+2)}^{l-2,m}=\gamma_{(+2)}^{l-2,m}=
%\epsilon_{(+1)}^{l-1,m}=\delta_{(+1)}^{l-1,m}=\zeta_{(+1)}^{l-1,m}=
%\xi_{(+1)}^{l-1,m}=0$ for $m=l$. This automatically guarantees that no
%functions $f_{n}^{\alpha lm}$ with $l<0$ ever occur in Eqs.\ (\ref{calF}).

\subsection{Summary of prescription for constructing the Lorenz-gauge SF}

Start by calculating the Lorenz-gauge perturbation fields $\bar h^{(i)lm}$
(10 of which for each $l\geq 2,m$), by solving (numerically) the field equations (\ref{FE}).
Obtain the modes $l=0,1$ of $\bar h^{(i)lm}$ through the procedure described at the
end of Sec.\ \ref{subsec:MP}.
Construct the functions $f_n^r$ and $f_n^t$ using the formulas in Appendix \ref{AppB},
and then construct the quantities $\cal F$ through Eqs.\ (\ref{calF}).
Use these in Eq.\ (\ref{Flfull}) to obtain
the scalar-harmonic $l$-modes of the full force, $\left[F_{\rm full}^{\alpha l}(x_0)\right]_{\pm}$.
Incorporate the full force modes in the mode-sum formula (\ref{ModeSum})
[or (\ref{ModeSum2})] to obtain the SF.

%%%%%%%%%%%%%%%%%%%%%%%%%%%%%%%%%%%%%%%%%%%%%%%%%%%%%
\section{Numerical implementation}\label{Sec:numerics}
%%%%%%%%%%%%%%%%%%%%%%%%%%%%%%%%%%%%%%%%%%%%%%%%%%%%%

In this section we summarize the numerical method used to calculate the SF
through the mode-sum formula (\ref{ModeSum}). The evolution of the Lorenz-gauge
field equations (\ref{FE}) was described in BL, but we will briefly review this
method here---mainly in order to supplement a few details of the finite difference scheme
left out in BL. We then describe the numerical construction of the regularized SF
modes, and the technique used to evaluated the infinite sum over $l$.
There are three main sources of numerical error in our calculation:
(i) error from the finite-grid discretization (which, in the procedure described
below, comes from the error in a Richardson-type extrapolation to zero grid size);
(ii) error from estimation of the large-$l$ tail of the mode-sum series; and
(iii) error from residual spurious waves resulting from the imperfection of initial data.
We explain how all these errors are monitored and controlled in our
calculation.

%To derive the self-force on a particle, we need to
%calculate the metric perturbations induced by itself.
%For $\ell \ge 2$, we solve the field equations
%in the time domain under the Lorenz gauge condition.
%To do that, we apply the Lousto-Price's scheme
%\cite{Lousto:1997wf} based on the finite difference method.
%In the implementation of this scheme in the field equations
%under the Lorenz gauge condition,
%we have to pay attention to the terms containing
%$v$-derivatives \cite{BL2005}.
%On the other hand, we deal with the $\ell=0,1$ modes separately.
%Once we obtain the solution of the field equations,
%we calculate the self-force according to the mode-sum
%scheme shown in previous section.
%In this section, we briefly summarize the scheme
%which we used to solve the field equations
%and to calculate the self-force in our code.

\subsection{Metric perturbation: finite-difference scheme for $l \geq 2$}
\label{sec:solve-FE}

To solve Eqs.\ (\ref{FE}) for the various modes $l\geq 2$ we use characteristic
time-domain evolution on a fixed 2-D staggered double-null grid based
on $v,u$ coordinates. The numerical domain is depicted in Fig.\ 2 of BL. The evolution
starts with characteristic initial data on two initial ``rays'' $v=v_0$ and
$u=u_0$, taken such that the vortex $v_0,u_0$ corresponds to $r=r_0$ with
initial time $t_0=0$. The circular orbit worldline then traces a straight vertical
line through the grid, connecting the ``lower'' and ``upper'' vertices
(see Fig.\ 2 of BL). In this setup the worldline cuts through grid points. This does
not cause problem, since the Lorenz-gauge perturbation modes are continuous
at the worldline.
For initial data we simply take $\bar h^{(i)lm}=0$ along $v=v_0$ and $u=u_0$, for
all $i$. This sparks a burst of spurious radiation at the initial vortex,
which, however, dies off efficiently over time and leaves very little trace
after 1--2 orbital periods of evolution (we demonstrate this below).

\begin{figure}[htb]
\includegraphics[width=5cm]{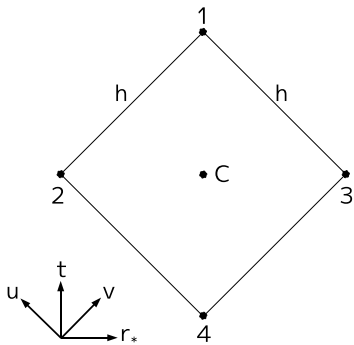}
\caption{A numerical grid cell, of dimensions $h\times h$ (see description in the text).
Our 2-D grid is based on characteristic (Eddington--Finkelstein) coordinates $v$ and $u$.
These are related to the Schwarzschild coordinates through $v=t+r_*$ and $u=t-r_*$,
where $r_*=r+2M\ln[r/(2M)-1]$.}
\label{fig:GridCell}
\end{figure}

Our finite-difference scheme (particularly the handling of the delta function
source term) is based on the method first introduced by Lousto and Price
\cite{Lousto:1997wf} and later implemented by a number of authors
\cite{Barack:2000zq,Barack:2002ku,Martel:2001yf,Martel:2003jj}.
In this method, the finite-difference equation is obtained by approximating
the (2-D) {\em integral} of both sides of the field equation over a grid cell,
at a suitable accuracy. This automatically deals with the delta-function
singularity on the right-hand side of the equation.
For the following discussion, consider Fig.\ \ref{fig:GridCell}:
Suppose that we have already solved for all $\bar h^{(i)lm}$'s at the grid points
numbered $2$, $3$, and $4$. Denote the values calculated at these points by
$\bar h^{(i)}_2$, $\bar h^{(i)}_3$, and $\bar h^{(i)}_4$, respectively (we omit here the
indices $l,m$ for brevity), and let the sides of the grid cell be $\Delta v=\Delta
u=h$. We are interested in obtaining $\bar h^{(i)}_1$, the value of the
$\bar h^{(i)}$'s at point 1.
%~~~~~~~~~~~~~~~~~~~~~~~~~~~~~~~~~~~~~~~~~~~~~~~~~~~~~~~~~~~~~~~~~~~~~~~~~~~~~~
%~~~~~~~~~~~~~~~~~~~~~~~~~~~~~~~~~~~~~~~~~~~~~~~~~~~~~~~~~~~~~~~~~~~~~~~~~~~~~~

%The field equations are given by (\ref{FE}),
%\begin{equation}
%\frac{\partial^2 \bar{h}^{(i)}}{\partial v \partial u}
%+{\cal M}_{(j)}^{(i)}(r) \bar{h}^{(j)}
%= S^{(i)},
%\end{equation}
%where we omit $\ell$ and $m$ for simplicity.
%We solve these equations in the numerical domain
%with fixed grid size (like Fig.2 in Barack-Lousto's paper).
%To use the finite difference method, we integrate them over
%a grid cell.
Consider first the principal part of the field equations (\ref{FE}).
For any of the ten $i$'s, it reads $\bar h^{(i)}_{,uv}$. This term is integrated
exactly over the grid cell, to give
%~~~~~~~~~~~~~~~~~~~~~~~~~~~~~~~~~~~~~~~~~~~~~~~~~~~~~~~~~~~~~~~~~~~~~~~
\begin{equation}
\int\int_{\rm cell} \bar h^{(i)}_{,uv}\, dudv
=
\bar{h}_1^{(i)} - \bar{h}_2^{(i)}
- \bar{h}_3^{(i)} + \bar{h}_4^{(i)}.
\label{eq:FDM-uv}
\end{equation}
%~~~~~~~~~~~~~~~~~~~~~~~~~~~~~~~~~~~~~~~~~~~~~~~~~~~~~~~~~~~~~~~~~~~~~~~
Since the $\bar h^{(i)}$'s are continuous at the worldline, the above integral
holds even for grid cells crossed by the particle.
%where the lower indecies of $\bar{h}_n^{(i)}$ show
%the location in Fig.~\ref{fig:grid1}.
The remaining part of the left-hand side of the field equations (\ref{FE}) includes
three types of terms, of the form $V_1(r)\bar{h}^{(i)}$,
$V_2(r) \bar{h}^{(i)}_{,r}$, and $V_3(r) \bar{h}^{(i)}_{,v}$, where
the $V(r)$'s are known radial functions.
As for terms of the first two types, we can approximate their integrals
over the grid cell as
%~~~~~~~~~~~~~~~~~~~~~~~~~~~~~~~~~~~~~~~~~~~~~~~~~~~~~~~~~~~~~~~~~~~~~~~
\begin{equation} \label{eq:FDM-pot}
\int\int_{\rm cell} V_1(r)\bar{h}^{(i)}\, dudv
=
\frac{1}{2}h^2 V_1(r_c)
( \bar h_2^{(i)} + \bar h_3^{(i)} )
+ \left\{ \begin{array}{ll}
O(h^3) & \mbox{(particle)}, \\
O(h^4) & \mbox{(no particle)},
\end{array} \right.
\end{equation}
%~~~~~~~~~~~~~~~~~~~~~~~~~~~~~~~~~~~~~~~~~~~~~~~~~~~~~~~~~~~~~~~~~~~~~~~
%~~~~~~~~~~~~~~~~~~~~~~~~~~~~~~~~~~~~~~~~~~~~~~~~~~~~~~~~~~~~~~~~~~~~~~~
\begin{equation}
\int\int_{\rm cell} V_2(r)\bar{h}^{(i)}_{,r}\, dudv
=
h f^{-1}(r_c)V_2(r_c)( \bar h_3^{(i)} - \bar h_2^{(i)} )
+ \left\{ \begin{array}{ll}
O(h^3) & \mbox{(particle)}, \\
O(h^4) & \mbox{(no particle)},
\end{array} \right. \label{eq:FDM-rdev}
\end{equation}
%~~~~~~~~~~~~~~~~~~~~~~~~~~~~~~~~~~~~~~~~~~~~~~~~~~~~~~~~~~~~~~~~~~~~~~~
where
$r_c$ is the value of $r$ at point C in the middle of the cell
(see Fig.\ \ref{fig:GridCell}). The case indicated as ``particle'' refers
to grid cells crossed by the worldline. ``No particle'' refers to
grid cells not crossed by the worldline. The difference in the error
terms arises because the $\bar h^{(i)}$'s generally have discontinues
$r$ derivatives across the worldline. We can accommodate a local error
term of $O(h^3)$ along the worldline, as the worldline is crossed
only once in each evolution time step, and so such an error accumulates
over time to give a global error of only $O(h^2)$.

%~~~~~~~~~~~~~~~~~~~~~~~~~~~~~~~~~~~~~~~~~~~~~~~~~~~~~~~~~~~~~~~~~~~~~~~~~~
\begin{figure}[htb]
\includegraphics[width=12cm]{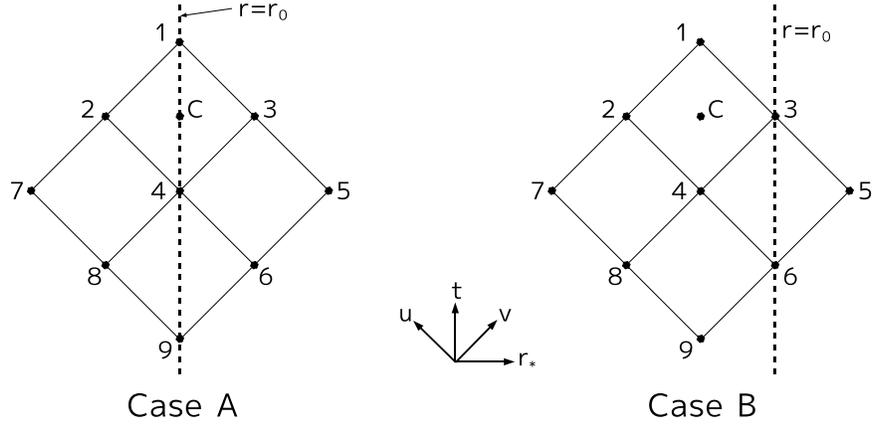}
\caption{Finite-difference scheme for terms in the field equations involving
single $v$ derivatives (diagrams to go with the description in the text).
The dashed line represents the worldline, with two cases shown.}
\label{fig:grid1}
\end{figure}
%~~~~~~~~~~~~~~~~~~~~~~~~~~~~~~~~~~~~~~~~~~~~~~~~~~~~~~~~~~~~~~~~~~~~~~~~~~
Terms of the form $V_3(r) \bar{h}^{(i)}_{,v}$ require a more careful treatment.
For these we will need information from outside the grid cell of Fig.\ \ref{fig:GridCell}.
Consider Fig.\ \ref{fig:grid1}, showing an extended area around the central point C,
now including also points 5--9. We assume all functions $\bar h^{(i)}_{2}$--$\bar h^{(i)}_{9}$
have been calculated before, and we need to obtain $\bar h^{(i)}_{1}$. The figure shows
two special cases: In ``case A'' the worldline crosses the point C; in ``case B''
it crosses ``just to the right'' of point C, through points 3 and 6.
The integral over the term $V_3(r) \bar{h}^{(i)}_{,v}$ can be approximated,
in the various cases, through
%~~~~~~~~~~~~~~~~~~~~~~~~~~~~~~~~~~~~~~~~~~~~~~~~~~~~~~~~~~~~~~~~~~~~~~~
\begin{eqnarray}
\int\int_{\rm cell} V_3(r) \bar{h}^{(i)}_{,v}\, dudv =
\frac{1}{2} h V_3(r_c)\times
\left\{ \begin{array}{ll}
 2\bar{h}_3^{(i)} - \bar{h}_4^{(i)}
 - \frac{1}{2}\bar{h}_5^{(i)}
 - \frac{1}{2}\bar{h}_6^{(i)}
 + O(h^3)
& \mbox{(Case A)}, \\
 \bar{h}_3^{(i)} - \bar{h}_2^{(i)}
 + 3\bar{h}_4^{(i)}
 - 2 \bar{h}_6^{(i)} -2 \bar{h}_8^{(i)}
 + \bar{h}_9^{(i)}
 + O(h^3)
& \mbox{(Case B)}, \\
 3\bar{h}_3^{(i)} - 3\bar{h}_4^{(i)}
 - \bar{h}_5^{(i)} + \bar{h}_6^{(i)}
 + O(h^4)
& \mbox{(no particle)}, \\
\end{array} \right. \label{eq:FDM-vdev}
\end{eqnarray}
%~~~~~~~~~~~~~~~~~~~~~~~~~~~~~~~~~~~~~~~~~~~~~~~~~~~~~~~~~~~~~~~~~~~~~~~
where ``no particle'' refers to the most common case, in which the particle passes
through neither point C nor point $3$. There are, of course, alternative schemes
which approximate this integral at the same order of accuracy. We have chosen
the particular scheme in (\ref{eq:FDM-vdev}) as we found it experimentally
stable.

Finally, we need to integrate the source term $S^{(i)lm}$ on the right-hand side
of the field equations [the explicit form of the source term
is given in Eq.\ (\ref{Si}) of Appendix \ref{AppA}]. Thanks to the delta function
in $S^{(i)}$ we can work out the integral over the grid cell exactly. We find
%~~~~~~~~~~~~~~~~~~~~~~~~~~~~~~~~~~~~~~~~~~~~~~~~~~~~~~~~~~~~~~~~~~~~~~~
\begin{eqnarray}
\int\int_{\rm cell} S^{(i)lm}\, dudv
&=&
8\pi {\cal E}_0\alpha^{(i)}f^{-1}(r_c)\times h\times
{\rm sinc}\left(m\Omega_0 h/2\right)  \nonumber\\
&& \times\, e^{-im\Omega_0 t_c}\times\left\{ \begin{array}{ll}
Y^{lm*}(\pi/2,0), &  \text{for $i=1$--$7$}, \\
Y_{,\theta}^*(\pi/2,0), & \text{for $i=8$--$10$}, \\
\end{array} \right.
\label{eq:FDM-source}
\end{eqnarray}
%~~~~~~~~~~~~~~~~~~~~~~~~~~~~~~~~~~~~~~~~~~~~~~~~~~~~~~~~~~~~~~~~~~~~~~~
where $t_c$ is the value of $t$ at point C, an asterix denotes complex conjugation,
the coefficients $\alpha^{(i)}$ are those given in Eqs.\ (\ref{alphai}), and
sinc$(x)\equiv(\sin x)/x$ for any $x\ne 0$, with sinc$(0)=1$.

Integrating the field equations (\ref{FE}) using above Eqs.\
(\ref{eq:FDM-uv})--(\ref{eq:FDM-source}),
we can solve for the $\bar h^{(i)}$'s at point 1 given the values calculated in
previous steps of the evolution. In this scheme we have to keep two
$v$=const data vectors at each (advanced) time step.
The local finite-differentiating error at each grid point scales as $\sim h^4$,
except for points belonging to Cases A or B (Fig.\ \ref{fig:grid1}), for which
the local error scales as $\sim h^3$. Since the total number of steps scales as
$\sim h^{-2}$, and the number of steps belonging to Cases A or B scales as $h^{-1}$,
we expect the error accumulated over the entire evolution to scale as
$\sim h^2$.

%In the rest of this subsection, we estimate the accumulated
%error in integrateing the field equations.
%In case A and B, these three terms have the error of $O(h^3)$.
%This error is accumulated only along the particle's
%trajectory or the line next to the particle's
%after integrating over the whole grid space.
%The number of grid points for one dimension is propotional
%to the inverse of the step size, $O(1/h)$.
%Therefore the accumulated error in case A and B amount to
%$O(h^2)$ after integration.
%In other cases, in which the particle does not cross
%between the grid points used in the finite difference
%scheme, the error of $O(h^4)$ is produced in each step.
%After integrating over the whole grid space,
%the accumulated error in these cases is $O(h^2)$.

\subsection{Metric perturbation: Monopole and dipole modes}

The monopole and dipole modes are dealt with separately.
For $l=0$ we use the analytic solution for the $\bar h^{(i)}$'s
from Sec.\ III-D of BL. For the mode $l=1$, $m=0$ we use the analytic
solution from Sec.\ III-C therein. For the mode $l=m=1$ we follow the
method of Detweiler and Poisson \cite{Detweiler:2003ci}, which involves solving
(numerically) a coupled set of 3 ordinary differential equations, with boundary
conditions at infinity and along the horizon, and with matching conditions
across the particle.

One may attempt to compute the modes $l=0,1$ too using the evolution
equations $(\ref{FE})$ (which reduce to four equations for $l=0$, six
equations for $l=m=1$, and one equation for $l=1$, $m=0$).
In practice, however, the system $(\ref{FE})$, in its present form, does not seem to
evolve stably for these modes: For $l<2$, some of the potential functions in these
equations turn negative for some $r$ values outside the black hole,
apparently rendering the evolution unstable. This is not a serious problem
in our present analysis, as we simply derive these two modes using other methods.
The problem will have to be address when extending the analysis to non-circular
orbits, for which analytic (or semi-analytic) solutions are not yet at hand.
One may then either attempt to derive analytic (or semi-analytic) $l=0,1$
solutions for eccentric orbits; or, alternatively, attempt to find a
re-formulation of the evolution equations suitable for $l=0,1$.

\subsection{Taking derivatives of the metric perturbation}

\begin{figure}[htb]
\includegraphics[width=10cm]{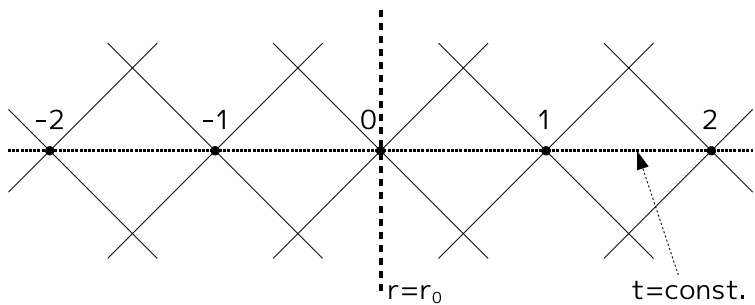}
\caption{Diagram to explain how $r$ derivatives are taken at the worldline
(see description in the text). The dashed line represents the particle's
trajectory on the numerical grid. The SF is calculated at the point labeled
`0'.}
\label{fig:rdiff}
\end{figure}

The construction formulas for the $r$ and $t$ components of the SF require the
derivatives $\bar h^{(i)}_{,r}$ and $\bar h^{(i)}_{,t}$, both evaluated at the
particle. For the $t$ derivatives we can simply make the substitution
$\partial_t\to -im\Omega_0$, since in the
circular orbit case the fields $\bar h^{(i)}$ are stationary and depend upon
$t$ only through $[e^{im\varphi_0(t)}]^*\propto e^{-im\Omega_0 t}$.
The $r$ derivatives are taken numerically, using the finite-difference formula
%~~~~~~~~~~~~~~~~~~~~~~~~~~~~~~~~~~~~~~~~~~~~~~~~~~~~~~~~~~~~~~~~~~~~~~~
\begin{equation} \label{dr}
\left. \bar{h}^{(i)}_{,r} \right|_{\pm} =
\mp \frac{3\bar{h}_0^{(i)}-4\bar{h}_{\pm 1}^{(i)}
          +\bar{h}_{\pm 2}^{(i)}}
         {2h f(r_0)}
+O(h^2),
\end{equation}
%~~~~~~~~~~~~~~~~~~~~~~~~~~~~~~~~~~~~~~~~~~~~~~~~~~~~~~~~~~~~~~~~~~~~~~~
where the various quantities refer to the diagram in Fig.\ \ref{fig:rdiff}:
$\bar{h}_0^{(i)}$ is the value calculated on the worldline, at a given time,
and $\bar{h}_{\pm 1}$ and $\bar{h}_{\pm 2}$ are the values at points $\pm 1$
and $\pm 2$, respectively. Recall $f(r_0)=1-2M/r_0$. The subscripts $+/-$ refer
to one-sided derivatives taken from $r_0^+$ or $r_0^-$. This scheme allows for
discontinues $r$ derivatives, which we expect some of the $\bar{h}^{(i)}$'s to
have. The scheme gives the local derivative with an error that scales as
$\sim h^2$---the same as the accumulated error in the
$\bar{h}^{(i)}$.
%\footnote{Readers may be worried that, since the quantities
%$\bar{h}_{0,\pm 1,\pm 2}^{(i)}$ in Eq.\ (\ref{dr}) each carries along an accumulated
%error of $\sim h^2$, the error in $\bar{h}^{(i)}_{,r}$ will actually scale
%as $\sim h$. This is not the case:
Hence, we expect the total finite-differentiation error in the SF to scale
as $\sim h^2$.

%taking
%$r$ and $t$ derivatives (respectively) of the functions $h^(i)$---
%To calculate the self-force, we need the derivatives
%to $r$ ($r_*$)
%of the metric perturbation at the location of the particle.
%By using the values of MP functions on the $t=$const. surface
%(see Fig.~\ref{fig:rdiff}),
%we can derive their derivatives at the accuracy of
%$O(h^2)$ as
%where $|_{+}$ mean the value evaluated on the $r>r_0$ side
%and $|_{-}$ on the $r<r_0$ side.
%~~~~~~~~~~~~~~~~~~~~~~~~~~~~~~~~~~~~~~~~~~~~~~~~~~~~~~~~~~~~~~~~~~~~~~~
%~~~~~~~~~~~~~~~~~~~~~~~~~~~~~~~~~~~~~~~~~~~~~~~~~~~~~~~~~~~~~~~~~~~~~~~

Once we have at hand the various $\bar h^{(i)}$'s and their derivatives
at the particle, we can construct the various scalar-harmonic
modes of the full force, $[F_{\rm full}^{t l}(x_0)]_{\pm}$
and $[F_{\rm full}^{r l}(x_0)]_{\pm}$, using Eq.\ (\ref{Flfull}).
[Recall that to obtain a single scalar-harmonic mode
of the full force we need to calculate all tensor-harmonic modes of the
perturbation with multipole numbers between $l-2$ and $l+2$ ($l-3$ to $l+3$ for
the $t$ component).] We then construct the regularized modes $F^{\alpha l}_{\rm reg}$
through Eqs.\ (\ref{Freg}) and (\ref{Ftreg}).
%As a test, we repeat the
%calculation of $F^{rl}_{\rm reg}$ with both right-hand side (RHS) and LHS
%values [i.e., using both $(F_{\rm full}^{rl})_{+}$ and $(F_{\rm
%full}^{rl})_{-}$].
We use this procedure to calculate all modes up to
$l=15$ for the radial component, and up to $l=4$--$9$ (depending on $r_0$) for the
temporal component. (For the two radii $r_0=6M$ and $r_0=100M$ we obtain all modes
of the radial component up to $l=25$; we use the extra mode information in testing
the validity of our numerical procedure---see below.)

Of course, the stationarity of our problem allows us to choose any point along the orbit
for calculating the force. We need to make sure, though, that this point
is taken late enough in the evolution, where the effect of initial spurious waves
is negligible. To monitor any residual effect from the initial waves we repeat
the calculation at two different evolution times. We will give more details on
this procedure in Sec.\ \ref{subsec:initial} below.

\subsection{Extrapolation to zero step size}

%$F_\ell^\alpha$ calculated in our code are expected
%to have second order convergence with respect to
%the step size, $h$
%(we will check it in Sec.~\ref{sec:n-conv}).

To obtain $F^{\alpha l}_{\rm reg}$ with good accuracy requires to calculate
$[F_{\rm full}^{\alpha l}(x_0)]_{\pm}$ with
an even better accuracy, which, in turn, requires to evolve the field
equations with a sufficiently fine grid. This can become very demanding
computationally, especially for large $l$. To reduce computational cost we
use a Richardson extrapolation to $h\to 0$.
The idea is to extrapolate the value of $F^{\alpha l}_{\rm reg}$ (using a
rational function) to the limit of vanishing step size, using a sequence of
values obtained with progressively decreasing step sizes. Specifically, we
employ the Bulirsch--Stoer method \cite{NumRec}, which utilizes the sequence
%~~~~~~~~~~~~~~~~~~~~~~~~~~~~~~~~~~~~~~~~~~~~~~~~~~~~~~~~~~~~~~~~~~~~~~~
\begin{equation}\label{hi}
h_i = \frac{1}{n_i},
\end{equation}
%~~~~~~~~~~~~~~~~~~~~~~~~~~~~~~~~~~~~~~~~~~~~~~~~~~~~~~~~~~~~~~~~~~~~~~~
where $i=1,2,3,\ldots$ and
%~~~~~~~~~~~~~~~~~~~~~~~~~~~~~~~~~~~~~~~~~~~~~~~~~~~~~~~~~~~~~~~~~~~~~~~
\begin{equation}
n_i = \{ 2, 4, 6, 8, 12, \cdots \} \quad
(n_i=2n_{i-2}).
\end{equation}
%~~~~~~~~~~~~~~~~~~~~~~~~~~~~~~~~~~~~~~~~~~~~~~~~~~~~~~~~~~~~~~~~~~~~~~~
Namely, we repeat the calculation of $F^{\alpha l}_{\rm reg}$
for all step sizes $h_1,h_2,\ldots,h_{i_{\rm max}}$, and then extrapolate the
resulting series of values to $i\to\infty$ ($h\to\ 0$). To control the error
in this procedure we introduce the estimator
%~~~~~~~~~~~~~~~~~~~~~~~~~~~~~~~~~~~~~~~~~~~~~~~~~~~~~~~~~~~~~~~~~~~~~~~
\begin{equation}
\Delta^{\alpha l}[i_{\rm max}] \equiv
2\left|
\frac{F^{\alpha l}_{\rm reg}[i_{\rm max}]-F^{\alpha l}_{\rm reg}[i_{\rm max}-1]}
{F^{\alpha l}_{\rm reg}[i_{\rm max}]+F^{\alpha l}_{\rm reg}[i_{\rm max}-1]}
\right|,
\end{equation}
%~~~~~~~~~~~~~~~~~~~~~~~~~~~~~~~~~~~~~~~~~~~~~~~~~~~~~~~~~~~~~~~~~~~~~~~
where $F^{\alpha l}_{\rm reg}[i_{\rm max}]$ is the value extrapolated from
the sequence of $i_{\rm max}$ values of $F^{\alpha l}_{\rm reg}$ obtained with
$h_1,\ldots,h_{i_{\rm max}}$. We repeat our calculation with increasing values
of $i_{\rm max}$,
until $\Delta^{\alpha l}[i_{\rm max}]$ is smaller than a prescribed threshold,
$\Delta^{\alpha l}_{\rm thresh}$.

What value should $\Delta^{\alpha l}_{\rm thresh}$ be set to? This requires some
consideration. Since the contribution to the mode sum from individual $l$ modes
decreases with $l$ (at large $l$), it makes sense to relax the threshold for large
$l$ modes. This is certainly true for the $t$ component, for which the mode
sum converges {\em exponentially} at large $l$. Accordingly, for the $t$ component we
set the tight threshold $\Delta^{t l}_{\rm thresh}=10^{-4}$ for each of
the modes $l\leq 3$, but for $l>3$ we set
$
\Delta^{t l}_{\rm thresh} = 10^{-4} \times
\left|
\left(\sum_{l'=0}^{l-1}F^{tl'}_{\rm reg}\right)/F^{t,l-1}_{\rm reg}
\right|.
$
This is slightly larger than $10^{-4}$ for $l=4$, but grows exponentially at large $l$.
Experimentally, it yields $\Delta^{t l}_{\rm thresh}\sim 1$ for $l=5$--$9$ (depending
on $r_0$: higher $l$ for smaller $r_0$).
In fact, for the $t$ component we use $\Delta^{t l}_{\rm thresh}$ also as an indicator
to tell us when it is appropriate to terminate the sum over modes: We sum up to
the first $l$ mode for which $\Delta^{t l}_{\rm thresh}> 1$. This guarantees an
overall truncation error less than a few $\times 10^{-4}$ in the $t$ component of the SF.
As for the $r$ component: Here the mode sum converges only as $\sim l^{-2}$,
and determining $\Delta^{r l}_{\rm thresh}$ requires more caution.
As we describe below, we estimate the contribution from the truncated $l>15$ tail
by extrapolating the numerical data from $l\leq 15$; Assigning an $l$-dependent
threshold could make it difficult to control the error from such an estimation.
For the $r$ component we therefore conservatively set a fixed threshold of
$\Delta^{r l}_{\rm thresh}=10^{-2}$ for each individual $l$ mode computed.

%for the temporal self-force, we take the threshold as
%$\Delta_*=10^{-4}$ for $\ell\le 3$ and
%\[
% \Delta_* =
%10^{-4} \times
%\left|
%\frac{\sum_{\ell'=0}^{\ell-1}F_{\ell'}^t}{F_{\ell-1}^t}
%\right|,
%\]
%for $\ell\ge 4$.
%This method of determining the threshold garantees the
%accuracy of $0.01$\% in calculating the temporal self-force.
%On the other hand, for the radial self-force,
%we take the threshold as $\Delta_*=0.01$ for all modes
%because we need the same accuracy for them to derive
%the fitting formula of the high multipole modes
%as we explain in the next subsection.

For both $r$ and $t$ components, we estimate the overall (fractional)
discretization error in the SF as
%~~~~~~~~~~~~~~~~~~~~~~~~~~~~~~~~~~~~~~~~~~~~~~~~~~~~~~~~~~~~~~~~~~~~~~~
\begin{equation} \label{Deltah}
\Delta^{\alpha}_{\rm discr} \sim\frac{
\sum_{l=0}^{l_{\rm max}}
\left|\Delta^{\alpha l}[i_{\rm thresh}]\,F_{\rm reg}^{\alpha l}[i_{\rm thresh}] \right|}
{\left|\sum_{l=0}^{l_{\rm max}} F_{\rm reg}^{\alpha l}[i_{\rm thresh}]\right| },
\end{equation}
%~~~~~~~~~~~~~~~~~~~~~~~~~~~~~~~~~~~~~~~~~~~~~~~~~~~~~~~~~~~~~~~~~~~~~~~
where $i_{\rm thresh}$ (depending on $l$ and $r_0$) is the smallest
$i_{\rm max}$ for which $\Delta^{\alpha l}[i_{\rm max}]<\Delta^{\alpha l}_{\rm thresh}$.
Note that we take here the fractional total error as the {\em sum} of fractional errors
from the individual $l$-modes, rather than their root-mean-square value. This makes sense
because the individual errors are not randomly distributed but rather reflect a systematic
extrapolation error.
In practice, to reach the above thresholds, we needed $i_{\rm thresh}$ values of
up to 8 for the $r$ component, and up to 9 for the $t$ component (depending on $l$
and $r_0$: larger $i_{\rm thresh}$ is generally required for larger $l$ and smaller $r_0$).

For each value of $r_0$ and $l$ we started the above procedure with
$i_{\rm max}=4$ for the $r$ component and $i_{\rm max}=3$ for the $t$ component;
namely, we used at least 4(3) terms for the Richardson extrapolation. For some of the low-$l$ modes
this already yielded $\Delta^{\alpha l}[i_{\rm max}]$ smaller than $\Delta^{\alpha l}_{\rm thresh}$
(this is partly because the low-$l$ modes have large $l=0,1$ tensorial-harmonic components,
which are available analytically). Since the total error $\Delta^{\alpha}_{\rm discr}$ is dominated by
errors from these low-$l$ modes, we generally find $\Delta^{\alpha}_{\rm discr}$ values {\em much smaller}
than the above set thresholds. The actual values obtained for $\Delta^{\alpha}_{\rm discr}$ will
be stated in the Results section.

\subsection{Estimation of contribution from large-$l$ tail} \label{subsec:tail}

The mode-sum formula requires summation over all regularized modes
$F^{\alpha l}_{\rm reg}$ from $l=0$ to $l=\infty$. In practice, of course, we
calculate only a finite number of modes, $0\leq l\leq l_{\rm max}$.
It is then necessary to estimate the contribution from the remaining, truncated
part of the series. This is straightforward in the case of the $t$ component, where the
mode sum converges exponentially, and thus the contribution from the truncated tail
drops exponentially with $l_{\rm max}$. We find experimentally that it is sufficient
to take $l_{\rm max}=4$--$9$ (depending on $r_0$; larger $l_{\rm max}$ is needed
for smaller $r_0$) for the contribution from the truncated tail to drop below our
standard discretization error ($\sim 10^{-4}$).

%In principle, we can calculate all modes with the scheme
%introduced in the previous sections.
%In practical calculation, however,
%we truncate the summation over $\ell$ at the finite length.
%For the temporal component of the self-force,
%since the force by modes decreases exponentially with
%respect to $\ell$, the high multipole contribution can be
%less than the numerical error by calculating
%the low multipole modes up to the reasonably finite $\ell$
%(less than ten modes for the accuracy of $0.01\%$).

The situation is different with the $r$ component, where the mode sum converges
as $\propto l^{-2}$, and the contribution from the truncated high-$l$ tail
scales as $1/l_{\rm max}$. For computationally realistic values of $l_{\rm max}$
(here we take $l_{\rm max}=15$) this contribution cannot be neglected. To evaluate
it we apply the following strategy.
Let
%~~~~~~~~~~~~~~~~~~~~~~~~~~~~~~~~~~~~~~~~~~~~~~~~~~~~~~~~~~~~~~~~~~~~~~~
\begin{equation}
F^r = F^{r}_{l\leq l_{\rm max}} + F^r_{l> l_{\rm max}},
\end{equation}
%~~~~~~~~~~~~~~~~~~~~~~~~~~~~~~~~~~~~~~~~~~~~~~~~~~~~~~~~~~~~~~~~~~~~~~~
where
%~~~~~~~~~~~~~~~~~~~~~~~~~~~~~~~~~~~~~~~~~~~~~~~~~~~~~~~~~~~~~~~~~~~~~~~
\begin{equation}\label{high&low}
F^{r}_{l\leq l_{\rm max}} \equiv
\sum_{l=0}^{l_{\rm max}} F_{\rm reg}^{rl} \qquad \text{and} \quad
F^r_{l> l_{\rm max}} \equiv
\sum_{l=l_{\rm max}+1}^{\infty} F_{\rm reg}^{rl}.
\end{equation}
%~~~~~~~~~~~~~~~~~~~~~~~~~~~~~~~~~~~~~~~~~~~~~~~~~~~~~~~~~~~~~~~~~~~~~~~
The part $F^{r}_{l\leq l_{\rm max}}$ is computed numerically. To evaluate
$F^r_{l> l_{\rm max}}$ we extrapolate the last few modes in
$F^{r}_{l\leq l_{\rm max}}$ using the fitting formula
%~~~~~~~~~~~~~~~~~~~~~~~~~~~~~~~~~~~~~~~~~~~~~~~~~~~~~~~~~~~~~~~~~~~~~~~
\begin{equation}
F_{\rm reg}^{rl} \simeq
\sum_{n=1}^{N} \frac{D^r_{2n}}{L^{2n}}
\label{eq:highL-formula}
\end{equation}
%~~~~~~~~~~~~~~~~~~~~~~~~~~~~~~~~~~~~~~~~~~~~~~~~~~~~~~~~~~~~~~~~~~~~~~~
(for some $N\geq 1$), where, recall, $L=l+1/2$, and $D^r_{2n}$ are $l$-independent
coefficients, which serve here as fitting parameters. In practice we have
used the last 6 modes of $F^{r}_{l\leq l_{\rm max}}$ (i.e., $10\leq l\leq15$) for
the fitting, but have checked that fitting using a different number of modes
does not change the result significantly (we demonstrate this below).
Given the coefficients $D^r_{2n}$, the large-$l$ tail is estimated as
%~~~~~~~~~~~~~~~~~~~~~~~~~~~~~~~~~~~~~~~~~~~~~~~~~~~~~~~~~~~~~~~~~~~~~~~
\begin{equation}
F^r_{l> l_{\rm max}} \simeq
\sum_{n=1}^{N} D^r_{2n} \sum_{l=l_{\rm max}+1}^{\infty} L^{-2n}=
\sum_{n=1}^{N}
\frac{D^r_{2n}}{(2n-1)!} \Psi(2n-1,l_{\rm max}+3/2),
\end{equation}
%~~~~~~~~~~~~~~~~~~~~~~~~~~~~~~~~~~~~~~~~~~~~~~~~~~~~~~~~~~~~~~~~~~~~~~~
where $\Psi(n,x)$ is the polygamma function of order $n$, defined as
%~~~~~~~~~~~~~~~~~~~~~~~~~~~~~~~~~~~~~~~~~~~~~~~~~~~~~~~~~~~~~~~~~~~~~~~
\begin{equation}
\Psi(n,x) = \frac{d^{n+1} [\log\Gamma(x)]}{dx^{n+1}},
\end{equation}
%~~~~~~~~~~~~~~~~~~~~~~~~~~~~~~~~~~~~~~~~~~~~~~~~~~~~~~~~~~~~~~~~~~~~~~~
in which $\Gamma(x)$ is the standard gamma function.

To determine how many terms it is necessary to include in the fitting formula
(\ref{eq:highL-formula}) requires some experimentation. The data in Tables
\ref{table:highL-fit} demonstrate the effect of varying $N$:
It shows the values obtained for $F^r_{l> l_{\rm max}}$ (both external and
internal values) using $N$ in the range 1--4. We display data for the two sample
radii $r_0=6M$ and $r_0=100M$.
We find that taking $N=3$ or $N=4$ gives the same value of $F^r_{l> l_{\rm max}}$
as taking $N=2$, with a fractional difference of merely $\lesssim 10^{-4}$ at most.
Since $F^r_{l> l_{\rm max}}$ itself contributes at most $\sim 2\%$ of the total
force (see Tables \ref{table:highL-effect},\ref{table:highL-effect2} in Appendix
\ref{AppD}), we conclude that it is sufficient to take
$N=2$. Taking only $N=1$ would produce a fitting error similar in magnitude to
$\Delta^{\alpha}_{\rm discr}$; So, taking $N=2$ effectively eliminates the
large-$l$ fitting as a source of error in our calculation. We find similar
numbers for other values of $r_0$, and so we take $N=2$ in all cases.
%~~~~~~~~~~~~~~~~~~~~~~~~~~~~~~~~~~~~~~~~~~~~~~~~~~~~~~~~~~~~~~~~~~~~~~~
\begin{table}[htb]
\begin{tabular}{c||c|c||c|c}
\hline\hline
 & $\left[F^r_{l>15}\right]_-$
 & Relative difference
 & $\left[F^r_{l>15}\right]_+$
 & Relative difference \\
$N$ & (fit using $10\leq l\leq 15$)
    & w.r.t $N=2$
    & (fit using $10\leq l\leq 15$)
    & w.r.t $N=2$ \\
\hline\hline
\multicolumn{5}{c}{$r_0=6M$} \\
\hline\hline
$1$ & $-5.117158$ [$2\times 10^{-3}$]
    & $-1.4\times 10^{-2}$
    & $-5.117176$ [$2\times 10^{-3}$]
    & $-1.4\times 10^{-2}$ \\
$2$ & $-5.046144$ [$3\times 10^{-5}$]
    & $0$
    & $-5.046189$ [$3\times 10^{-5}$]
    & $0$ \\
$3$ & $-5.046984$ [$6\times 10^{-5}$]
    & $-1.7\times 10^{-4}$
    & $-5.046812$ [$2\times 10^{-4}$]
    & $-1.2\times 10^{-4}$ \\
$4$ & $-5.046316$ [$4\times 10^{-4}$]
    & $-3.4\times 10^{-5}$
    & $-5.046133$ [$9\times 10^{-4}$]
    & $-1.1\times 10^{-5}$ \\
\hline\hline
\multicolumn{5}{c}{$r_0=100M$} \\
\hline\hline
$1$ & $-5.904165$ [$1\times 10^{-3}$]
    & $-7.5\times 10^{-3}$
    & $-5.904159$ [$1\times 10^{-3}$]
    & $-7.5\times 10^{-3}$ \\
$2$ & $-5.859951$ [$2\times 10^{-5}$]
    & $0$
    & $-5.859925$ [$2\times 10^{-5}$]
    & $0$ \\
$3$ & $-5.860397$ [$8\times 10^{-6}$]
    & $-7.6\times 10^{-5}$
    & $-5.860450$ [$6\times 10^{-6}$]
    & $-9.0\times 10^{-5}$ \\
$4$ & $-5.860430$ [$7\times 10^{-5}$]
    & $-8.2\times 10^{-5}$
    & $-5.860380$ [$5\times 10^{-5}$]
    & $-7.8\times 10^{-5}$ \\
\hline\hline
\end{tabular}
\caption{Data demonstrating how sensitive the estimation of the large-$l$
contribution is to the number of terms $N$ included in
the fitting formula (\ref{eq:highL-formula}). We display here results for the
strong-field case ($r_0=6M$) and for the weak-field case ($r_0=100M$), and
in both cases use the six numerically-computed modes $10\le l\le 15$ to fit for
the part of the mode-sum with $l>15$. $[F^r_{l>15}]_{-}$ and $[F^r_{l>15}]_{+}$
are the estimated contributions from $l>15$, obtained using internal and external
data, respectively. For convenience, the values of $[F^r_{l>15}]_{\pm}$ are
given multiplied by $10^{4}(M/\mu)^2$ for $r_0=6M$, and by $10^{8}(M/\mu)^2$
for $r_0=100M$. Values in square brackets indicate the fractional fitting
error in $[F^r_{l>15}]_{\pm}$. The values in the 3rd and 5th columns show the
relative differences in $[F^r_{l>15}]_{\pm}$ with respect to the reference case
$N=2$, which we adopt in the rest of this work. Note that the differences
indicated are relative to the large-$l$ contribution $[F^r_{l>15}]_{\pm}$
only; the differences relative to the total SF are at least 2 orders of
magnitude smaller.}
\label{table:highL-fit}
\end{table}
%~~~~~~~~~~~~~~~~~~~~~~~~~~~~~~~~~~~~~~~~~~~~~~~~~~~~~~~~~~~~~~~~~~~~~~~

%To validate our fitting formula,
%we investigate the dependence of the estimate of the high
%multipole contribution on the fitting function and
%the fitting region.
%In Table~\ref{table:highL-fit},
%we list the estimated values of $F^r_{\rm high}$ by fitting
%six modes in $10\le\ell\le 15$ to
%\begin{equation}
%F^{(N)} \equiv
%\sum_{n=1}^{N} \frac{D^r_{2n}}{L^{2n}},
%\end{equation}
%where $N$ correspond to the number of fitting parameters.
%We can find that, for $N\ge 2$,
%the change in $F^r(N)$ is less than $1$\%.

As a further robustness test for the above scheme, we check how the
estimation of $F^r_{l> l_{\rm max}}$ would change by using higher multipole
modes for the fitting. For this, we calculated numerically all modes up to $l=25$
for $r_0=6M,100M$. Results from this experiment are shown in Table~\ref{table:highL-fit2}.
Once again we take $F^r_{l> l_{\rm max}}$ as the sum of all modes $l>15$,
but this time we obtain the fitting parameters $D_{2n}$ based on all {\em sixteen}
modes $10\le l\le 25$. Again, we check how the results depend on $N$.
%We also show the values of $F^r(N)$ estimated by fitting
%with sixteen modes in $10\le\ell\le 25$
%in Table~\ref{table:highL-fit2}.
We find that the relative difference in the value of $F^r_{l> l_{\rm max}}$
with respect to our reference case [$N=2$ and fitting based on $10\le l\le 15$]
is $\lesssim 10^{-4}$ in all cases, as long as we take $N\ge 2$.
This reassures us that it is sufficient to base the fitting on $10\le l\le 15$,
as we do in the rest of this analysis.
%~~~~~~~~~~~~~~~~~~~~~~~~~~~~~~~~~~~~~~~~~~~~~~~~~~~~~~~~~~~~~~~~~~~~~~~
\begin{table} [htb]
\begin{tabular}{c||c|c||c|c}
\hline\hline
 & $\left[F^r_{l>15}\right]_-$
 & Relative difference
 & $\left[F^r_{l>15}\right]_+$
 & Relative difference \\
$N$ & (fit using $10\leq l\leq 25$)
    & w.r.t reference case
    & (fit using $10\leq l\leq 25$)
    & w.r.t reference case \\
\hline\hline
\multicolumn{5}{c}{$r_0=6M$} \\
\hline\hline
$1$ & $-5.106648$ [$1\times 10^{-3}$]
    & $-1.2\times 10^{-2}$
    & $-5.106701$ [$1\times 10^{-3}$]
    & $-1.2\times 10^{-2}$ \\
$2$ & $-5.046340$ [$2\times 10^{-5}$]
    & $-3.9\times 10^{-5}$
    & $-5.046533$ [$3\times 10^{-5}$]
    & $-6.8\times 10^{-5}$ \\
$3$ & $-5.046634$ [$3\times 10^{-5}$]
    & $-9.7\times 10^{-5}$
    & $-5.046985$ [$4\times 10^{-5}$]
    & $-1.6\times 10^{-4}$ \\
$4$ & $-5.046500$ [$1\times 10^{-4}$]
    & $-7.1\times 10^{-5}$
    & $-5.047089$ [$1\times 10^{-4}$]
    & $-1.8\times 10^{-4}$ \\
$5$ & $-5.046769$ [$4\times 10^{-4}$]
    & $-1.2\times 10^{-4}$
    & $-5.047224$ [$5\times 10^{-4}$]
    & $-2.1\times 10^{-4}$ \\
\hline\hline
\multicolumn{5}{c}{$r_0=100M$} \\
\hline\hline
$1$ & $-5.897633$ [$8\times 10^{-4}$]
    & $-6.4\times 10^{-3}$
    & $-5.897625$ [$8\times 10^{-4}$]
    & $-6.4\times 10^{-3}$ \\
$2$ & $-5.860129$ [$1\times 10^{-5}$]
    & $-3.0\times 10^{-5}$
    & $-5.860106$ [$1\times 10^{-5}$]
    & $-3.1\times 10^{-5}$ \\
$3$ & $-5.860367$ [$7\times 10^{-6}$]
    & $-7.1\times 10^{-5}$
    & $-5.860349$ [$8\times 10^{-6}$]
    & $-7.2\times 10^{-5}$ \\
$4$ & $-5.860371$ [$2\times 10^{-5}$]
    & $-7.2\times 10^{-5}$
    & $-5.860304$ [$2\times 10^{-5}$]
    & $-6.5\times 10^{-5}$ \\
$5$ & $-5.860407$ [$9\times 10^{-5}$]
    & $-7.8\times 10^{-5}$
    & $-5.860308$ [$1\times 10^{-4}$]
    & $-6.5\times 10^{-5}$ \\
\hline\hline
\end{tabular}
\caption{Data demonstrating how sensitive the estimation of the high-$l$
contribution is to the number of modes used for the high-$l$ fitting.
The structure of this table is the same as that of Table \ref{table:highL-fit}.
Again we show data for $r_0=6M$ and $r_0=100M$, and $[F^r_{l>15}]_{\pm}$
describes the contribution from $l>15$; however, we now use all 16 modes
$10\leq l\leq 25$ for fitting. The values in the 3rd and 5th columns show
the relative differences with respect to the reference case, i.e., $N=2$ and
fitting using $10\le l\le 15$---the case displayed in Table
\ref{table:highL-fit}. This demonstrates that, taking $N=2$, it is sufficient
to base our large-$l$ fitting on data from $10\le l\le 15$.} \label{table:highL-fit2}
\end{table}
%~~~~~~~~~~~~~~~~~~~~~~~~~~~~~~~~~~~~~~~~~~~~~~~~~~~~~~~~~~~~~~~~~~~~~~~

\begin{figure}[htb]
\includegraphics[width=8cm]{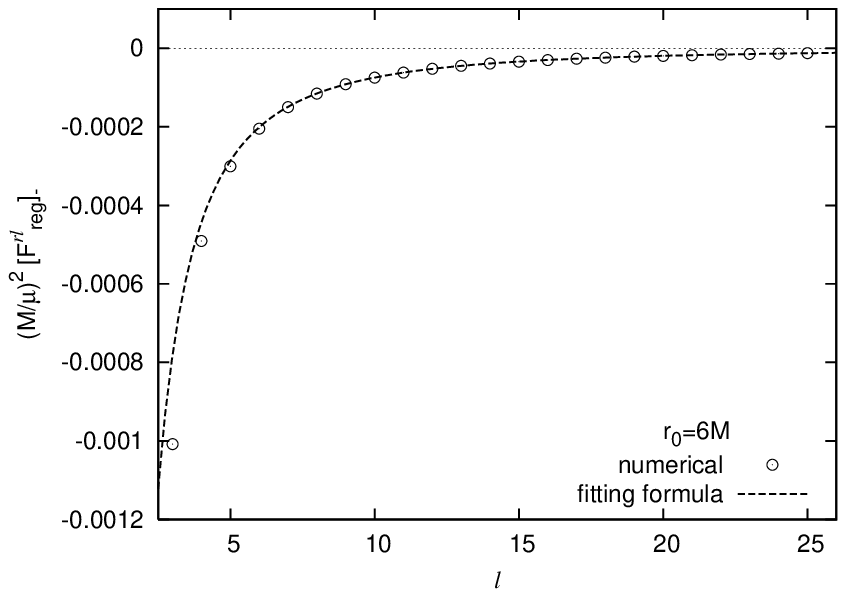}
\includegraphics[width=8cm]{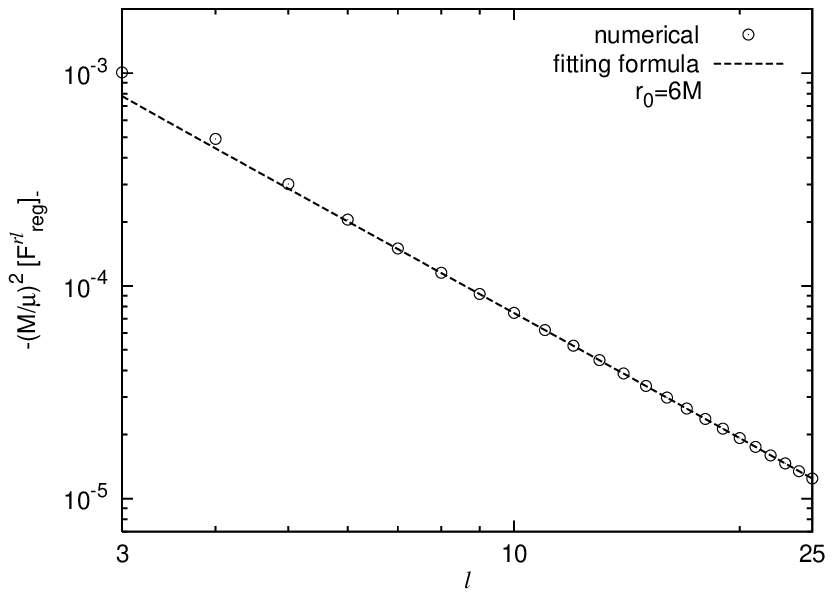}
\caption{Analytic fitting for the large-$l$ tail of the SF, exemplified here
for $r_0=6M$. We used the fitting formula (\ref{eq:highL-formula}) with $N=2$,
and based on the modes $10\le l\le 15$. Circles (`$\odot$') represent
actual data obtained for $F_{\rm reg}^{r l}$ (calculated from $r_0^-$),
for the various modes $3\leq l\leq 25$. The dashed line is the analytic fit.
Left/right panels show the same data on linear/log scales. The large-$l$ tail
of the mode-sum series shows the $l^{-2}$ fall-off expected from theory
(cf.\ Fig.\ \ref{fig:large l}).
} \label{fig:FrL-fit}
\end{figure}

Results from the above fitting procedure (with $N=2$ and $10\le l\le 15$) are
illustrated in Fig.\ \ref{fig:FrL-fit} for $r_0=6M$. As an indicative measure
of the calculation error in $F^r_{l> l_{\rm max}}$ we take the standard
(fractional) fitting error \cite{NumRec}, which we denote $\Delta_{\rm tail}$.
At least in the examples shown in Tables \ref{table:highL-fit} and
\ref{table:highL-fit2}, this has roughly the same magnitude as the error from
changing the fitting method, so $\Delta_{\rm tail}$ provides a realistic
estimate of the total (fractional) error in determining the tail contribution
$F^r_{l> l_{\rm max}}$. The {\em relative} error from determining the large-$l$
tail, expressed as a fraction of the total SF, is
%~~~~~~~~~~~~~~~~~~~~~~~~~~~~~~~~~~~~~~~~~~~~~~~~~~~~~~~~~~~~~~~~~~~~~~~
\begin{equation}\label{Deltafit}
\Delta_{\rm tail,rel}= \Delta_{\rm tail}\times
\left|\frac{F^r_{l> l_{\rm max}}} {F^r_{l\leq l_{\rm max}}}\right|.
\end{equation}
%~~~~~~~~~~~~~~~~~~~~~~~~~~~~~~~~~~~~~~~~~~~~~~~~~~~~~~~~~~~~~~~~~~~~~~~
In practice we find (see Appendix \ref{AppD})
$F^r_{l> l_{\rm max}}/F^{r}_{l\leq l_{\rm max}}\sim 10^{-2}$--$10^{-4}$
for $r_0$ in the range $6M$--$150M$. The relative fitting error $\Delta_{\rm tail,rel}$
is a mere $\sim 7\times 10^{-5}$ at $r_0=6M$, dropping down to $\sim 2\times 10^{-7}$
at $r_0=150M$.
%~~~~~~~~~~~~~~~~~~~~~~~~~~~~~~~~~~~~~~~~~~~~~~~~~~~~~~~~~~~~~~~~~~~~~~~
\section{Code validation} \label{Sec:validation}
%%%%%%%%%%%%%%%%%%%%%%%%%%%%%%%%%%%%%%%%%%%%%%%%%%

A few validation tests for our metric perturbation code were presented
in BL. These included (i) a demonstration that the numerical solutions
for the various $\bar h^{(i)lm}$'s converge quadratically as $h\to 0$;
(ii) a demonstration that these solutions satisfy the Lorenz gauge conditions;
(iii) a confirmation that the flux of energy radiated to null infinity
in the various modes, as calculated from our Lorenz gauge solutions,
compares well with the flux obtained using other methods/gauges.
Here we present some more
validation tests, focusing on the new ingredient of the analysis, i.e.,
the calculation of the SF. We will demonstrate (i) quadratic numerical
convergence of the computed SF; (ii) that the SF does not depend on our
choice of initial data; (iii) that the full-force modes have large-$l$
behavior as predicted in theory [Eq.\ (\ref{ModeSum})]; (iv) that the two
one-sided values obtained for the final SF (from $r_0^+$ and from $r_0^-$)
agree; and (v) that the total flux of energy to infinity and through the
horizon is consistent with the value obtained for the temporal component
of the SF.

\subsection{Numerical convergence} \label{sec:n-conv}

\begin{figure}[htb]
\includegraphics[width=8cm]{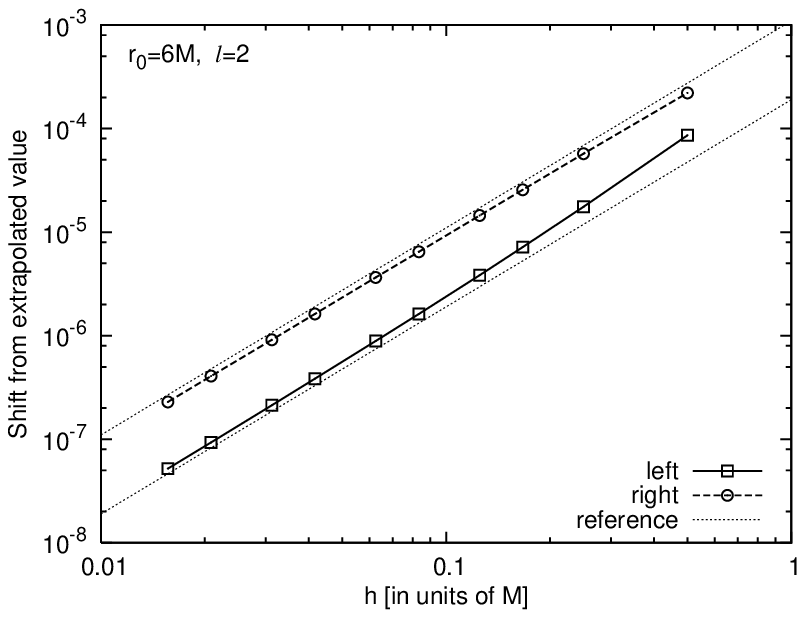}
\includegraphics[width=8cm]{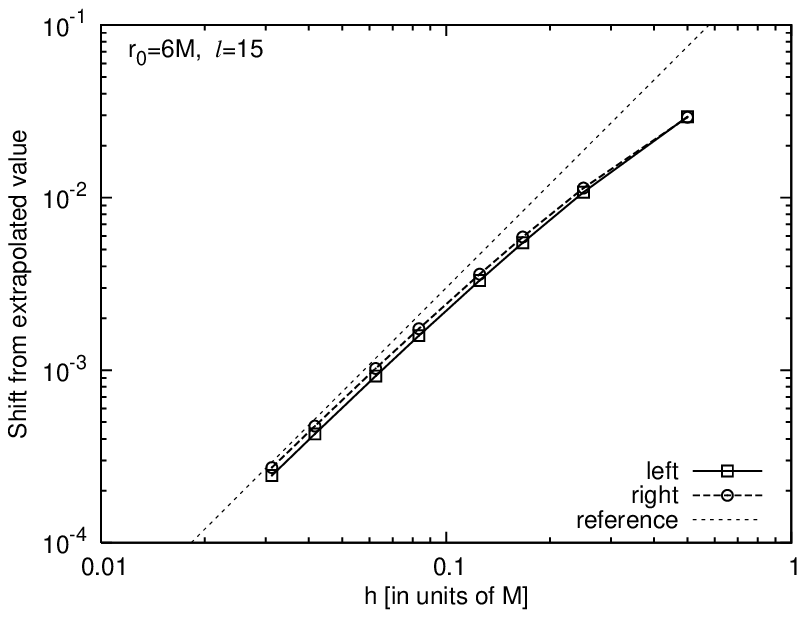}
\caption{Numerical convergence of the calculated SF, demonstrated here
for the $r$ component, for $r_0=6M$. The left and right panels show $l=2$
and $l=15$, respectively. Plotted is the difference
$F^{rl}_{\rm reg}[h_i]-F^{rl}_{\rm reg}$
between the value of the regularized mode computed
with step size $h_i$, and the value extrapolated to $h\to 0$.
Each panel displays both one-sided values of the force: ``left'' and ``right''
stand for $r_0^-$ and $r_0^+$ values, respectively.
%We have derived the extrapolated values by using
%ten points for $\ell=2$ and eight points for $\ell=15$.
The reference lines (dotted) have slops $\propto h^2$.
This demonstrates the quadratic convergence of the numerical calculation.
}
\label{fig:Fr-convergence}
\end{figure}

The scheme introduced in Sec.\ \ref{Sec:numerics} should yield the final
SF with a numerical error scaling as $\sim h^2$ (where $h$ is the step
size in both $u$ and $v$). To check this, we performed the following test,
for a selection of $r_0$ values in the range $6M$--$150M$ and $l$ values
in the range $0$--$15$. For given $r_0$ and $l$ we calculated the regularized
mode $F^{rl}_{\rm reg}$ through the scheme described in Sec.\ \ref{Sec:numerics}
(including the extrapolation to $h\to 0$). We recorded the values of the force
calculated with the different resolutions $h_{i}$ [see Eq.\ (\ref{hi})].
Denoting these by $F^{rl}_{\rm reg}[h_i]$, we then plotted the difference
$F^{rl}_{\rm reg}[h_i]-F^{rl}_{\rm reg}$ (where the second term is the extrapolated
force) as a function of $h_i$. Figure \ref{fig:Fr-convergence} shows the
two modes $l=2,15$ at $r_0=6M$. In each case we plot both
one-sided values of the difference $F^{rl}_{\rm reg}[h_i]-F^{rl}_{\rm reg}$.
In all cases we find
that this difference decreases approximately like $h^2$ at small $h$,
demonstrating quadratic numerical convergence. Similar convergence is
observed for the $t$ component.
%~~~~~~~~~~~~~~~~~~~~~~~~~~~~~~~~~~~~~~~~~~~~~~~~~~~~~~~~~~~~~~~~~~~~~~~
%~~~~~~~~~~~~~~~~~~~~~~~~~~~~~~~~~~~~~~~~~~~~~~~~~~~~~~~~~~~~~~~~~~~~~~~

\subsection{Dependence on initial conditions} \label{subsec:initial}

\begin{figure}[htb]
\includegraphics[width=8cm]{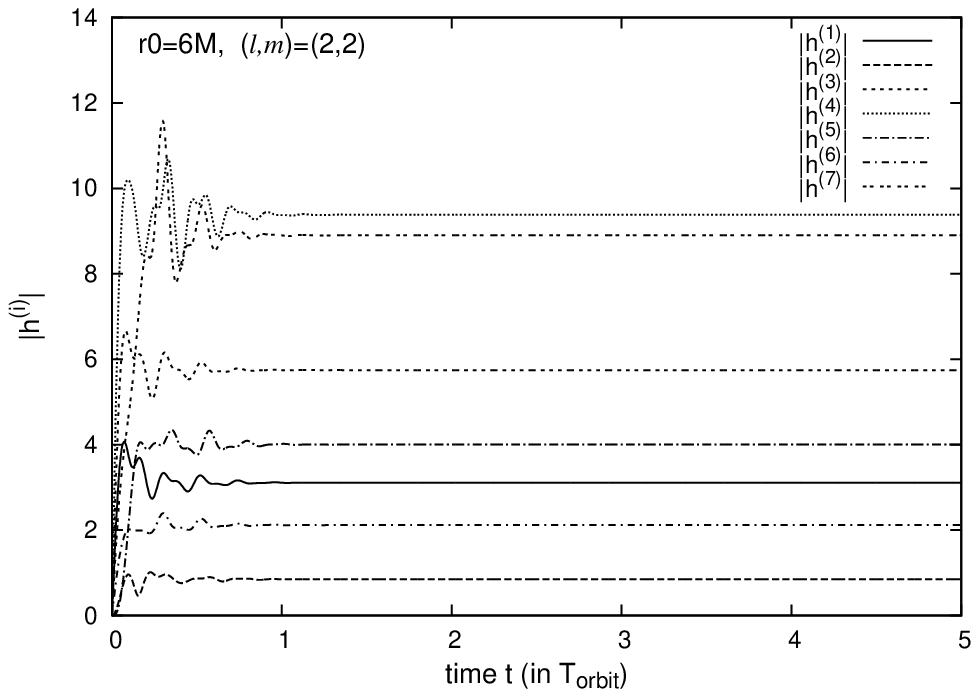}
\includegraphics[width=8cm]{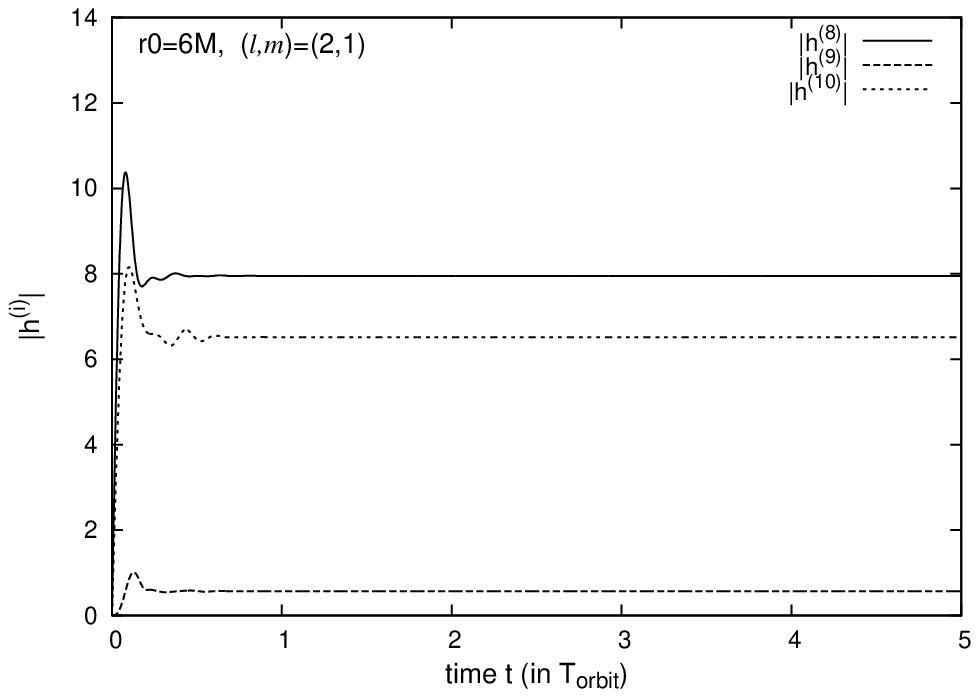} \\
\includegraphics[width=8cm]{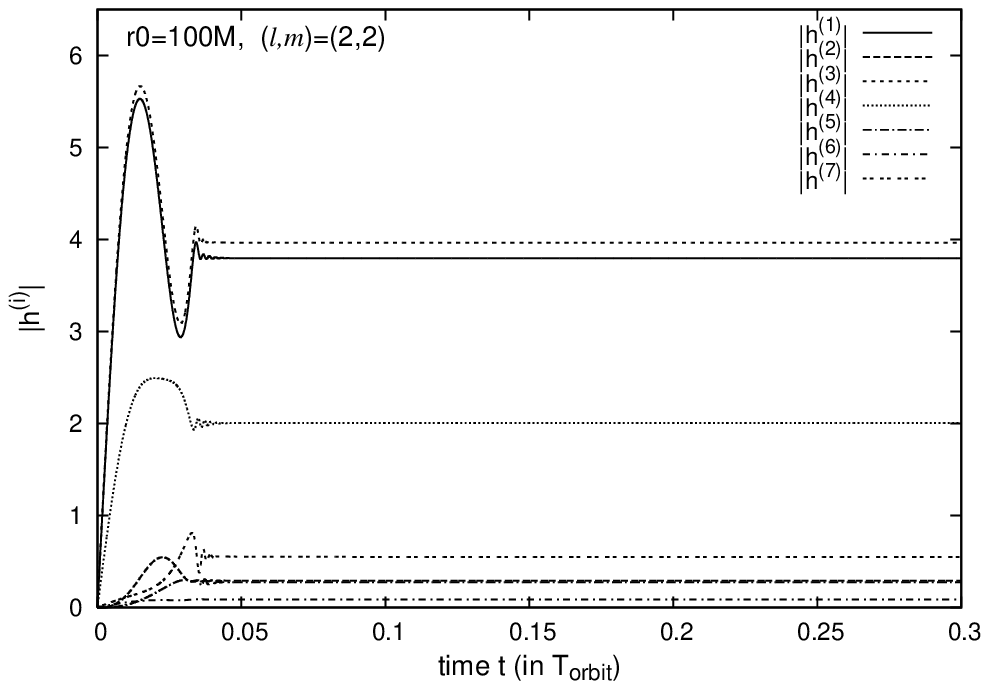}
\includegraphics[width=8cm]{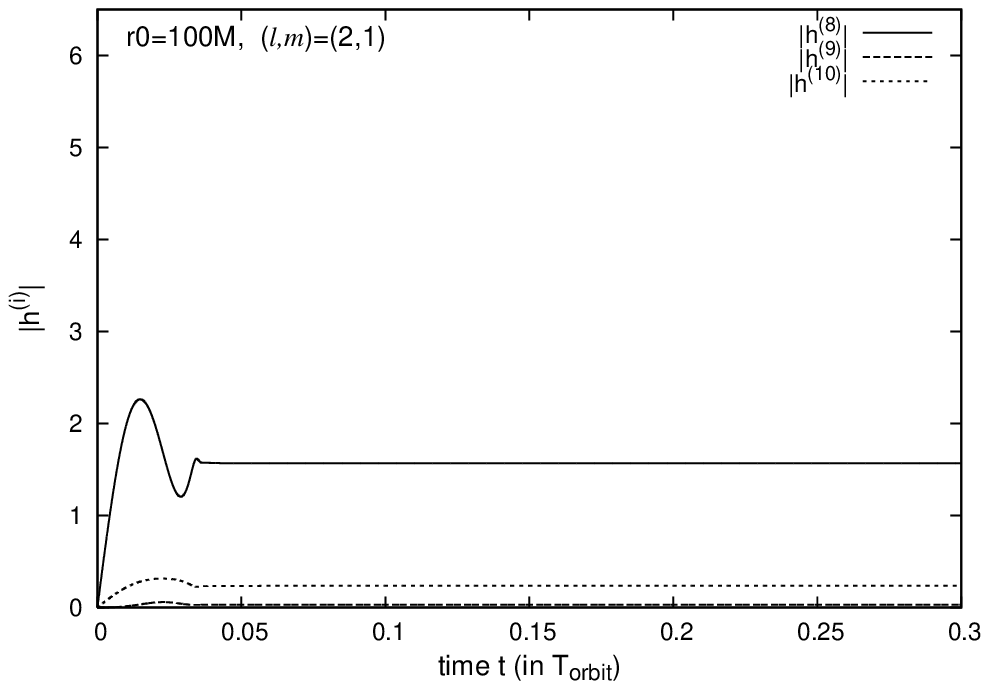}
\caption{Time evolution of the metric perturbation.
We plot the absolute values of the various functions $\bar h^{(i)lm}$, evaluated
at the location of the particle, as a function of $t$, for $l=2$ and $m=1,2$.
Top two figures are for $r_0=6M$; lower two figures are for $r_0=100M$.
The time $t$ is indicated along the horizontal axis in units of the orbital
period, $T_{\rm orb}$. The initial transient phase is due to the imperfection
of the initial conditions; these spurious waves dissipate rapidly, clearing the
stage for the physical, stationary solution. The SF is calculated at late
time, when the effect of the initial spurious waves is negligible.}
\label{fig:hb-stationary}
\end{figure}

As explained above, we start the evolution of each of the modes
$\bar h^{(i)lm}$ with null values along the initial surfaces, $v=v_0$
and $u=u_0$. This creates a `spark' of spurious radiation which propagates
through the grid, but dies off at late time. During the early transient
period the numerical solution is not stationary; As the spurious waves
die off, the numerical solution approaches its physical, stationary
value. This behavior is demonstrated in the plots in Fig.\
(\ref{fig:hb-stationary}). The SF has to be calculated at late enough time,
to assure that the error due to residual initial waves is negligible.
This sets a lower limit on the required evolution time $T_{\rm evo}$,
which in practice depends on $r_0$: Waves from an initial spark at $r_0=100M$
dissipate more efficiently (faster) than waves from an initial spark at
$r_0=6M$, since the former experience less scattering off spacetime curvature.
%~~~~~~~~~~~~~~~~~~~~~~~~~~~~~~~~~~~~~~~~~~~~~~~~~~~~~~~~~~~~~~~~~~~~~~~
%~~~~~~~~~~~~~~~~~~~~~~~~~~~~~~~~~~~~~~~~~~~~~~~~~~~~~~~~~~~~~~~~~~~~~~~

In our analysis we determined $T_{\rm evo}$ experimentally, for each
value of $r_0$ considered. To assess the effect of residual waves, we
compared the values obtained for the final SF at two different evolution
times, $T_{\rm evo}$ and $0.8\times T_{\rm evo}$. We regard this difference
as indicative of the error from non-stationarity in our calculation.
The errors calculated for the various values of $r_0$ are shown in Tables
\ref{table:compare-Edot} and \ref{table:result-Fr} of Sec.\ \ref{Sec:results} below.
The fractional error is smaller than
$10^{-4}$ in all cases, hence smaller than our standard discretization error.
Table \ref{table:Tevo} lists the evolution times $T_{\rm evo}(r_0)$
used in our analysis.
%~~~~~~~~~~~~~~~~~~~~~~~~~~~~~~~~~~~~~~~~~~~~~~~~~~~~~~~~~~~~~~~~~~~~~~~
\begin{table} [htb]
\begin{tabular}{|c|c|c|}
\hline\hline
 $r_0/M$                & $T_{\rm evo}/T_{\rm orb}$  & $T_{\rm evo}/T_{\rm orb}$   \\
\mbox{}                 & ($r$ component)            & ($t$ component) \\
\hline
6--10                   & 3                          &3                 \\
11--12                  & 2.5                        &2.8                 \\
13--14                  & 2                          &2.8                 \\
15                      & 1.5                        &2.5                 \\
20                      & 1                          &2                 \\
30                      & 0.8                        &1.8                 \\
40                      & 0.6                        &1.7                 \\
50                      & 0.5                        &1.5                 \\
60                      & 0.45                       &1.5                 \\
70                      & 0.4                        &1.2                 \\
80                      & 0.3                        &1.0                 \\
90                      & 0.25                       &1.0                 \\
100                     & 0.2                        &0.8                 \\
120                     & 0.15                       &0.8                 \\
150                     & 0.12                       &0.6                 \\
\hline\hline
\end{tabular}
\caption{Evolution times taken in our analysis. These were chosen (experimentally)
long enough to guarantee that any residual effect from the initial spurious
waves is negligible. $T_{\rm evo}$ is the numerical evolution time, and
$T_{\rm orb}=2\pi/\Omega_0$ is the orbital period.}
\label{table:Tevo}
\end{table}
%~~~~~~~~~~~~~~~~~~~~~~~~~~~~~~~~~~~~~~~~~~~~~~~~~~~~~~~~~~~~~~~~~~~~~~~

\subsection{Large $l$ behaviour of the full-force modes}

\begin{figure}[htb]
\includegraphics[width=8cm]{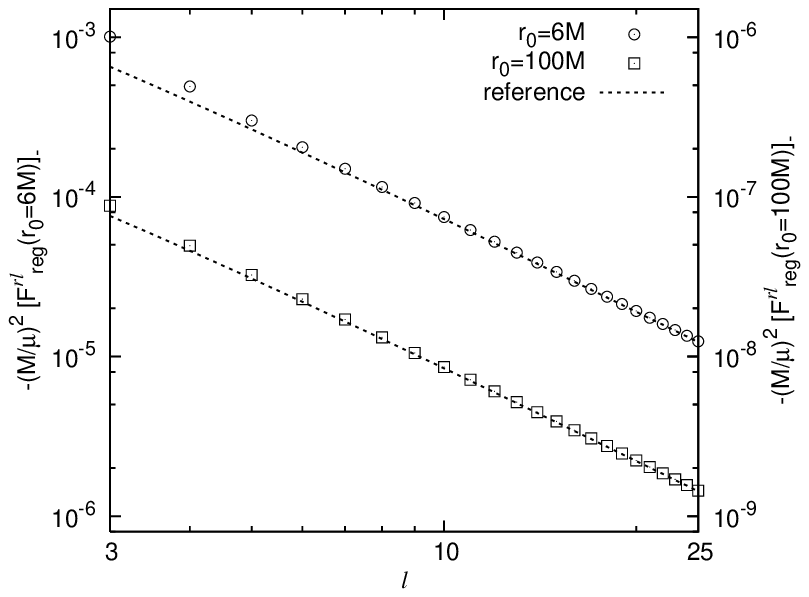}
\includegraphics[width=8cm]{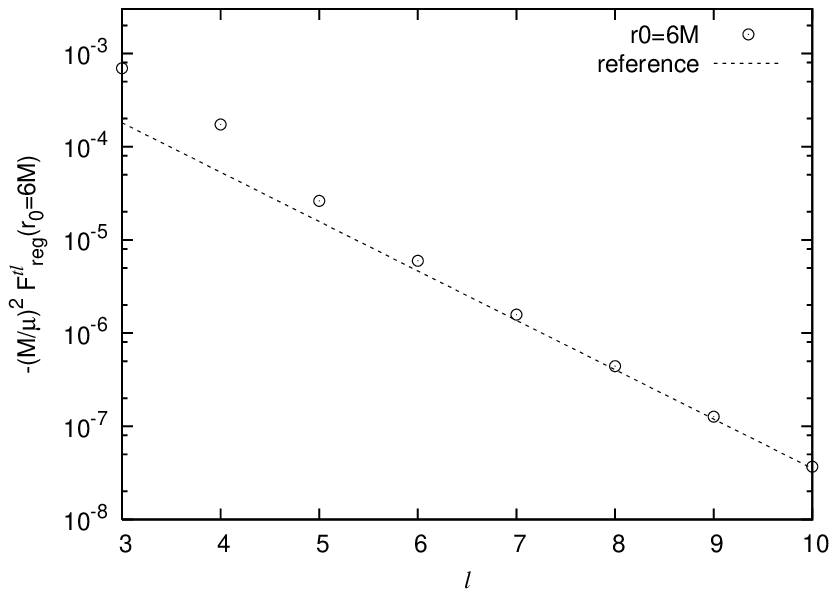} \\
\caption{Large-$l$ behavior of the regularized modes $F^{\alpha l}_{\rm reg}$.
The reference lines in the left panel are $\propto l^{-2}$; the reference line
in the right panel is $\propto \exp(-1.22l)$.
The left panel displays the radial component $F^{r l}_{\rm reg}$ (calculated using
internal derivatives) for $3\leq l\leq 25$ and for $r_0=6M,100M$. The regularized
modes appear to fall off as $\propto l^{-2}$ at large $l$, in agreement with theory.
The right panel shows the $l$-modes of the $t$
component (for $r_0=6M$); these are already ``regularized'', and show an exponential
fall-off at large $l$.
(Note the scale for the two panels is different: log-log for
the left panel, semi-log for the right.)}
\label{fig:large l}
\end{figure}

The fact that our numerically-derived full modes $[F_{\rm full}^{\alpha l}(x_0)]_{\pm}$
exhibit the right behaviour at large $l$, through {\em three} leading terms
in the $1/L$ expansion [$O(L)$ to $O(1/L)$] provides a very strong quantitative
check on our code. The plots in Fig.\ \ref{fig:large l} demonstrate that the regularized modes
$F_{\rm reg}^{\alpha l}\equiv [F_{\rm full}^{\alpha l}(x_0)]_{\pm}-A^{\alpha}_{\pm}L-B^{\alpha}$
fall off faster than $1/l$ at large $l$: The $r$ component falls off
as $\sim L^{-2}$, and the $t$ component falls off exponentially. For the $r$ component
this indicates that the ``singular'' (large $l$) part of the calculated full modes
is correctly described by the analytic regularization
function $A^{r}_{\pm}L+B^{r}$, through $O(1/L)$.
Of course, this agreement is necessary for a successful implementation of the mode-sum
formula: If the modes $[F_{\rm full}^{r l}(x_0)]_{\pm}$ were inconsistent with the
regularization function, the sum over modes would diverge.
Note also that the procedure for evaluating the large $l$ tail
(Sec.\ \ref{subsec:tail}) relies on the computed modes having the correct
behaviour at large $l$.
%~~~~~~~~~~~~~~~~~~~~~~~~~~~~~~~~~~~~~~~~~~~~~~~~~~~~~~~~~~~~~~~~~~~~~~~
%~~~~~~~~~~~~~~~~~~~~~~~~~~~~~~~~~~~~~~~~~~~~~~~~~~~~~~~~~~~~~~~~~~~~~~~

\subsection{Agreement between one-sided values of the SF}

In this analysis we applied the standard version of the mode-sum formula,
Eq.\ (\ref{ModeSum}), rather than the ``averaged'' version, Eq.\ (\ref{ModeSum2}).
For the $r$ component we carried out two independent calculations of the
SF, once using the internal values $[F_{\rm full}^{r l}(x_0)]_{-}$, and again
using the external values $[F_{\rm full}^{r l}(x_0)]_{+}$.
(The two one-sided values of the $t$ component
coincide automatically, as the full modes of this component are
continuous at the worldline.) This allows for an important consistency check:
The external and internal values of the final SF must agree. Any difference between
the two values is due to numerical error. Denoting the computational difference
between external and internal values of the final SF by $\Delta_{\pm}$, we find experimentally
%~~~~~~~~~~~~~~~~~~~~~~~~~~~~~~~~~~~~~~~~~~~~~~~~~~~~~~~~~~~~~~~~~~~~~~~
\begin{equation}\label{Deltapm}
\Delta_{\pm}/F^{r}\sim 10^{-5}\text{--}10^{-9}
\end{equation}
%~~~~~~~~~~~~~~~~~~~~~~~~~~~~~~~~~~~~~~~~~~~~~~~~~~~~~~~~~~~~~~~~~~~~~~~
(depending on $r_0$). This is well under the numerical error from discretisation.
The experimental values of $\Delta_{\pm}$, for the different
orbital radii considered, can be found in Table \ref{table:result-Fr} of
Sec.\ \ref{Sec:results}.

\subsection{Energy balance}\label{subsec:balance}

Equation (\ref{EdotLdot}) relates the temporal component of the SF to the
momentary rate of change of the (specific) orbital energy parameter $\cal E$.
In terms of time $t$, this relation becomes
$\dot{\cal E}=-(\mu u_0^t)^{-1}F_t$, where an overdot denotes $d/dt$,
and $u_0^t=(1-3M/r_0)^{-1/2}$.
If we assume that $\mu/M$ is small enough that radiation reaction is
negligible over an orbital period $T_{\rm orb}$ (``adiabatic approximation''),
then, for a circular orbit, $\dot{\cal E}$ also represents the average rate
of change of $\cal E$ over $T_{\rm orb}$. This must be balanced by the flux
of gravitational-wave energy radiated to infinity and through the horizon,
averaged over $T_{\rm orb}$. If we denote the former by $\dot{E}_{\infty}$
and the latter by $\dot{E}_{{\rm EH}}$ (both taken positive),
we have the energy balance formula
%~~~~~~~~~~~~~~~~~~~~~~~~~~~~~~~~~~~~~~~~~~~~~~~~~~~~~~~~~~~~~~~~~~~~~~~
\begin{equation}\label{balance}
\dot{E}_{\rm total}\equiv\dot{E}_{\infty}+\dot{E}_{\rm EH}=
-\mu \dot{\cal E}=F_t/u_0^t.
\end{equation}
%~~~~~~~~~~~~~~~~~~~~~~~~~~~~~~~~~~~~~~~~~~~~~~~~~~~~~~~~~~~~~~~~~~~~~~~
Both asymptotic fluxes $\dot{E}_{\infty}$ and $\dot{E}_{{\rm EH}}$ can be
constructed from the perturbation fields $\bar h^{(i)lm}$, evaluated at
the corresponding asymptotic domains. Validity of Eq.\ (\ref{balance})
then provides a strong qualitative test of our calculation.

We can readily express the asymptotic fluxes $\dot{E}_{\infty}$ and
$\dot{E}_{{\rm EH}}$ in terms of the $\bar h^{(i)lm}$'s, with the help of the
Weyl scalars, $\psi_0=-C_{\alpha\beta\gamma\delta}
l^{\alpha}m^{\beta}l^{\gamma}m^{\delta}$ and
$\psi_4=
-C_{\alpha\beta\gamma\delta}
n^{\alpha}m^{*\beta}n^{\gamma}m^{*\delta}$.
Here $C_{\alpha\beta\gamma\delta}$ is the Weyl tensor corresponding to the
perturbation $h_{\alpha\beta}$, and $l^{\alpha}$, $n^{\alpha}$, and $m^{\alpha}$
are the Kinnersley null vectors, given by
$l^{\alpha} = (f^{-1},1,0,0)$
$n^{\alpha} = \frac{1}{2}(1,-f,0,0)$, and
$m^{\alpha} =\frac{1}{\sqrt{2}r}(0,0,1,\frac{i}{\sin\theta})$.
Decomposing the perturbation as in Eq.\ (\ref{h construction1}),
we obtain the asymptotic relations
%~~~~~~~~~~~~~~~~~~~~~~~~~~~~~~~~~~~~~~~~~~~~~~~~~~~~~~~~~~~~~~~~~~~~~~~
\begin{eqnarray} \label{Weyl}
\psi_4(r\to\infty) &=&
\sum_{l=2}^{\infty}\sum_{m=-l}^{l}\frac{1}{4l(l+1)\lambda r}
\left( \ddot{\bar{h}}^{(7)} + i\ddot{\bar{h}}^{(10)} \right)
\left[ D_2Y_{lm}-i(\sin\theta)^{-1}D_1Y_{lm} \right], \nonumber\\
\psi_0 (r\to 2M)&=&
\sum_{l=2}^{\infty}\sum_{m=-l}^{l}
\frac{1}{4(2M)^3f^2} \left\{
\bar{h}^{(1)}+\bar{h}^{(2)}
+\frac{\bar{h}^{(4)}+\bar{h}^{(5)}
       -i(\bar{h}^{(8)}+\bar{h}^{(9)})}
      {l(l+1)} \right.
\nonumber \\ && \hspace*{1.5cm}
-\frac{4M}{l(l+1)}\left[
\dot{\bar{h}}^{(4)}+\dot{\bar{h}}^{(5)}
-i\left(\dot{\bar{h}}^{(8)}+\dot{\bar{h}}^{(9)}\right)
\right]
\nonumber \\ && \hspace*{1.5cm} \left.
-\frac{4M}{l(l+1)\lambda}
\left[
\dot{\bar{h}}^{(7)}-i\dot{\bar{h}}^{(10)}
-4M \left(\ddot{\bar{h}}^{(7)}-i\ddot{\bar{h}}^{(10)}\right)
\right]\right\}
\nonumber \\ && \hspace*{1.5cm} \times
\left( D_2Y_{lm}+i(\sin\theta)^{-1}D_1Y_{lm} \right).
\end{eqnarray}
%~~~~~~~~~~~~~~~~~~~~~~~~~~~~~~~~~~~~~~~~~~~~~~~~~~~~~~~~~~~~~~~~~~~~~~~
The first relation is valid at leading order in $1/r$, and the second at
leading order in $f$. In obtaining these relations we have made the
replacements $f\partial_r\to-\partial_t$ (for $\psi_4$ at null infinity)
and $f\partial_r\to\partial_t$ (for $\psi_0$ at the horizon).
For circular orbits, the asymptotic fluxes are given in terms of the Weyl
scalars as \cite{Teukolsky:1973ha,Teukolsky:1974yv}
%~~~~~~~~~~~~~~~~~~~~~~~~~~~~~~~~~~~~~~~~~~~~~~~~~~~~~~~~~~~~~~~~~~~~~~~
\begin{eqnarray}\label{WeylFlux}
\dot{E}_{\infty} &=&
\int d\tilde\Omega \frac{r^2}{4\pi m^2\Omega_0^2}
\left| \psi_4(r\to\infty) \right|^2, \\
\dot{E}_{\rm EH} &=&
\int d\tilde\Omega \frac{M^4 f^4}{\pi(1+16M^2m^2\Omega_0^2)}
\left| \psi_0(r\to 2M) \right|^2,
\end{eqnarray}
%~~~~~~~~~~~~~~~~~~~~~~~~~~~~~~~~~~~~~~~~~~~~~~~~~~~~~~~~~~~~~~~~~~~~~~~
where the integration is carried out over 2-spheres $r={\rm const}\to\infty$
and $r=2M$, respectively. Now proceed as follows:
(i) Substitute Eqs.\ (\ref{Weyl}) in Eqs.\ (\ref{WeylFlux}).
(ii) For $\dot{E}_{\rm EH}$ apply the asymptotic gauge conditions
$\bar{h}^{(2)}=\bar{h}^{(1)}$, $\bar{h}^{(4)}=\bar{h}^{(5)}$, and
$\bar{h}^{(8)}=\bar{h}^{(9)}$ [see Eqs.\ (\ref{gauge1})--(\ref{gauge4})
in Appendix \ref{AppA}].
(iii) Replace $d/dt\to -im\Omega_0$.
(iv) Integrate over the spheres using the formulas (A4) of BL.
This yields the final relations
%~~~~~~~~~~~~~~~~~~~~~~~~~~~~~~~~~~~~~~~~~~~~~~~~~~~~~~~~~~~~~~~~~~~~~~~
\begin{equation}\label{fluxinfty}
\dot{E}_{\infty}= \sum_{l=2}^{\infty}\sum_{m=-l}^{l}
\frac{\mu^2m^2\Omega_0^2}{64\pi\lambda l(l+1)}
\left|\bar h^{(7)}_{\infty}-i\bar h^{(10)}_{\infty}\right|^2,
\end{equation}
%~~~~~~~~~~~~~~~~~~~~~~~~~~~~~~~~~~~~~~~~~~~~~~~~~~~~~~~~~~~~~~~~~~~~~~~
%~~~~~~~~~~~~~~~~~~~~~~~~~~~~~~~~~~~~~~~~~~~~~~~~~~~~~~~~~~~~~~~~~~~~~~~
\begin{eqnarray}\label{fluxEH}
\dot{E}_{\rm EH}&=& \sum_{l=2}^{\infty}\sum_{m=-l}^{l}
\frac{\mu^2 \lambda l(l+1)}{256\pi M^2(1+16M^2m^2\Omega_0^2)}
\nonumber\\
&&\times
\left|
\bar h^{(1)}_{\rm EH}+
\frac{1+4iMm\Omega_0}{l(l+1)}\left[
\bar h^{(5)}_{\rm EH}-i\bar h^{(9)}_{\rm EH}
+2iMm\Omega_0\lambda^{-1}
\left(\bar h^{(7)}_{\rm EH}-i\bar h^{(10)}_{\rm EH}\right)
\right]\right|^2,
\end{eqnarray}
%~~~~~~~~~~~~~~~~~~~~~~~~~~~~~~~~~~~~~~~~~~~~~~~~~~~~~~~~~~~~~~~~~~~~~~~
where $\lambda=(l-1)(l+2)$,
$\bar h^{(i)}_{\infty}$ are the fields $\bar h^{(i)lm}$ evaluated
at null infinity ($u\ll v\to\infty$), and $\bar h^{(i)}_{\rm EH}$
are these fields evaluated at the event horizon ($v\ll u\to\infty$).
Eq.\ (\ref{fluxinfty}) for $\dot{E}_{\infty}$ agrees with Eq.\ (57) of BL,
which was derived directly from the Issacson effective energy-momentum tensor.

%BL do not
%give the expression for $\dot{E}_{\rm EH}$, but this can be obtained following
%the line of derivation in their Sec.\ IV-C.\footnote{To obtain
%$\dot{E}_{\rm EH}$, start with the Issacson effective energy-momentum tensor
%[generalized to a non-traceless gauge, Eq.\ (53) of BL], and observe that
%Eq.\ (54) of BL is still valid replacing $\dot{E}_{\infty}\to \dot{E}_{\rm EH}$
%and carrying the 2-sphere integral over the horizon. This integral
%still takes care automatically of the averaging over a large scale, as
%the wavelength of the blue-shifted gravitational waves at the horizon is
%much smaller than the background curvature length-scale there ($\sim M$).
%Hence, one recovers Eq.\ (55) of BL with `$\infty$'$\to$`EH'.
%Noticing that the asymptotic gauge conditions in BL's Eq. (56) are valid
%%at the horizon as well, one concludes that the horizon flux has the same
%functional form as the flux at infinity, Eq. (57) therein.
%}

To test our calculation of $F_t$, we used our evolution code to obtain the
energy fluxes at infinity and through the horizon, based on Eqs.\ (\ref{fluxinfty})
and (\ref{fluxEH}), and then checked consistency with the balance equation (\ref{balance}).
To extract $\dot{E}_{\infty}$ we evaluated the numerical solutions $\bar h^{(7,10)}$
at $v=5200M,9000M,15000M$ and $u=800M,3000M,5000M$, for orbital radii
in the ranges $6M\leq r_0<20M$, $20M\leq r_0\le 100M$, and $r_0> 100M$, respectively.
To derive $\dot{E}_{\rm EH}$ we evaluated $\bar h^{(1,5,7,9,10)}$
at $u=5200M,6500M,7500M$ and $v=800M,1500M,2500M$ for the above corresponding
values of $r_0$. These values were selected experimentally such that the fractional
error in the total flux (from the finite extraction distance and from the
spurious initial waves) is less than $10^{-4}$ for each of the $l$ modes.
We then used these values in Eqs.\ (\ref{fluxinfty}) and (\ref{fluxEH}),
summing from $l=2$ to $l=l_{\rm max}$, where $l_{\rm max}$ was determined
experimentally, requiring that the fractional truncation error in the total
flux (from omitting the modes $l>l_{\rm max}$) is $<10^{-4}$. This required
$l_{\rm max}$ values between $9$ (for $r_0=6M$) and 4 (for $r_0=150M$).

Following the above procedure, we obtained $\dot{E}_{\infty}$ and
$\dot{E}_{\rm EH}$ for a list of orbital radii between $6M$ and $150M$.
The ratio $\dot{E}_{\rm EH}/\dot{E}_{\infty}$ turns out very small for
all radii, decreasing monotonically with $r_0$ from $3.3\times 10^{-3}$ at
$r_0=6M$ to $2.4\times 10^{-9}$ at $r_0=150M$ (these values are consistent
with Martel's \cite{Martel:2003jj}). The values obtained for the {\em total} energy
flux, $\dot{E}_{\rm total}=\dot{E}_{\infty}+\dot{E}_{\rm EH}$, are listed
in Table~\ref{table:compare-Edot} of Sec.\ \ref{Sec:results} below.
We find that the fractional difference
$\left| u_0^t\dot{E}_{\rm total}/F_t-1\right|$
is less than $\sim 5\times 10^{-4}$ in all cases, providing a strong
quantitative check of our results.

%%%%%%%%%%%%%%%%%%%%%%%%%%%%%%%%%%%%%%%%%%%%%%%%%%%%%%%%%%%%%
\section{Results}
         \label{Sec:results}
%%%%%%%%%%%%%%%%%%%%%%%%%%%%%%%%%%%%%%%%%%%%%%%%%%%%%%%%%%%%%

\subsection{Temporal component}

We calculated the $t$ component of the SF for 29 values of the orbital
radius, in the range from $r_0=6M$ to $r_0=150M$, using the procedure
described in Sec.\ \ref{Sec:numerics}.  The results are displayed in
Table~\ref{table:compare-Edot}. The computation error in $F^t$ is estimated
at $\lesssim 10^{-4}$ for all radii considered. The table also shows, for each
of the radii considered, how the work done by the temporal component of the local
SF is balanced by the total flux of radiated energy.
%~~~~~~~~~~~~~~~~~~~~~~~~~~~~~~~~~~~~~~~~~~~~~~~~~~~~~~~~~~~~~~~~~~~~~~~
\begin{table}
\begin{tabular}{c|c||c|c|c}
\hline\hline
$r_0/M$ & $(M/\mu)^2F^t$
       & $(M/\mu)^2 F_t/u_0^t$
       & $(M/\mu)^2\dot{E}_{\rm total}$
       & rel. diff. \\
\hline
 6.0  & $-1.99476 \times 10^{-3}$ [$7 \times 10^{-5}$] [$6 \times 10^{-6}$]
       & $9.40338 \times 10^{-4}$
       & $9.40190 \times 10^{-4}$
       & $1.6 \times 10^{-4}$ \\
 6.2  & $-1.60515 \times 10^{-3}$ [$8 \times 10^{-5}$] [$1 \times 10^{-5}$]
       & $7.81183 \times 10^{-4}$
       & $7.81064 \times 10^{-4}$
       & $1.5 \times 10^{-4}$ \\
 6.4  & $-1.30550 \times 10^{-3}$ [$7 \times 10^{-5}$] [$1 \times 10^{-5}$]
       & $6.54180 \times 10^{-4}$
       & $6.54101 \times 10^{-4}$
       & $1.2 \times 10^{-4}$ \\
 6.6  & $-1.07197 \times 10^{-3}$ [$7 \times 10^{-5}$] [$1 \times 10^{-5}$]
       & $5.51794 \times 10^{-4}$
       & $5.51723 \times 10^{-4}$
       & $1.3 \times 10^{-4}$ \\
 6.8  & $-8.87844 \times 10^{-4}$ [$6 \times 10^{-5}$] [$1 \times 10^{-5}$]
       & $4.68497 \times 10^{-4}$
       & $4.68411 \times 10^{-4}$
       & $1.8 \times 10^{-4}$ \\
 7.0  & $-7.41101 \times 10^{-4}$ [$7 \times 10^{-5}$] [$1 \times 10^{-5}$]
       & $4.00157 \times 10^{-4}$
       & $4.00117 \times 10^{-4}$
       & $1.0 \times 10^{-4}$ \\
 7.2  & $-6.23065 \times 10^{-4}$ [$7 \times 10^{-5}$] [$1 \times 10^{-5}$]
       & $3.43687 \times 10^{-4}$
       & $3.43627 \times 10^{-4}$
       & $1.7 \times 10^{-4}$ \\
 7.4  & $-5.27271 \times 10^{-4}$ [$6 \times 10^{-5}$] [$1 \times 10^{-5}$]
       & $2.96692 \times 10^{-4}$
       & $2.96645 \times 10^{-4}$
       & $1.6 \times 10^{-4}$ \\
 7.6  & $-4.48905 \times 10^{-4}$ [$6 \times 10^{-5}$] [$1 \times 10^{-5}$]
       & $2.57336 \times 10^{-4}$
       & $2.57288 \times 10^{-4}$
       & $1.9 \times 10^{-4}$ \\
 7.8  & $-3.84324 \times 10^{-4}$ [$5 \times 10^{-5}$] [$1 \times 10^{-5}$]
       & $2.24184 \times 10^{-4}$
       & $2.24148 \times 10^{-4}$
       & $1.6 \times 10^{-4}$ \\
 8.0  & $-3.30740 \times 10^{-4}$ [$5 \times 10^{-5}$] [$1 \times 10^{-5}$]
       & $1.96105 \times 10^{-4}$
       & $1.96066 \times 10^{-4}$
       & $2.0 \times 10^{-4}$ \\
 9.0  & $-1.66810 \times 10^{-4}$ [$5 \times 10^{-5}$] [$1 \times 10^{-5}$]
       & $1.05933 \times 10^{-4}$
       & $1.05908 \times 10^{-4}$
       & $2.4 \times 10^{-4}$ \\
 10.0  & $-9.19067 \times 10^{-5}$ [$3 \times 10^{-5}$] [$9 \times 10^{-6}$]
       & $6.15158 \times 10^{-5}$
       & $6.15047 \times 10^{-5}$
       & $1.8 \times 10^{-4}$ \\
 11.0  & $-5.41605 \times 10^{-5}$ [$3 \times 10^{-5}$] [$2 \times 10^{-5}$]
       & $3.77904 \times 10^{-5}$
       & $3.77856 \times 10^{-5}$
       & $1.3 \times 10^{-4}$ \\
 12.0  & $-3.36587 \times 10^{-5}$ [$2 \times 10^{-5}$] [$2 \times 10^{-5}$]
       & $2.42911 \times 10^{-5}$
       & $2.42857 \times 10^{-5}$
       & $2.2 \times 10^{-4}$ \\
 13.0  & $-2.18388 \times 10^{-5}$ [$2 \times 10^{-5}$] [$2 \times 10^{-5}$]
       & $1.62071 \times 10^{-5}$
       & $1.62022 \times 10^{-5}$
       & $3.1 \times 10^{-4}$ \\
 14.0  & $-1.46851 \times 10^{-5}$ [$1 \times 10^{-4}$] [$2 \times 10^{-5}$]
       & $1.11574 \times 10^{-5}$
       & $1.11564 \times 10^{-5}$
       & $8.5 \times 10^{-5}$ \\
 15.0  & $-1.01772 \times 10^{-5}$ [$1 \times 10^{-4}$] [$2 \times 10^{-5}$]
       & $7.88904 \times 10^{-6}$
       & $7.88597 \times 10^{-6}$
       & $3.9 \times 10^{-4}$ \\
 20.0  & $-2.25549 \times 10^{-6}$ [$6 \times 10^{-5}$] [$6 \times 10^{-6}$]
       & $1.87151 \times 10^{-6}$
       & $1.87111 \times 10^{-6}$
       & $2.2 \times 10^{-4}$ \\
 30.0  & $-2.80813 \times 10^{-7}$ [$4 \times 10^{-5}$] [$4 \times 10^{-5}$]
       & $2.48643 \times 10^{-7}$
       & $2.48600 \times 10^{-7}$
       & $1.7 \times 10^{-4}$ \\
 40.0  & $-6.51219 \times 10^{-8}$ [$3 \times 10^{-5}$] [$3 \times 10^{-6}$]
       & $5.95007 \times 10^{-8}$
       & $5.94897 \times 10^{-8}$
       & $1.8 \times 10^{-4}$ \\
 50.0  & $-2.10849 \times 10^{-8}$ [$2 \times 10^{-5}$] [$4 \times 10^{-5}$]
       & $1.96249 \times 10^{-8}$
       & $1.96203 \times 10^{-8}$
       & $2.3 \times 10^{-4}$ \\
 60.0  & $-8.41306 \times 10^{-9}$ [$9 \times 10^{-5}$] [$3 \times 10^{-5}$]
       & $7.92670 \times 10^{-9}$
       & $7.92424 \times 10^{-9}$
       & $3.1 \times 10^{-4}$ \\
 70.0  & $-3.87411 \times 10^{-9}$ [$8 \times 10^{-5}$] [$4 \times 10^{-5}$]
       & $3.68189 \times 10^{-9}$
       & $3.68086 \times 10^{-9}$
       & $2.8 \times 10^{-4}$ \\
 80.0  & $-1.98069 \times 10^{-9}$ [$7 \times 10^{-5}$] [$8 \times 10^{-5}$]
       & $1.89462 \times 10^{-9}$
       & $1.89360 \times 10^{-9}$
       & $5.4 \times 10^{-4}$ \\
 90.0  & $-1.09654 \times 10^{-9}$ [$6 \times 10^{-5}$] [$6 \times 10^{-5}$]
       & $1.05415 \times 10^{-9}$
       & $1.05365 \times 10^{-9}$
       & $4.8 \times 10^{-4}$ \\
 100.0  & $-6.46305 \times 10^{-10}$ [$6 \times 10^{-5}$] [$4 \times 10^{-5}$]
       & $6.23806 \times 10^{-10}$
       & $6.23628 \times 10^{-10}$
       & $2.9 \times 10^{-4}$ \\
 120.0  & $-2.59096 \times 10^{-10}$ [$5 \times 10^{-5}$] [$3 \times 10^{-5}$]
       & $2.51573 \times 10^{-10}$
       & $2.51496 \times 10^{-10}$
       & $3.1 \times 10^{-4}$ \\
 150.0  & $-8.47172 \times 10^{-11}$ [$4 \times 10^{-5}$] [$6 \times 10^{-5}$]
       & $8.27475 \times 10^{-11}$
       & $8.27279 \times 10^{-11}$
       & $2.4 \times 10^{-4}$ \\
\hline\hline
\end{tabular}
\caption{
The temporal component of the SF, as a function of the orbital radius $r_0$.
Values in the first square brackets in the second column are estimates of the {\it fractional}
numerical error in $F^t$ from the finite-grid discretization, $\Delta^t_{\rm discr}$
[see Eq.\ (\ref{Deltah})].
Values in the second square brackets in the second column are estimates of the
{\it fractional} error from residual non-stationarity of the late-time numerical
evolution (which is mainly due to leftover spurious waves arising from the imperfect
initial data). The third and fourth columns compare between the work done by
the temporal SF and the total flux of energy radiated in gravitational waves,
the latter extracted from the numerical solutions using the procedure described
in Sec.\ \ref{subsec:balance}. The last column displays the
relative difference $\left|\dot{E}_{\rm total}/(F_t/u_0^t)-1\right|$, showing that
the balance equation (\ref{balance}) is satisfied within the numerical accuracy,
and providing a strong quantitative check of our results.}
\label{table:compare-Edot}
\end{table}
%~~~~~~~~~~~~~~~~~~~~~~~~~~~~~~~~~~~~~~~~~~~~~~~~~~~~~~~~~~~~~~~~~~~~~~~

\subsection{Radial component}

We calculated the radial component of the SF for 29 values of the orbital
radius, in the range from $r_0=6M$ to $t_0=150M$.
Tables~\ref{table:highL-effect} and \ref{table:highL-effect2} in Appendix
\ref{AppD}
present results for the internal and external values of the SF, respectively
(recall the two one-sided values are expected to agree with each other, within
numerical error). In each table we indicate separately the two contributions
$[F^r_{l\leq 15}]_{\pm}$ and $[F^r_{l>15}]_{\pm}$,
where the former is the part calculated directly using our evolution code,
and the latter is the extrapolated contribution from $l>15$, calculated
as explained in Sec.\ \ref{subsec:tail} above. The large-$l$ tail contributes
at most 2\% of the total SF (depending on $r_0$). The relative fractional
error in $[F^r_{l>15}]_{\pm}$ [i.e., $\Delta_{\rm tail,rel}$, calculated through
Eq.\ (\ref{Deltafit})] is at most comparable to (and mostly much smaller than)
the fractional discretization error in $[F^r_{l\leq 15}]_{\pm}$,
which itself is at most $\sim 10^{-3}$.

In Table~\ref{table:result-Fr} we present our final results for the radial
component: the SF as a function of the orbital radius $r_0$. As the `final'
result we quote the {\em average} between the two (nearly identical)
one-sided values. The third column displays the magnitude of the {\em difference}
between the two one-sided values, which is entirely due to computational error.
The relative magnitude of this error (given in square brackets in the third
column) is in all cases much smaller than the fractional discretization
error in $F^r$. The latter error, given in square bracket in the second column,
is taken as the average between the two one-sided discterization errors
$\Delta^r_{\rm discr}$ estimated from Eq.\ (\ref{Deltah}).

The last column of Table~\ref{table:result-Fr} displays the estimated
error from residual non-stationarity of the numerical solutions (from
residual spurious initial waves). Displayed is the difference in the
values $F^r$ obtained at different evolution times (as explained in
more detail in Sec.\ \ref{subsec:initial} above). The values in square brackets
in the last column show the fractional error from non-stationarity relative
to the total SF. This error is in all cases much smaller than the discretization
error.

Thus, the dominant source of error in our analysis is associated with the finite-grid
discretization of the field equations. We estimate our final results for
$F^r$ to be correct to within at least $\sim 0.1\%$ for all orbital radii considered.
The results for $10M\lesssim r_0\lesssim 20M$ are likely to be correct to within
$\sim 0.01\%$, and the results for $r_0\gtrsim 20M$ to within mere $\sim 0.001\%$.
%~~~~~~~~~~~~~~~~~~~~~~~~~~~~~~~~~~~~~~~~~~~~~~~~~~~~~~~~~~~~~~~~~~~~~~~
\begin{table}
\begin{tabular}{c|c|c|c}
\hline\hline
  &   & Error from disagreement & Error from \\
$r_0/M$ & ${F}^r\times(M/\mu)^2$
       & $[F^r]_+\leftrightarrow[F^r]_-$
       & non-stationarity\\
\hline
 6.0  & $2.44661 \times 10^{-2}$ [$9\times 10^{-4}$]
       & $1.20 \times 10^{-7}$ [$5 \times 10^{-6}$]
       & $1.83 \times 10^{-8}$ [$7 \times 10^{-7}$] \\
 6.2  & $2.39651 \times 10^{-2}$ [$9\times 10^{-4}$]
       & $5.01 \times 10^{-9}$ [$2 \times 10^{-7}$]
       & $5.92 \times 10^{-8}$ [$2 \times 10^{-6}$] \\
 6.4  & $2.33954 \times 10^{-2}$ [$8\times 10^{-4}$]
       & $8.54 \times 10^{-8}$ [$4 \times 10^{-6}$]
       & $3.91 \times 10^{-8}$ [$2 \times 10^{-6}$] \\
 6.6  & $2.27829 \times 10^{-2}$ [$7\times 10^{-4}$]
       & $1.56 \times 10^{-7}$ [$7 \times 10^{-6}$]
       & $4.24 \times 10^{-8}$ [$2 \times 10^{-6}$] \\
 6.8  & $2.21462 \times 10^{-2}$ [$7\times 10^{-4}$]
       & $2.11 \times 10^{-7}$ [$1 \times 10^{-5}$]
       & $3.21 \times 10^{-8}$ [$1 \times 10^{-6}$] \\
 7.0  & $2.14989 \times 10^{-2}$ [$6\times 10^{-4}$]
       & $2.42 \times 10^{-7}$ [$1 \times 10^{-5}$]
       & $2.64 \times 10^{-8}$ [$1 \times 10^{-6}$] \\
 7.2  & $2.08504 \times 10^{-2}$ [$6\times 10^{-4}$]
       & $2.75 \times 10^{-7}$ [$1 \times 10^{-5}$]
       & $2.24 \times 10^{-8}$ [$1 \times 10^{-6}$] \\
 7.4  & $2.02078 \times 10^{-2}$ [$6\times 10^{-4}$]
       & $3.02 \times 10^{-7}$ [$1 \times 10^{-5}$]
       & $1.93 \times 10^{-8}$ [$1 \times 10^{-6}$] \\
 7.6  & $1.95761 \times 10^{-2}$ [$5\times 10^{-4}$]
       & $3.18 \times 10^{-7}$ [$2 \times 10^{-5}$]
       & $1.59 \times 10^{-8}$ [$8 \times 10^{-7}$] \\
 7.8  & $1.89586 \times 10^{-2}$ [$5\times 10^{-4}$]
       & $3.28 \times 10^{-7}$ [$2 \times 10^{-5}$]
       & $1.32 \times 10^{-8}$ [$7 \times 10^{-7}$] \\
 8.0  & $1.83577 \times 10^{-2}$ [$5\times 10^{-4}$]
       & $3.34 \times 10^{-7}$ [$2 \times 10^{-5}$]
       & $1.13 \times 10^{-8}$ [$6 \times 10^{-7}$] \\
 9.0  & $1.56369 \times 10^{-2}$ [$4\times 10^{-4}$]
       & $3.23 \times 10^{-7}$ [$2 \times 10^{-5}$]
       & $5.70 \times 10^{-9}$ [$4 \times 10^{-7}$] \\
 10.0  & $1.33895 \times 10^{-2}$ [$8 \times 10^{-5}$]
       & $1.00 \times 10^{-9}$ [$7 \times 10^{-8}$]
       & $2.89 \times 10^{-9}$ [$2 \times 10^{-7}$] \\
 11.0  & $1.15518 \times 10^{-2}$ [$6 \times 10^{-5}$]
       & $1.55 \times 10^{-9}$ [$1 \times 10^{-7}$]
       & $1.49 \times 10^{-11}$ [$1 \times 10^{-9}$] \\
 12.0  & $1.00463 \times 10^{-2}$ [$5 \times 10^{-5}$]
       & $9.79 \times 10^{-10}$ [$1 \times 10^{-7}$]
       & $7.84 \times 10^{-12}$ [$8 \times 10^{-10}$] \\
 13.0  & $8.80489 \times 10^{-3}$ [$4 \times 10^{-5}$]
       & $3.38 \times 10^{-9}$ [$4 \times 10^{-7}$]
       & $1.96 \times 10^{-9}$ [$2 \times 10^{-7}$] \\
 14.0  & $7.77307 \times 10^{-3}$ [$1 \times 10^{-5}$]
       & $1.50 \times 10^{-9}$ [$2 \times 10^{-7}$]
       & $1.62 \times 10^{-9}$ [$2 \times 10^{-7}$] \\
 15.0  & $6.90815 \times 10^{-3}$ [$3 \times 10^{-5}$]
       & $1.17 \times 10^{-9}$ [$2 \times 10^{-7}$]
       & $1.25 \times 10^{-9}$ [$2 \times 10^{-7}$] \\
 20.0  & $4.15706 \times 10^{-3}$ [$1 \times 10^{-5}$]
       & $1.76 \times 10^{-10}$ [$4 \times 10^{-8}$]
       & $7.11 \times 10^{-10}$ [$2 \times 10^{-7}$] \\
 30.0  & $1.96982 \times 10^{-3}$ [$5 \times 10^{-6}$]
       & $8.39 \times 10^{-11}$ [$4 \times 10^{-8}$]
       & $2.86 \times 10^{-10}$ [$1 \times 10^{-7}$] \\
 40.0  & $1.14288 \times 10^{-3}$ [$2 \times 10^{-6}$]
       & $1.65 \times 10^{-11}$ [$1 \times 10^{-8}$]
       & $8.78 \times 10^{-11}$ [$8 \times 10^{-8}$] \\
 50.0  & $7.44949\times 10^{-4}$ [$1 \times 10^{-6}$]
       & $3.03 \times 10^{-12}$ [$4 \times 10^{-9}$]
       & $4.98 \times 10^{-11}$ [$7 \times 10^{-8}$] \\
 60.0  & $5.23613\times 10^{-4}$ [$2 \times 10^{-5}$]
       & $4.86 \times 10^{-10}$ [$9 \times 10^{-7}$]
       & $1.57 \times 10^{-11}$ [$3 \times 10^{-8}$] \\
 70.0  & $3.88010\times 10^{-4}$ [$1 \times 10^{-5}$]
       & $2.47 \times 10^{-10}$ [$6 \times 10^{-7}$]
       & $1.79 \times 10^{-12}$ [$5 \times 10^{-9}$] \\
 80.0  & $2.98979\times 10^{-4}$ [$8 \times 10^{-6}$]
       & $1.36 \times 10^{-10}$ [$5 \times 10^{-7}$]
       & $1.39 \times 10^{-11}$ [$5 \times 10^{-8}$] \\
 90.0  & $2.37406\times 10^{-4}$ [$7 \times 10^{-6}$]
       & $8.01 \times 10^{-11}$ [$3 \times 10^{-7}$]
       & $1.58 \times 10^{-11}$ [$7 \times 10^{-8}$] \\
 100.0  & $1.93063\times 10^{-4}$ [$5 \times 10^{-6}$]
       & $4.90 \times 10^{-11}$ [$3 \times 10^{-7}$]
       & $1.53 \times 10^{-11}$ [$8 \times 10^{-8}$] \\
 120.0  & $1.34868\times 10^{-4}$ [$4 \times 10^{-6}$]
       & $2.17 \times 10^{-11}$ [$2 \times 10^{-7}$]
       & $1.22 \times 10^{-12}$ [$9 \times 10^{-9}$] \\
 150.0  & $8.68274 \times 10^{-5}$ [$2 \times 10^{-6}$]
       & $7.62 \times 10^{-12}$ [$9 \times 10^{-8}$]
       & $1.48 \times 10^{-11}$ [$2 \times 10^{-7}$] \\
\hline\hline
\end{tabular}
\caption{Final results for the radial component of the SF.
The second column lists the values obtained for the various orbital
radii $r_0$, taken as the average between internal and external values:
$([F^r]_++[F^r]_-)/2$.
Values in square brackets in the second column are estimates of the {\it fractional}
numerical error in $F^r$ from the finite-grid discretization (see the text for details).
The third and fourth columns display estimates of the magnitude of error from two other sources:
The third column shows the magnitude of the difference $[F^r]_+-[F^r]_-$,
which is entirely due to numerical error; the values in square brackets
give the fractional error $2([F^r]_+-[F^r]_-)/([F^r]_++[F^r]_-)$.
The fourth column shows the estimated error from residual non-stationarity of
the late-time numerical evolution, which is mainly due to leftover spurious waves
arising from the imperfect initial data; values in square brackets again describe
the fractional error. Both sources of error contribute negligibly to the
overall error in the SF, which is therefore dominated by the discretization error.}
\label{table:result-Fr}
\end{table}
%~~~~~~~~~~~~~~~~~~~~~~~~~~~~~~~~~~~~~~~~~~~~~~~~~~~~~~~~~~~~~~~~~~~~~~~

\begin{figure}[tbp]
\includegraphics[width=10cm]{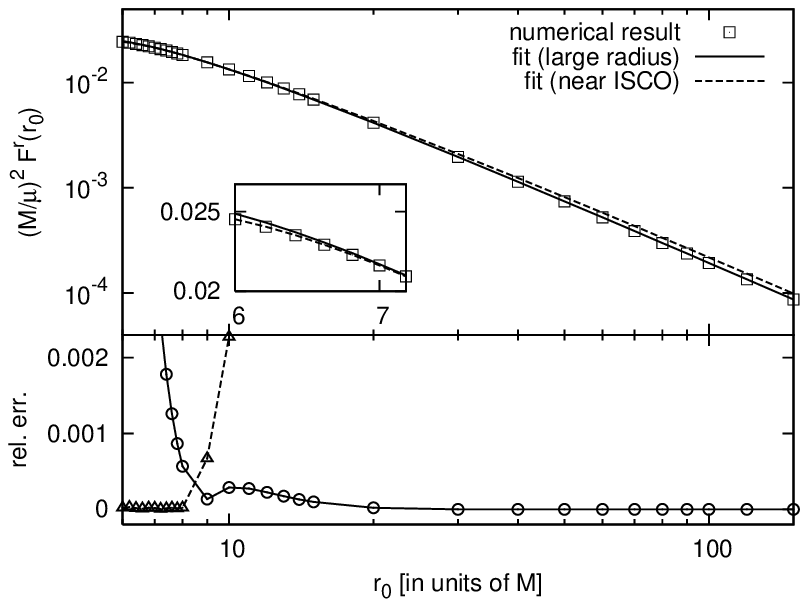}
\caption{The radial component of the SF.
The data in the upper panel correspond to the second column in Table
\ref{table:result-Fr}. % The values of $F^r$ are normalized by $(\mu/M)^2$.
The inset shows an expansion of the ISCO area. The solid line is a plot
of the large-$r_0$ analytic fit given in Eq.\ (\ref{eq:Fr-PNform}).
The dashed line represents the complementary near-ISCO fit (\ref{eq:Fr-xform}).
The lower panel shows the relative difference between the numerical data and
values obtained using the analytic fit formulas. Using the large-$r_0$ fit for
$r_0\geq 8M$ and the near-ISCO fit for $6M \leq r_0\leq 8M$, one recovers
all numerical data to within the numerical accuracy.}
\label{fig:r0vsFr}
\end{figure}

We plot $F^r(r_0)$ in Fig.\ \ref{fig:r0vsFr}. The radial SF is ``repulsive''
(i.e, acting outward, away from the central black hole) for all $r_0$.
At large orbital radii the numerical data can be fitted analytically as
%~~~~~~~~~~~~~~~~~~~~~~~~~~~~~~~~~~~~~~~~~~~~~~~~~~~~~~~~~~~~~~~~~~~~~~~
\begin{equation}
F^r (r_0\gg M) \simeq
\frac{\mu^2}{r_0^2}
\left[
a_0 + a_1 \frac{M}{r_0}
+ a_2 \left( \frac{M}{r_0} \right)^2
+ a_3 \left( \frac{M}{r_0} \right)^3
\right],
\label{eq:Fr-PNform}
\end{equation}
%~~~~~~~~~~~~~~~~~~~~~~~~~~~~~~~~~~~~~~~~~~~~~~~~~~~~~~~~~~~~~~~~~~~~~~~
with
%~~~~~~~~~~~~~~~~~~~~~~~~~~~~~~~~~~~~~~~~~~~~~~~~~~~~~~~~~~~~~~~~~~~~~~~
\begin{equation} \label{a}
a_0=1.999991,\quad
a_1=-6.9969,\quad
a_2=6.29,\quad
a_3=-24.6.
\end{equation}
%~~~~~~~~~~~~~~~~~~~~~~~~~~~~~~~~~~~~~~~~~~~~~~~~~~~~~~~~~~~~~~~~~~~~~~~
This formula reproduces the numerical data within the numerical accuracy
[$\lesssim 10^{-3}$] for all $r_0\geq 8M$. The leading-order term,
$F^r \simeq a_0\mu^2/r_0^2\simeq 2\mu^2/r_0^2$ is consistent with the
``Keplerian'' SF describing the back-reaction effect from the motion of the
black hole about the system's center of mass (we discuss this below, when
considering the SF correction to the orbital frequency).
Near the Innermost Stable Circular Orbit (ISCO), $r_0=6M$, we fit the numerical data
analytically as
%~~~~~~~~~~~~~~~~~~~~~~~~~~~~~~~~~~~~~~~~~~~~~~~~~~~~~~~~~~~~~~~~~~~~~~~
\begin{equation}
F^r(r_0\gtrsim 6M) \simeq
\frac{\mu^2}{r_0^2}(1-2M/r_0)
\left(
b_0 + b_1 x_0 + b_2 x_0^2 +b_3 x_0^3
\right),
\label{eq:Fr-xform}
\end{equation}
%~~~~~~~~~~~~~~~~~~~~~~~~~~~~~~~~~~~~~~~~~~~~~~~~~~~~~~~~~~~~~~~~~~~~~~~
where $x_0\equiv 1-6M/r_0$ and the `best fit' parameters (based on data in
$6M\leq r_0\leq 8M$) are given by
%~~~~~~~~~~~~~~~~~~~~~~~~~~~~~~~~~~~~~~~~~~~~~~~~~~~~~~~~~~~~~~~~~~~~~~~
\begin{equation}
b_0=1.32120, \quad
b_1=1.2391, \quad
b_2=-1.297, \quad
b_3=1.07.
\end{equation}
%~~~~~~~~~~~~~~~~~~~~~~~~~~~~~~~~~~~~~~~~~~~~~~~~~~~~~~~~~~~~~~~~~~~~~~~
This reproduces the numerical data within the numerical accuracy
for all $r_0\leq 8M$.\footnote{The analytic-fit formula (\ref{eq:Fr-xform})
should not be regarded as representing the first few terms in a convergent
expansion in $x\equiv 1-6M/r_0$; In fact, it is evident from Fig.\ \ref{fig:r0vsFr}
(lower panel) that such a series is likely to have a very small radius of
convergence around $r_0=6M$. We give the analytic fit (\ref {eq:Fr-xform})
merely for practical reasons, as it correctly gives the value of the SF
anywhere in the strong-field regime $6M\leq r_0\leq 8M$ to within our
numerical error ($\lesssim 10^{-3}$).}

\subsection{Conservative shift in the orbital parameters}

\begin{figure}[htb]
\includegraphics[width=8cm]{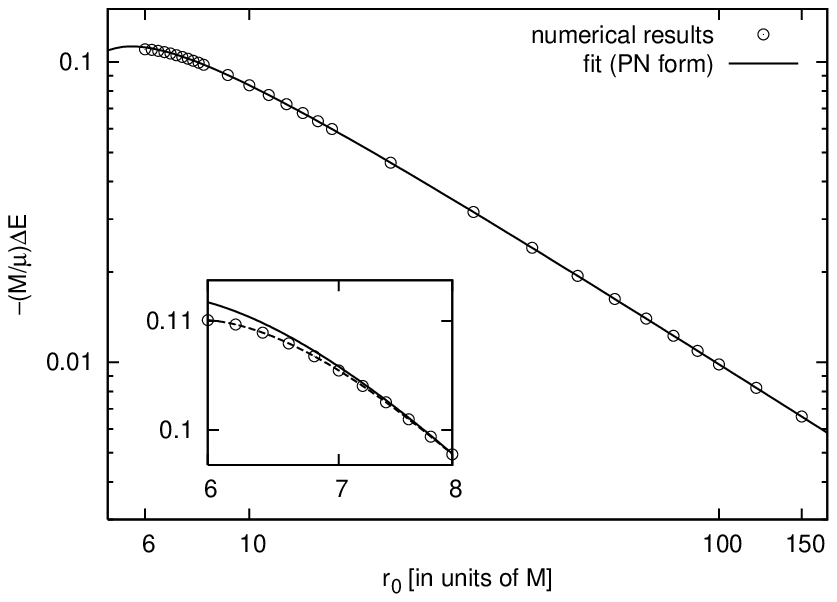}
\includegraphics[width=8cm]{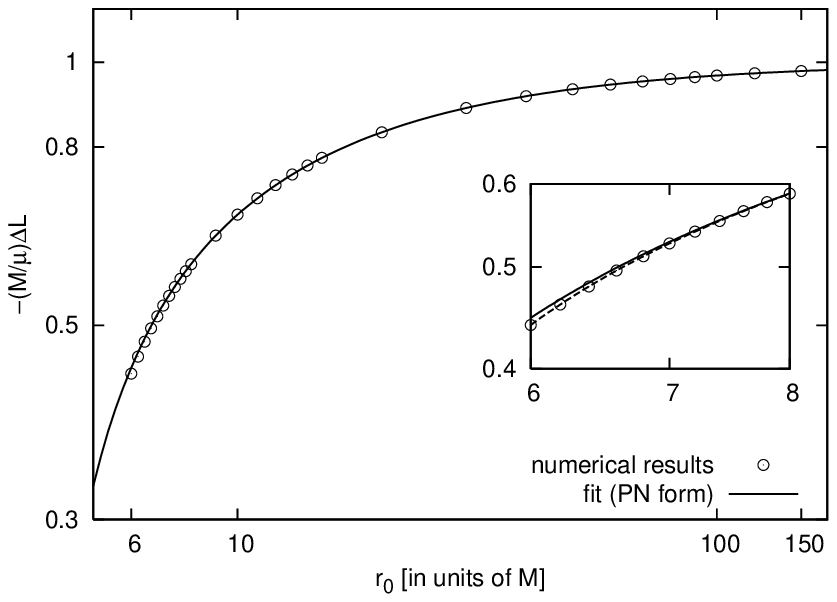}
\caption{The shift in the energy (left panel) and angular momentum
(right panel) parameters, due to the conservative SF effect.
The plots show the relative shifts $\Delta {\cal E}\equiv ({\cal E}-{\cal E}_0)/{\cal E}_0$
and $\Delta {\cal L}\equiv ({\cal L}-{\cal L}_0)/{\cal L}_0$, where
${\cal E}_0$ and ${\cal L}_0$ are the geodesic values.
Solid and dashed lines correspond to the large-$r_0$ and near-ISCO analytic
fits, Eqs.\ (\ref{eq:Fr-PNform}) and (\ref{eq:Fr-xform}), respectively.
Recall $\Delta {\cal E}$ and $\Delta {\cal L}$ are gauge dependent; the values
shown here are the Lorenz-gauge values.}
\label{fig:ELvsr0}
\end{figure}

Given $F^r$, we can calculate the shift in the orbital energy and
angular momentum parameters using Eq.\ (\ref{EandL}). The relative
shifts $\Delta {\cal E}\equiv ({\cal E}-{\cal E}_0)/{\cal E}_0$
and $\Delta {\cal L}\equiv ({\cal L}-{\cal L}_0)/{\cal L}_0$
are plotted in Fig.\ \ref{fig:ELvsr0}. Recall that this effect is
gauge-dependent; the values computed here for $\Delta {\cal E}$
and $\Delta {\cal L}$ are the Lorenz-gauge values.
%~~~~~~~~~~~~~~~~~~~~~~~~~~~~~~~~~~~~~~~~~~~~~~~~~~~~~~~~~~~~~~~~~~~~~~~
%~~~~~~~~~~~~~~~~~~~~~~~~~~~~~~~~~~~~~~~~~~~~~~~~~~~~~~~~~~~~~~~~~~~~~~~

\begin{figure}[htb]
\includegraphics[width=10cm]{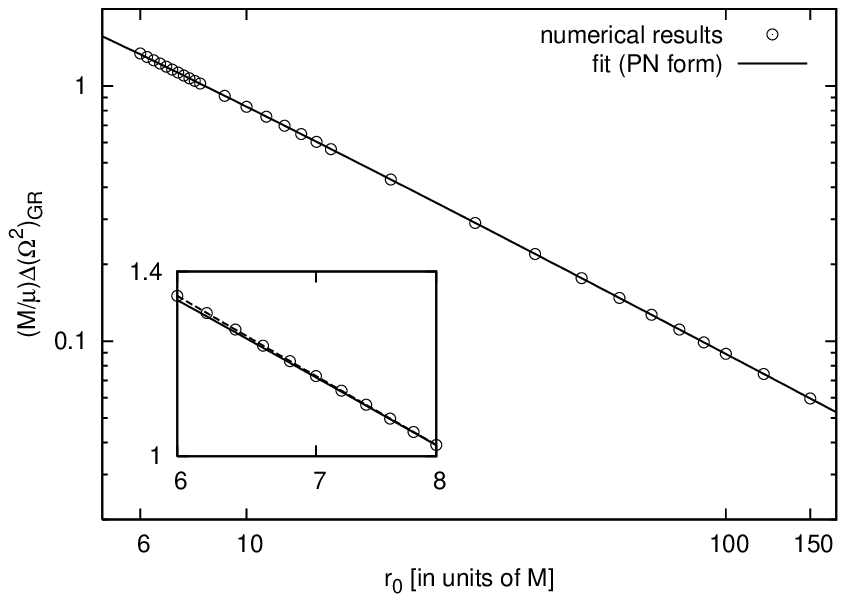}
\caption{Shift in the orbital frequency due to the conservative SF effect. Shown is
the quantity
$\Delta (\Omega^2)_{\rm GR}\equiv (\Omega^2 - \Omega_0^2)/\Omega_0^2 - (-2 \mu/M)$,
describing the relative shift in $\Omega^2$, minus the leading-order Keplerian term.
The solid line shows the large-$r_0$ (3PN) analytic fit, Eq.\ (\ref{OmegaPN}).
The dashed line corresponds to the near-ISCO fit (\ref{eq:Fr-xform}).}
\label{fig:r0vsOmg2}
\end{figure}

The shift in the orbital frequency $\Omega_0$ can be derived from Eq.\ (\ref{Omega}).
At large $r_0$ we obtain, using Eq.\ (\ref{eq:Fr-PNform}),
%~~~~~~~~~~~~~~~~~~~~~~~~~~~~~~~~~~~~~~~~~~~~~~~~~~~~~~~~~~~~~~~~~~~~~~~
\begin{equation} \label{OmegaPN}
\Omega^2 (r_0\gg M)\simeq
\Omega_0^2 \left\{ 1 + \frac{\mu}{M}
\left[-a_0+c_1\frac{M}{r_0}+c_2\left(\frac{M}{r_0}\right)^2
+c_3\left(\frac{M}{r_0}\right)^3\right]\right\},
\end{equation}
%~~~~~~~~~~~~~~~~~~~~~~~~~~~~~~~~~~~~~~~~~~~~~~~~~~~~~~~~~~~~~~~~~~~~~~~
where $c_1=3a_0-a_1$, $c_2=3a_1-a_2$, and $c_3=3a_2-a_3$, with the coefficients
$a_n$ given in Eq.\ (\ref{a}). The term proportional to $a_0(\simeq 2)$ is precisely
the ``Newtonian'' SF [see, e.g., Eq.\ (2) of \cite{Detweiler:2005kq}], which dominates
the SF effect at $r_0\gg M$.
This piece of the force is simply the $O(\mu)$ difference between the
standard Keplerian frequency $\Omega^2=(M+\mu)/R^3$ (expressed in terms of
the separation $R$) and $\Omega_0^2=M/r_0^3$, with the separation $R$ related
to the ``center-of-mass'' distance $r_0$ through $M(R-r_0)=\mu r_0$.
The rest of the terms in Eq.\ (\ref{OmegaPN})
are general relativistic (3PN) corrections. We define
$\Delta (\Omega^2)_{\rm GR} \equiv
(\Omega^2 - \Omega_0^2)/\Omega_0^2 + 2 \mu/M$, and in Fig.~\ref{fig:r0vsOmg2}
plot $\Delta (\Omega^2)_{\rm GR}$ as a function of $r_0$.
%~~~~~~~~~~~~~~~~~~~~~~~~~~~~~~~~~~~~~~~~~~~~~~~~~~~~~~~~~~~~~~~~~~~~~~~
%~~~~~~~~~~~~~~~~~~~~~~~~~~~~~~~~~~~~~~~~~~~~~~~~~~~~~~~~~~~~~~~~~~~~~~~

It is natural to ask how our result for $\Omega$ compares, at large $r_0$,
with results from PN literature. We must first note that, although $\Omega_0$
is gauge-invariant, $r_0$ itself is not, and so the functional form of $\Omega(r_0)$
in Eq.\ (\ref{OmegaPN}) is gauge dependent. This makes it difficult
to compare with the standard (non-perturbative) PN result \cite{Blanchet:2002av},
which is given in a
particular coordinate gauge (the ``harmonic'' gauge) that, in a perturbative
context, does not coincide with the Lorenz gauge employed here.
Another calculation of the conservative PN SF was carried out recently by
Nakano \cite{Nakano:Capra9}, within perturbation theory, using a modified version
of the Regge-Wheelar gauge. The results from this calculation, too, cannot be
directly compared to ours, because of the different gauges used.

One may attempt to circumvent the gauge ambiguity problem by writing
down an expression (in a PN form) for one gauge-invariant quantity in terms of
a second gauge-invariant quantity---such an expression would be `truly' gauge
invariant and would allow direct comparison between calculations done in
different gauges. For circular orbits, both $\Omega$ and
$S\equiv {\cal E}-\Omega {\cal L}$ are gauge invariant (see Sec.\
\ref{subsec:gauge}), and we may attempt an expression of the form $S(\Omega)$.
Introducing the new gauge-invariant variable $x\equiv (M\Omega)^{1/3}$, we obtain,
at 3PN,
%~~~~~~~~~~~~~~~~~~~~~~~~~~~~~~~~~~~~~~~~~~~~~~~~~~~~~~~~~~~~~~~~~~~~~~~
\begin{equation}
S =
1-\frac{3}{2}x^2
-\frac{9}{8}x^4
-\frac{27}{16}x^6 + O(\mu^2),
\end{equation}
%~~~~~~~~~~~~~~~~~~~~~~~~~~~~~~~~~~~~~~~~~~~~~~~~~~~~~~~~~~~~~~~~~~~~~~~
with a vanishing $O(\mu)$ term. Hence, $S(x)$ is not useful for comparing
the conservative SF effect in the case of a circular orbit. Comparison of our
results with results from PN literature is not at all straightforward,
and we leave it for future work.

\section{Concluding remarks and future applications}
          \label{Sec:summary}
%%%%%%%%%%%%%%%%%%%%%%%%%%%%%%%%%%%%%%%%%%%%%%%%%%%%%%

This work marks a minor milestone in a long-term program aimed to
develop the theoretical and practical tools for computing EMRI orbits
(and, eventually, their gravitational waveforms). We compute here for the
first time the gravitational SF in an example of a particle orbiting a
black hole, demonstrating the applicability of our approach, whose main
elements are
(i) direct solution for the metric perturbation, in the Lorenz gauge;
(ii) numerical evolution in the time domain; and
(iii) use of the mode-sum scheme to derive the local SF.
In the case of a strictly circular orbit, the analysis of the local SF provides
us with little new physics: The radiative effect is well known from
energy-balance analysis, and the conservative force does not have a strict
gauge-invariant significance. Calculation of gauge invariant conservative effects
(like the shift in the ISCO frequency, or the correction to the rate of
perihelion precession) requires analysis of (at least slightly) non-circular
orbits. In follow-up work we intend to extend our analysis to eccentric
orbits (see below), which would gain us access to this more interesting physics.

Self-force calculations bring about major issues of computational cost and
computational efficiency. All computations in this work were carried out on
a standard desktop computer (3GHz dual-processor, with 4Gb of RAM). Calculation
of the SF at a single strong-field point, with fractional accuracy
$\lesssim 10^{-3}$, took $\sim 2$ hours of CPU time. This is practical enough
for studying the simple one-parameter family of circular orbits, but may not
be practical for studying more general orbits. There are a few obvious ways by
which one may improve the efficiency of the numerical algorithm:
(1) Our evolution code currently utilizes a uniform grid. This is very
inefficient, since the resolution requirement near the worldline is much
higher than anywhere else on the 2-dimensional grid. Our problem naturally
calls for a mesh-refinement treatment. This is a standard technique in numerical
relativity, but its application would require a major modification of our code.
(2) We may try to improve the rate of decay of the initial spurious waves,
by using the stationary numerical solutions obtained with low resolution as
initial conditions for the evolution at higher resolution. This will allow
to evolve for shorter periods, hence saving computation time. It should be
straightforward to implement such a procedure.
(3) In the present analysis we conservatively set the same accuracy threshold
for each individual $l$ mode of the force. Since the contribution of the individual
modes to the total SF vary over a few orders of magnitude, this procedure is not
very economic. It would be better to use an algorithm which incorporates a
threshold on the total force. This, too, could be implemented rather easily.

Since our code is based on time-domain evolution (with no frequency
decomposition), it is readily extensible to deal with any orbit in
Schwarzschild spacetime. The finite-difference algorithm would change
slightly (it would resemble the algorithm used for radial plunge trajectories
\cite{Barack:2000zq,Barack:2002ku}), but the stability features and resolution requirements
of the code would not change. Work to extend our analysis to eccentric orbits
is now in progress.

It is more challenging to apply our approach for orbits in Kerr spacetime.
In this case we may no longer rely on a spherical-harmonic decomposition
of the field equations, and---insisting on a time-domain analysis in the
Lorenz gauge---we would have to apply time evolution in 2+1-D.
The challenge here is two-fold: Firstly, the solutions to the 2+1-D
field equations are no longer continuous along the worldline (as in the
1+1-D case), but rather diverge there logarithmically. Secondly, a stable
numerical scheme for evolution of Lorenz-gauge perturbations in $2+1$D
is yet to be developed. A numerical scheme for dealing with the first of
the above difficulties had been outlined in Sec.\ V of BL, and was recently
implemented for a scalar-field toy model \cite{BG07}.

\section*{ACKNOWLEDGEMENTS}

This work was supported by PPARC through grant number PP/D001110/1.
LB also gratefully acknowledges financial support from the Nuffield
Foundation. NS thanks Hiroyuki Nakano and Misao Sasaki for helpful
discussions.

\appendix

\section{Field equations and gauge conditions for the perturbation
functions $\bar h^{(i)lm}(r,t)$}
\label{AppA}

We give here explicit expressions for the various terms appearing in
our basic set of mode-decomposed field equations (\ref{FE}).
We use the notation $f=1-2M/r$, $f'=2M/r^2$, $f_0=1-2M/r_0$,
and $\lambda=(l+2)(l-1)$.
$\partial_r$ is taken with fixed $t$, and $\partial_v$ is taken with fixed $u$
($v$ and $u$ are the standard Eddington-Finkelstein coordinates).
For brevity we occasionally omit the indices $l,m$.

The terms ${\cal M}^{(i)}_{\;(j)}\bar h^{(j)}$ in Eqs.\ (\ref{FE})
are given by
%~~~~~~~~~~~~~~~~~~~~~~~~~~~~~~~~~~~~~~~~~~~~~~~~~~~~~~~~~~~~~~~~~~~~~~~
\begin{mathletters}\label{M}
\begin{equation} \label{M1}
{\cal M}^{(1)}_{\;(j)}\bar h^{(j)}=
\frac{1}{2}f^2f'\bar h^{(3)}_{,r}
+\frac{f}{2r^2}(1-4M/r)\left(\bar h^{(1)}-\bar h^{(5)}-f\bar h^{(3)}\right)
-\frac{f^2}{2r^2}(1-6M/r)\bar h^{(6)},
\end{equation}
\begin{equation} \label{M2}
{\cal M}^{(2)}_{\;(j)}\bar h^{(j)}=
\frac{1}{2}f^2f'\bar h^{(3)}_{,r}
+f'\left(\bar h^{(2)}_{,v}-\bar h^{(1)}_{,v}\right)
+\frac{f^2}{2r^2}\left(\bar h^{(2)}-\bar h^{(4)}\right)
-\frac{ff'}{2r}\left(\bar h^{(1)}-\bar h^{(5)}-f\bar h^{(3)}
-2f\bar h^{(6)}\right),
\end{equation}
\begin{eqnarray} \label{M3}
{\cal M}^{(3)}_{\;(j)}\bar h^{(j)}=
-\frac{f}{2r^2}\left[\bar h^{(1)}-\bar h^{(5)}-(1-4M/r)
\left(\bar h^{(3)}+\bar h^{(6)}\right)\right],
\end{eqnarray}
\begin{equation} \label{M4}
{\cal M}^{(4)}_{\;(j)}\bar h^{(j)}=
\frac{1}{2}f'\left(\bar h^{(4)}_{,v}-\bar h^{(5)}_{,v}\right)
-\frac{1}{2}\,l(l+1)\,(f/r^2)\bar h^{(2)}
-\frac{1}{4}f'f/r\left[3\bar h^{(4)}+2\bar h^{(5)}-\bar h^{(7)}+l(l+1)\bar h^{(6)}\right],
\end{equation}
\begin{eqnarray} \label{M5}
{\cal M}^{(5)}_{\;(j)}\bar h^{(j)}=
\frac{f}{r^2}\left[
(1-4.5M/r)\bar h^{(5)}-\frac{1}{2}l(l+1)\left(\bar h^{(1)}-f\bar h^{(3)}\right)+
\frac{1}{2}(1-3M/r)\left(l(l+1)\bar h^{(6)}-\bar h^{(7)}\right)
\right],
\end{eqnarray}
\begin{equation} \label{M6}
{\cal M}^{(6)}_{\;(j)}\bar h^{(j)}=
-\frac{f}{2r^2}\left[\bar h^{(1)}-\bar h^{(5)}
-(1-4M/r)\left(\bar h^{(3)}+\bar h^{(6)}\right)\right],
\end{equation}
\begin{equation} \label{M7}
{\cal M}^{(7)}_{\;(j)}\bar h^{(j)}=
-\frac{f}{2r^2}\left(\bar h^{(7)}
+\lambda\,\bar h^{(5)}\right),
\end{equation}
\begin{equation} \label{M8}
{\cal M}^{(8)}_{\;(j)}\bar h^{(j)}=
\frac{1}{2}f'\left(\bar h^{(8)}_{,v}-\bar h^{(9)}_{,v}\right)
-\frac{1}{4}f'f/r\left(3\bar h^{(8)}+2\bar h^{(9)}-\bar h^{(10)}
\right),
\end{equation}
\begin{equation} \label{M9}
{\cal M}^{(9)}_{\;(j)}\bar h^{(j)}=
\frac{f}{r^2}\left(1-4.5M/r\right)\bar h^{(9)}
-\frac{f}{2r^2}\left(1-3M/r\right)\,\bar h^{(10)},
\end{equation}
\begin{equation} \label{M10}
{\cal M}^{(10)}_{\;(j)}\bar h^{(j)}=
-\frac{f}{2r^2}\left(\bar h^{(10)}+\lambda\,\bar h^{(9)}\right).
\end{equation}
\end{mathletters}
%~~~~~~~~~~~~~~~~~~~~~~~~~~~~~~~~~~~~~~~~~~~~~~~~~~~~~~~~~~~~~~~~~~~~~~~
For a circular equatorial geodesic orbit with $r=r_0$ [hence with $t$-frequency
$\Omega_0=(M/r_0^3)^{1/2}$ and specific energy
${\cal E}_0=(1-2M/r_0)(1-3M/r_0)^{-1/2}$], the source terms $S^{(i)lm}$ in
Eqs.\ (\ref{FE}) read
%~~~~~~~~~~~~~~~~~~~~~~~~~~~~~~~~~~~~~~~~~~~~~~~~~~~~~~~~~~~~~~~~
\begin{equation} \label{Si}
S^{(i)lm}(r,t)=4\pi{\cal E}_0\alpha^{(i)}\delta(r-r_0)\times\left\{
\begin{array}{ll}
Y^{lm*}(\pi/2,\Omega_0 t),            & i=1\text{---}7  \text{ (even parity modes)}, \\
Y^{lm*}_{,\theta}(\pi/2,\Omega_0 t),  & i=8\text{---}10 \text{ (odd parity modes)}.
\end{array} \right.
\end{equation}
%~~~~~~~~~~~~~~~~~~~~~~~~~~~~~~~~~~~~~~~~~~~~~~~~~~~~~~~~~~~~~~~~
The coefficients $\alpha^{(i)}$ are given by
%~~~~~~~~~~~~~~~~~~~~~~~~~~~~~~~~~~~~~~~~~~~~~~~~~~~~~~~~~~~~~~~~~~~~~~~
\begin{eqnarray}\label{alphai}
\alpha^{(1)} &=& f_0^2/r_0,                   \nonumber\\
\alpha^{(3)} &=& f_0/r_0,                     \nonumber\\
\alpha^{(2)} &=& \alpha^{(5)}=\alpha^{(9)}=0, \nonumber\\
\alpha^{(4)} &=& 2if_0m\Omega_0,                 \nonumber\\
\alpha^{(6)} &=& r_0\Omega_0^2,                 \nonumber\\
\alpha^{(7)} &=& r_0\Omega_0^2[l(l+1)-2m^2],    \nonumber\\
\alpha^{(8)} &=& 2f_0\Omega_0,                  \nonumber\\
\alpha^{(10)}&=& 2imr_0\Omega_0^2.
\end{eqnarray}
%~~~~~~~~~~~~~~~~~~~~~~~~~~~~~~~~~~~~~~~~~~~~~~~~~~~~~~~~~~~~~~~~~~~~~~~

%Note $\bar h^{(3)}-\bar h^{(6)}$ (the trace) satisfied a scalar-field equation,
%$\square_{\rm sc}^{2d} \bar h=S$.
%Also note above equation can't be used for $l=1$:
%At $l=1$ the potential for $h^{(5)}$ gets negative for $r<8/3$, and solutions
%diverge.

The four Lorenz gauge conditions $\bar h_{\alpha\beta}{}^{\!;\beta}=0$ translate,
upon decomposing in tensor harmonics, to four constraints on the
time-radial functions $\bar h^{(i)}$. These read
%%~~~~~~~~~~~~~~~~~~~~~~~~~~~~~~~~~~~~~~~~~~~~~~~~~~~~~~~~~~~~~~~~~~~~~~~
\begin{mathletters} \label{gauge}
\begin{equation}\label{gauge1}
%H_{1}^{lm}(r,t)\equiv
-\bar h^{(1)}_{,t}+f\left(-\bar h^{(3)}_{,t}+
\bar h^{(2)}_{,r}+\bar h^{(2)}/r- \bar h^{(4)}/r\right)=0,
\end{equation}
\begin{eqnarray}\label{gauge2}
%H_{2}^{lm}(r,t)\equiv
\bar h^{(2)}_{,t}
-f \bar h^{(1)}_{,r}+f^2\bar h^{(3)}_{,r}
-(f/r)\left(\bar h^{(1)}-\bar h^{(5)}-f\bar h^{(3)}-2f \bar h^{(6)}\right)=0,
\end{eqnarray}
\begin{eqnarray}\label{gauge3}
%H_3^{lm}(r,t)\equiv
\bar h^{(4)}_{,t}-f\left( \bar h^{(5)}_{,r}+2\bar h^{(5)}/r+l(l+1)\,
\bar h^{(6)}/r-\bar h^{(7)}/r\right)=0,
\end{eqnarray}
\begin{eqnarray}\label{gauge4}
%H_4^{lm}(r,t)\equiv
\bar h^{(8)}_{,t}-f\left(\bar h^{(9)}_{,r}
+2\bar h^{(9)}/r-\bar h^{(10)}/r \right)=0.
\end{eqnarray}
\end{mathletters}
%%~~~~~~~~~~~~~~~~~~~~~~~~~~~~~~~~~~~~~~~~~~~~~~~~~~~~~~~~~~~~~~~~~~~~~~~

\section{Formulas for the coefficients $f^r_{n\pm}$ and $f^t_{n\pm}$}
\label{AppB}

We give here formulas for constructing the various coefficients
$f^{\alpha}_{n\pm}$ appearing in Eq.\ (\ref{Ffull2}),
using the Lorenz-gauge metric perturbation fields $\bar h^{(i)lm}(r,t)$
and their derivatives. For brevity we omit the superscripts $l,m$
from both the $f_{n\pm}^{\alpha lm}$'s and the $\bar h^{(i)lm}$'s.
We use here the notation $\tilde{\cal L}_0={\cal L}_0/r_0$, where, recall,
${\cal L}_0$ is given in Eq.\ (\ref{L0}). Also recall ${\cal E}_0$ is given in
Eq.\ (\ref{E0}), $f_0=(1-2M/r_0)$, and $\lambda=(l+2)(l-1)$.
$r$ derivatives are taken with fixed $t$, and $t$ derivatives are taken
with fixed $r$. All functions $\bar h^{(i)lm}$ and their derivatives in
the expressions below are evaluated at $r=r_0$ and
$t=t_0$. Subscripts `$\pm$' refer to taking $r$ derivatives from $r\to r_0^{\pm}$
(we omit these subscripts whenever $f^{\alpha}_{n+}=f^{\alpha}_{n-}$).

%In the expression for $\Phi^r_0$ we have made use of the fact that the
%following combination vanishes for any $i$:
%\begin{equation}
%{\cal E}ir_0\bar h^{(i)}_{,t} -m{\cal L} f_0 h^{(i)}=0.
%\end{equation}
%The reason: For a stationary field $\bar h^{(i)}_{,t}=-im\omega \bar h^{(i)}$, where
%$\omega=d\varphi/dt=u^{\varphi}/u^t=(f_0/r_0)({\cal L}/{\cal E})$.
For the $r$ component we have
\begin{mathletters} \label{fr}
\begin{eqnarray}
f_{0\pm}^r&=&{\cal E}_0^2(M/r_0)f_0^{-2}\bar h^{(1)}
+\frac{1}{4}{\cal E}_0^2f_0^{-2}\left(r_0 f_0 \bar h^{(1)}_{,r}-\bar h^{(1)}\right)
+\frac{1}{4}\tilde{\cal L}_0^2f_0\left(r_0 \bar h^{(3)}_{,r}-\bar h^{(3)}\right)
+\frac{1}{4}f_0\left(r_0 \bar h^{(6)}_{,r}-\bar h^{(6)}\right) \nonumber\\
%&&+\frac{i}{2} {\cal E}f_0^{-2}\left({\cal E}ir_0\bar h^{(2)}_{,t}
%-m{\cal L} f_0 h^{(2)}\right)
%-\frac{m{\cal L}f_0^{-1}}{2l(l+1)}\left({\cal E}ir_0\bar h^{(5)}_{,t}
%-m{\cal L} f_0 h^{(5)}\right)       \nonumber\\
&&+\frac{i m{\cal E}_0\tilde{\cal L}_0 r_0}{2l(l+1)}\bar h^{(4)}_{,r}
-\frac{m^2\tilde{\cal L}_0^2f_0}{4l(l+1)\lambda}\left(r_0 \bar h^{(7)}_{,r}+ \bar
h^{(7)}\right),
\end{eqnarray}
\begin{equation}
f_{1\pm}^r=\frac{1}{4}\tilde{\cal L}_0^2
\left(-2\bar h^{(1)}+2f_0\bar h^{(3)}+f_0\bar h^{(6)}+r_0f_0\bar
h^{(6)}_{,r}\right),
\end{equation}
\begin{equation}
f_{2\pm}^r=\frac{\tilde{\cal L}_0^2}{4l(l+1)}
\left[-2\bar h^{(5)}+(f_0/\lambda)\left(
r_0\bar h^{(7)}_{,r}+\bar h^{(7)}\right)\right],
\end{equation}
\begin{equation}
f_{3\pm}^r=-\frac{\tilde{\cal L}_0^2f_0}{4l(l+1)\lambda}
\left(r_0 \bar h^{(7)}_{,r}+\bar h^{(7)}\right),
\end{equation}
\begin{equation}
f_{4\pm}^r=\frac{im\tilde{\cal L}_0^2}{2 l(l+1)}\left[-\bar h^{(9)}+(f_0/\lambda)
\left(r_0\bar h^{(10)}_{,r}+\bar h^{(10)}\right)\right],
\end{equation}
\begin{equation}
f_{5\pm}^r=\frac{\tilde{\cal L}_0{\cal E}_0r_0}{2f_0
l(l+1)}\left(\bar h^{(9)}_{,t}-f_0\bar h^{(8)}_{,r}\right),
\end{equation}
\begin{equation}
f_6^r=f_7^r=0.
\end{equation}
\end{mathletters}
For the $t$ component we have
\begin{mathletters} \label{ft}
\begin{eqnarray}
f_0^t&=&
-\frac{1}{4}{\cal E}_0^2 \tilde{\cal L}_0^2 f_0^{-3}r_0            \bar h^{(1)}_{,t}
+\frac{1}{4}im{\cal E}_0 \tilde{\cal L}_0(2f_0-{\cal E}_0^2)f_0^{-3}   \bar h^{(1)}
+\frac{1}{2}{\cal E}_0^2 \tilde{\cal L}_0^2f_0^{-3}(M/r_0)          \bar h^{(2)} \nonumber\\
&&-\frac{1}{4}f_0^{-2}r_0 \tilde{\cal L}_0^2({\cal E}_0^2+f_0)        \bar h^{(3)}_{,t}
-\frac{1}{4}im\tilde{\cal L}_0^3 {\cal E}_0f_0^{-1}                \bar h^{(3)}
-\frac{im{\cal E}_0^3\tilde{\cal L}_0f_0^{-3}r_0}{2l(l+1)}         \bar h^{(4)}_{,t}
+\frac{m^2 \tilde{\cal L}_0^4f_0^{-1}}{2l(l+1)}                   \bar h^{(4)} \nonumber\\
&&+\frac{im\tilde{\cal L}_0{\cal E}_0^3f_0^{-3}(M/r_0)}{2l(l+1)}      \bar h^{(5)}
+\frac{1}{4}\tilde{\cal L}_0^2f_0^{-1} r_0                        \bar h^{(6)}_{,t}
+\frac{1}{4}im{\cal E}_0\tilde{\cal L}f_0^{-1}                    \bar h^{(6)}
+\frac{m^2 \tilde{\cal L}_0^2({\cal E}_0^2+f_0)f_0^{-2}r_0}{4l(l+1)\lambda}   \bar h^{(7)}_{,t}
\nonumber\\
&&+\frac{im {\cal E}_0\tilde{\cal L}_0^3(m^2+4)f_0^{-1}}{4l(l+1)\lambda} \bar h^{(7)},
\end{eqnarray}
\begin{equation}
f_1^t=-\frac{1}{2}\tilde{\cal L}_0^4f_0^{-1}\bar h^{(2)}
-\frac{im{\cal E}_0\tilde{\cal L}_0^3f_0^{-1}}{2l(l+1)}\bar h^{(5)}
-\frac{1}{4}\tilde{\cal L}_0^2(\tilde{\cal E}_0^2+f_0)f_0^{-2}r_0\bar h^{(6)}_{,t}
-\frac{1}{4}im{\cal E}_0\tilde{\cal L}_0^3f_0^{-1}\bar h^{(6)}
-\frac{im{\cal E}_0\tilde{\cal L}_0^3f_0^{-1}}{l(l+1)\lambda}\bar h^{(7)},
\end{equation}
\begin{equation}
f_2^t=-\frac{\tilde{\cal L}_0^4f_0^{-1}}{2l(l+1)}\bar h^{(4)}
-\frac{\tilde{\cal L}_0^2({\cal E}_0^2+f_0)f_0^{-2}r_0}{4l(l+1)\lambda}\bar h^{(7)}_{,t}
-\frac{5im{\cal E}_0\tilde{\cal L}_0^3f_0^{-1}}{4l(l+1)\lambda}\bar h^{(7)},
\end{equation}
\begin{equation}
f_3^t=\frac{\tilde{\cal L}_0^2({\cal E}_0^2+f_0)r_0f_0^{-2}}{4l(l+1)\lambda}\bar h^{(7)}_{,t}
+\frac{im{\cal E}_0\tilde{\cal L}_0^3f_0^{-1}}{4l(l+1)\lambda}\bar h^{(7)},
\end{equation}
\begin{equation}
f_4^t=-\frac{im\tilde{\cal L}_0^4f_0^{-1}}{2l(l+1)}\bar h^{(8)}
-\frac{im\tilde{\cal L}_0^2({\cal E}_0^2+f_0)f_0^{-2}r_0}{2l(l+1)\lambda}\bar h^{(10)}_{,t}
+\frac{m^2\tilde{\cal L}_0^3{\cal E}_0f_0^{-1}}{l(l+1)\lambda}\bar h^{(10)}
\end{equation}
\begin{equation}
f_5^t=\frac{{\cal E}_0^3\tilde{\cal L}_0f_0^{-3}r_0}{2l(l+1)}\bar h^{(8)}_{,t}
-\frac{{\cal E}_0^3\tilde{\cal L}_0(M/r_0)f_0^{-3}}{2l(l+1)}\bar h^{(9)}
+\frac{(m^2-1){\cal E}_0\tilde{\cal L}_0^3f_0^{-1}}{2l(l+1)\lambda}\bar h^{(10)},
\end{equation}
\begin{equation}
f_6^t=\frac{{\cal E}_0\tilde{\cal L}_0^3f_0^{-1}}{2l(l+1)}
\left(\bar h^{(9)}+\lambda^{-1}\bar h^{(10)}\right),
\end{equation}
\begin{equation}
f_7^t=\frac{{\cal E}_0\tilde{\cal L}^3f_0^{-1}}{2l(l+1)\lambda}\bar h^{(10)}.
\end{equation}
\end{mathletters}
%where $\lambda=(l+2)(l-1)$, ${\cal E}=-u_t$ and ${\cal L}=u_{\varphi}/r_0$
%(both dimensionless). Recall we have
%\begin{equation}
%{\cal E}^2=f_0(1+{\cal L}^2), \quad\quad
%{\cal E}=f_0(1-3M/r_0)^{-1/2}, \quad\quad
%{\cal L}=(M/r_0)^{1/2}(1-3M/r_0)^{-1/2}.
%\end{equation}

\section{Formulas for the coupling coefficients}
\label{AppC}

We give here formulas for re-expanding all angular functions in Eq.\ (\ref{Ffull2})
in spherical harmonics $Y^{lm}(\theta,\varphi)$. The following identities
hold for all $l,m$ and any $\theta,\varphi$. We have
\begin{eqnarray}
\sin^2\theta Y^{lm} &=& \alpha^{lm}_{(+2)}Y^{l+2,m}+\alpha^{lm}_{(0)} Y^{lm}+\alpha^{lm}_{(-2)}Y^{l-2,m}, \nonumber\\
\cos\theta\sin\theta Y^{lm}_{,\theta} &=& \beta^{lm}_{(+2)}Y^{l+2,m}+\beta^{lm}_{(0)} Y^{lm}+\beta^{lm}_{(-2)}Y^{l-2,m}, \nonumber\\
\sin^2\theta Y^{lm}_{,\theta\theta} &=& \gamma^{lm}_{(+2)}Y^{l+2,m}+\gamma^{lm}_{(0)} Y^{lm}+\gamma^{lm}_{(-2)}Y^{l-2,m}, \nonumber\\
\sin\theta Y^{lm}_{,\theta} &=& \delta^{lm}_{(+1)}Y^{l+1,m}+\delta^{lm}_{(-1)}Y^{l-1,m}, \nonumber\\
\cos\theta Y^{lm}-\sin\theta Y^{lm}_{,\theta} &=&
\epsilon^{lm}_{(+1)}Y^{l+1,m}+\epsilon^{lm}_{(-1)}Y^{l-1,m}, \nonumber\\
\sin^3\theta Y^{lm}_{,\theta} &=&
\zeta^{lm}_{(+3)}Y^{l+3,m}+\zeta^{lm}_{(+1)}Y^{l+1,m}+\zeta^{lm}_{(-1)}Y^{l-1,m}+\zeta^{lm}_{(-3)}Y^{l-3,m}, \nonumber\\
\cos\theta\sin^2\theta Y^{lm}_{,\theta\theta} &=&
\xi^{lm}_{(+3)}Y^{l+3,m}+\xi^{lm}_{+}Y^{l+1,m}+\xi^{lm}_{(-1)}Y^{l-1,m}+\xi^{lm}_{(-3)}Y^{l-3,m}.
\end{eqnarray}
The coefficients are all constructed from
\begin{equation}
C_{lm}=\left[\frac{l^2-m^2}{(2l+1)(2l-1)}\right]^{1/2},
\end{equation}
using
\begin{mathletters}
\begin{equation}
\alpha^{lm}_{(+2)}=-C_{l+1,m}C_{l+2,m}, \quad\quad
\alpha^{lm}_{(0)}=1-C_{lm}^2-C_{l+1,m}^2, \quad\quad
\alpha^{lm}_{(-2)}=-C_{lm} C_{l-1,m},
\end{equation}
\begin{equation}
\beta^{lm}_{(+2)}=lC_{l+1,m}C_{l+2,m}, \quad\quad
\beta^{lm}_{(0)}=lC_{l+1,m}^2-(l+1)C_{lm}^2, \quad\quad
\beta^{lm}_{(-2)}=-(l+1)C_{lm} C_{l-1,m},
\end{equation}
\begin{equation}
\gamma^{lm}_{(+2)}=l^2C_{l+1,m}C_{l+2,m}, \quad\quad
\gamma^{lm}_{(0)}=m^2-l(l+1)+l^2C_{l+1,m}^2+(l+1)^2C_{lm}^2, \quad\quad
\gamma^{lm}_{(-2)}=(l+1)^2C_{lm} C_{l-1,m},
\end{equation}
\begin{equation}
\delta^{lm}_{(+1)}=lC_{l+1,m}, \quad\quad
\delta^{lm}_{(-1)}=-(l+1)C_{lm},
\end{equation}
\begin{equation}
\epsilon^{lm}_{(+1)}=(1-l)C_{l+1,m}, \quad\quad
\epsilon^{lm}_{(-1)}=(l+2)C_{lm},
\end{equation}
\begin{eqnarray}
\zeta^{lm}_{(+3)}&=&-lC_{l+1,m}C_{l+2,m}C_{l+3,m},\nonumber\\
\zeta^{lm}_{(+1)}&=&C_{l+1,m}\left[l\left(1-C^2_{l+1,m}-C^2_{l+2,m}\right)+(l+1)C^2_{l,m}\right],
\nonumber\\
\zeta^{lm}_{(-1)}&=&-C_{l,m}\left[(l+1)\left(1-C^2_{l-1,m}-C^2_{l,m}\right)+lC^2_{l+1,m}\right],
\nonumber\\
\zeta^{lm}_{(-3)}&=&(l+1)C_{l,m}C_{l-1,m}C_{l-2,m},
\end{eqnarray}
\begin{eqnarray}
\xi^{lm}_{(+3)}&=&l^2C_{l+1,m}C_{l+2,m}C_{l+3,m}, \nonumber\\
\xi^{lm}_{(+1)}&=&C_{l+1,m}\left[m^2-l(l+1)+l^2 C^2_{l+1,m}+(l+1)^2 C^2_{l,m}+l^2 C^2_{l+2,m}\right],
\nonumber\\
\xi^{lm}_{(-1)}&=&C_{l,m}\left[m^2-l(l+1)+l^2 C^2_{l+1,m}+(l+1)^2 C^2_{l,m}+(l+1)^2 C^2_{l-1,m}\right],
\nonumber\\
\xi^{lm}_{(-3)}&=&(l+1)^2C_{l,m}C_{l-1,m}C_{l-2,m}.
\end{eqnarray}
\end{mathletters}

\section{Details of numerical results for $F^r$}
\label{AppD}

In this appendix we tabulate data obtained for the radial component $F^r$,
braking it up into low-$l$ and high-$l$ (extrapolated) contributions, and
displaying separately internal and external values. These data is used
in Sec.\ \ref{Sec:results} for error analysis.
%~~~~~~~~~~~~~~~~~~~~~~~~~~~~~~~~~~~~~~~~~~~~~~~~~~~~~~~~~~~~~~~~~~~~~~~
\begin{table} [htb]
\begin{tabular}{c|c|c|c|c}
\hline\hline
$r_0/M$ & $\left[F^r_{l\leq 15}\right]_- \times (M/\mu)^2$
       & $\left[F^r_{l> 15}\right]_- \times (M/\mu)^2$
       & $\left|F^r_{l> 15}/F^r_{l\leq 15}\right|$
       & $\Delta_{\rm tail,rel}$ \\
\hline
 6.0  & $2.49707 \times 10^{-2}$ [$5 \times 10^{-4}$]
       & $-5.04614 \times 10^{-4}$
       & $2.0 \times 10^{-2}$
       & $7 \times 10^{-7}$ \\
 6.2  & $2.44097 \times 10^{-2}$ [$5 \times 10^{-4}$]
       & $-4.44591 \times 10^{-4}$
       & $1.8 \times 10^{-2}$
       & $8 \times 10^{-7}$ \\
 6.4  & $2.37894 \times 10^{-2}$ [$4 \times 10^{-4}$]
       & $-3.93950 \times 10^{-4}$
       & $1.7 \times 10^{-2}$
       & $8 \times 10^{-7}$ \\
 6.6  & $2.31339 \times 10^{-2}$ [$4 \times 10^{-4}$]
       & $-3.50880 \times 10^{-4}$
       & $1.5 \times 10^{-2}$
       & $9 \times 10^{-7}$ \\
 6.8  & $2.24603 \times 10^{-2}$ [$3 \times 10^{-4}$]
       & $-3.13987 \times 10^{-4}$
       & $1.4 \times 10^{-2}$
       & $9 \times 10^{-7}$ \\
 7.0  & $2.17812 \times 10^{-2}$ [$3 \times 10^{-4}$]
       & $-2.82179 \times 10^{-4}$
       & $1.3 \times 10^{-2}$
       & $9 \times 10^{-7}$ \\
 7.2  & $2.11051 \times 10^{-2}$ [$2 \times 10^{-4}$]
       & $-2.54592 \times 10^{-4}$
       & $1.2 \times 10^{-2}$
       & $9 \times 10^{-7}$ \\
 7.4  & $2.04385 \times 10^{-2}$ [$2 \times 10^{-4}$]
       & $-2.30539 \times 10^{-4}$
       & $1.1 \times 10^{-2}$
       & $8 \times 10^{-7}$ \\
 7.6  & $1.97857 \times 10^{-2}$ [$2 \times 10^{-4}$]
       & $-2.09460 \times 10^{-4}$
       & $1.1 \times 10^{-2}$
       & $8 \times 10^{-7}$ \\
 7.8  & $1.91496 \times 10^{-2}$ [$2 \times 10^{-4}$]
       & $-1.90904 \times 10^{-4}$
       & $1.0 \times 10^{-2}$
       & $8 \times 10^{-7}$ \\
 8.0  & $1.85323 \times 10^{-2}$ [$2 \times 10^{-4}$]
       & $-1.74500 \times 10^{-4}$
       & $9.4 \times 10^{-3}$
       & $8 \times 10^{-7}$ \\
 9.0  & $1.57527 \times 10^{-2}$ [$1 \times 10^{-4}$]
       & $-1.15633 \times 10^{-4}$
       & $7.3 \times 10^{-3}$
       & $6 \times 10^{-7}$ \\
 10.0  & $1.34701 \times 10^{-2}$ [$2 \times 10^{-5}$]
       & $-8.06327 \times 10^{-5}$
       & $6.0 \times 10^{-3}$
       & $5 \times 10^{-7}$ \\
 11.0  & $1.16103 \times 10^{-2}$ [$1 \times 10^{-5}$]
       & $-5.84861 \times 10^{-5}$
       & $5.0 \times 10^{-3}$
       & $4 \times 10^{-7}$ \\
 12.0  & $1.00900 \times 10^{-2}$ [$9 \times 10^{-6}$]
       & $-4.37783 \times 10^{-5}$
       & $4.3 \times 10^{-3}$
       & $3 \times 10^{-7}$ \\
 13.0  & $8.83852 \times 10^{-3}$ [$7 \times 10^{-6}$]
       & $-3.36246 \times 10^{-5}$
       & $3.8 \times 10^{-3}$
       & $2 \times 10^{-7}$ \\
 14.0  & $7.79945 \times 10^{-3}$ [$8 \times 10^{-7}$]
       & $-2.63866 \times 10^{-5}$
       & $3.4 \times 10^{-3}$
       & $2 \times 10^{-7}$ \\
 15.0  & $6.92924 \times 10^{-3}$ [$4 \times 10^{-6}$]
       & $-2.10875 \times 10^{-5}$
       & $3.0 \times 10^{-3}$
       & $2 \times 10^{-7}$ \\
 20.0  & $4.16544 \times 10^{-3}$ [$1 \times 10^{-6}$]
       & $-8.38746 \times 10^{-6}$
       & $2.0 \times 10^{-3}$
       & $1 \times 10^{-7}$ \\
 30.0  & $1.97216 \times 10^{-3}$ [$5 \times 10^{-7}$]
       & $-2.34690 \times 10^{-6}$
       & $1.2 \times 10^{-3}$
       & $4 \times 10^{-8}$ \\
 40.0  & $1.14385 \times 10^{-3}$ [$2 \times 10^{-7}$]
       & $-9.62617 \times 10^{-7}$
       & $8.4 \times 10^{-4}$
       & $2 \times 10^{-8}$ \\
 50.0  & $7.45433 \times 10^{-4}$ [$6 \times 10^{-8}$]
       & $-4.84666 \times 10^{-7}$
       & $6.5 \times 10^{-4}$
       & $2 \times 10^{-8}$ \\
 60.0  & $5.23891 \times 10^{-4}$ [$2 \times 10^{-6}$]
       & $-2.77374 \times 10^{-7}$
       & $5.3 \times 10^{-4}$
       & $1 \times 10^{-8}$ \\
 70.0  & $3.88183 \times 10^{-4}$ [$1 \times 10^{-6}$]
       & $-1.73293 \times 10^{-7}$
       & $4.5 \times 10^{-4}$
       & $9 \times 10^{-9}$ \\
 80.0  & $2.99094 \times 10^{-4}$ [$9 \times 10^{-7}$]
       & $-1.15405 \times 10^{-7}$
       & $3.9 \times 10^{-4}$
       & $7 \times 10^{-9}$ \\
 90.0  & $2.37487 \times 10^{-4}$ [$6 \times 10^{-7}$]
       & $-8.06802 \times 10^{-8}$
       & $3.4 \times 10^{-4}$
       & $5 \times 10^{-9}$ \\
 100.0  & $1.93121 \times 10^{-4}$ [$5 \times 10^{-7}$]
       & $-5.85995 \times 10^{-8}$
       & $3.0 \times 10^{-4}$
       & $5 \times 10^{-9}$ \\
 120.0  & $1.34902 \times 10^{-4}$ [$3 \times 10^{-7}$]
       & $-3.37248 \times 10^{-8}$
       & $2.5 \times 10^{-4}$
       & $4 \times 10^{-9}$ \\
 150.0  & $8.68446 \times 10^{-5}$ [$2 \times 10^{-7}$]
       & $-1.71721 \times 10^{-8}$
       & $2.0 \times 10^{-4}$
       & $2 \times 10^{-9}$ \\
\hline\hline
\end{tabular}
\caption{
Results for the radial component of the SF, broken up into the contribution
from $l\leq 15$ (computed directly using our numerical evolution code), and the
contribution from the $l>15$ tail (estimated as described in Sec.\ \ref{subsec:tail}).
Presented here are ``internal'' values of the SF (i.e., those obtained through
taking one-sided derivatives of the perturbation from $r\to r_0^-$). The ``external''
values (which should be the same within numerical error) are given in Table
\ref{table:highL-effect2} below.
%The contributions $\left[F^r_{l\leq 15}\right]_-$ and
%$\left[F^r_{l> 15}\right]_-$ are referred to in the text as
%$[F^r_{[{\rm low}\,l]}]_{-}$ and $[F^r_{[{\rm high}\,l]}]_{-}$,
%respectively.
The values in square brackets in the second column represent
the {\em fractional} discretization error $\Delta^r_{\rm discr}$,
as estimated using Eq.\ (\ref{Deltah}).
$\Delta_{\rm tail,rel}$ is the relative fractional error in
$F^r_{l>15}$ (i.e., error expressed at a fraction of the total SF),
which is estimated using Eq.\ (\ref{Deltafit}). In our analysis we made sure
that the error from estimating the contribution from the large-$l$ tail
is kept smaller than the discretization error.
}
\label{table:highL-effect}
\end{table}
%~~~~~~~~~~~~~~~~~~~~~~~~~~~~~~~~~~~~~~~~~~~~~~~~~~~~~~~~~~~~~~~~~~~~~~~
%~~~~~~~~~~~~~~~~~~~~~~~~~~~~~~~~~~~~~~~~~~~~~~~~~~~~~~~~~~~~~~~~~~~~~~~
\begin{table}[htb]
\begin{tabular}{c|c|c|c|c}
\hline\hline
$r_0/M$ & $\left[F^r_{l\leq 15}\right]_+ \times (M/\mu)^2$
       & $\left[F^r_{l> 15}\right]_+ \times (M/\mu)^2$
       & $\left|F^r_{l> 15}/F^r_{l\leq 15}\right|$
       & $\Delta_{\rm tail,rel}$ \\
\hline
 6.0  & $2.49708 \times 10^{-2}$ [$1 \times 10^{-3}$]
       & $-5.04619 \times 10^{-4}$
       & $2.0 \times 10^{-2}$
       & $6 \times 10^{-7}$ \\
 6.2  & $2.44097 \times 10^{-2}$ [$1 \times 10^{-3}$]
       & $-4.44594 \times 10^{-4}$
       & $1.8 \times 10^{-2}$
       & $7 \times 10^{-7}$ \\
 6.4  & $2.37893 \times 10^{-2}$ [$1 \times 10^{-3}$]
       & $-3.93953 \times 10^{-4}$
       & $1.7 \times 10^{-2}$
       & $8 \times 10^{-7}$ \\
 6.6  & $2.31337 \times 10^{-2}$ [$1 \times 10^{-3}$]
       & $-3.50883 \times 10^{-4}$
       & $1.5 \times 10^{-2}$
       & $8 \times 10^{-7}$ \\
 6.8  & $2.24601 \times 10^{-2}$ [$1 \times 10^{-3}$]
       & $-3.13989 \times 10^{-4}$
       & $1.4 \times 10^{-2}$
       & $8 \times 10^{-7}$ \\
 7.0  & $2.17809 \times 10^{-2}$ [$1 \times 10^{-3}$]
       & $-2.82181 \times 10^{-4}$
       & $1.3 \times 10^{-2}$
       & $8 \times 10^{-7}$ \\
 7.2  & $2.11049 \times 10^{-2}$ [$9 \times 10^{-4}$]
       & $-2.54594 \times 10^{-4}$
       & $1.2 \times 10^{-2}$
       & $8 \times 10^{-7}$ \\
 7.4  & $2.04382 \times 10^{-2}$ [$9 \times 10^{-4}$]
       & $-2.30543 \times 10^{-4}$
       & $1.1 \times 10^{-2}$
       & $8 \times 10^{-7}$ \\
 7.6  & $1.97854 \times 10^{-2}$ [$8 \times 10^{-4}$]
       & $-2.09463 \times 10^{-4}$
       & $1.1 \times 10^{-2}$
       & $8 \times 10^{-7}$ \\
 7.8  & $1.91493 \times 10^{-2}$ [$8 \times 10^{-4}$]
       & $-1.90907 \times 10^{-4}$
       & $1.0 \times 10^{-2}$
       & $8 \times 10^{-7}$ \\
 8.0  & $1.85320 \times 10^{-2}$ [$8 \times 10^{-4}$]
       & $-1.74503 \times 10^{-4}$
       & $9.4 \times 10^{-3}$
       & $7 \times 10^{-7}$ \\
 9.0  & $1.57524 \times 10^{-2}$ [$6 \times 10^{-4}$]
       & $-1.15636 \times 10^{-4}$
       & $7.3 \times 10^{-3}$
       & $6 \times 10^{-7}$ \\
 10.0  & $1.34701 \times 10^{-2}$ [$1 \times 10^{-4}$]
       & $-8.06338 \times 10^{-5}$
       & $6.0 \times 10^{-3}$
       & $5 \times 10^{-7}$ \\
 11.0  & $1.16103 \times 10^{-2}$ [$1 \times 10^{-4}$]
       & $-5.84873 \times 10^{-5}$
       & $5.0 \times 10^{-3}$
       & $4 \times 10^{-7}$ \\
 12.0  & $1.00900 \times 10^{-2}$ [$9 \times 10^{-5}$]
       & $-4.37789 \times 10^{-5}$
       & $4.3 \times 10^{-3}$
       & $3 \times 10^{-7}$ \\
 13.0  & $8.83852 \times 10^{-3}$ [$7 \times 10^{-5}$]
       & $-3.36281 \times 10^{-5}$
       & $3.8 \times 10^{-3}$
       & $1 \times 10^{-7}$ \\
 14.0  & $7.79946 \times 10^{-3}$ [$3 \times 10^{-5}$]
       & $-2.63886 \times 10^{-5}$
       & $3.4 \times 10^{-3}$
       & $1 \times 10^{-7}$ \\
 15.0  & $6.92924 \times 10^{-3}$ [$5 \times 10^{-5}$]
       & $-2.10886 \times 10^{-5}$
       & $3.0 \times 10^{-3}$
       & $1 \times 10^{-7}$ \\
 20.0  & $4.16544 \times 10^{-3}$ [$3 \times 10^{-5}$]
       & $-8.38736 \times 10^{-6}$
       & $2.0 \times 10^{-3}$
       & $2 \times 10^{-7}$ \\
 30.0  & $1.97216 \times 10^{-3}$ [$1 \times 10^{-5}$]
       & $-2.34682 \times 10^{-6}$
       & $1.2 \times 10^{-3}$
       & $5 \times 10^{-8}$ \\
 40.0  & $1.14385 \times 10^{-3}$ [$4 \times 10^{-6}$]
       & $-9.62603 \times 10^{-7}$
       & $8.4 \times 10^{-4}$
       & $2 \times 10^{-8}$ \\
 50.0  & $7.45433 \times 10^{-4}$ [$2 \times 10^{-6}$]
       & $-4.84663 \times 10^{-7}$
       & $6.5 \times 10^{-4}$
       & $1 \times 10^{-8}$ \\
 60.0  & $5.23891 \times 10^{-4}$ [$3 \times 10^{-5}$]
       & $-2.77373 \times 10^{-7}$
       & $5.3 \times 10^{-4}$
       & $1 \times 10^{-8}$ \\
 70.0  & $3.88183 \times 10^{-4}$ [$2 \times 10^{-5}$]
       & $-1.73293 \times 10^{-7}$
       & $4.5 \times 10^{-4}$
       & $9 \times 10^{-9}$ \\
 80.0  & $2.99094 \times 10^{-4}$ [$2 \times 10^{-5}$]
       & $-1.15405 \times 10^{-7}$
       & $3.9 \times 10^{-4}$
       & $7 \times 10^{-9}$ \\
 90.0  & $2.37487 \times 10^{-4}$ [$1 \times 10^{-5}$]
       & $-8.06802 \times 10^{-8}$
       & $3.4 \times 10^{-4}$
       & $4 \times 10^{-9}$ \\
 100.0  & $1.93121 \times 10^{-4}$ [$1 \times 10^{-5}$]
       & $-5.85992 \times 10^{-8}$
       & $3.0 \times 10^{-4}$
       & $5 \times 10^{-9}$ \\
 120.0  & $1.34902 \times 10^{-4}$ [$7 \times 10^{-6}$]
       & $-3.37249 \times 10^{-8}$
       & $2.5 \times 10^{-4}$
       & $4 \times 10^{-9}$ \\
 150.0  & $8.68446 \times 10^{-5}$ [$4 \times 10^{-6}$]
       & $-1.71721 \times 10^{-8}$
       & $2.0 \times 10^{-4}$
       & $4 \times 10^{-9}$ \\
\hline\hline
\end{tabular}
\caption{Same as Table \ref{table:highL-effect2}, here for the external values of the SF.}
\label{table:highL-effect2}
\end{table}
%~~~~~~~~~~~~~~~~~~~~~~~~~~~~~~~~~~~~~~~~~~~~~~~~~~~~~~~~~~~~~~~~~~~~~~~

%\subsection{Monopole and dipole modes}

%For $l=|m|=1$ mode we use the numerical data as tabulated by
%Detweiler and Poisson \cite{DP2004}
%(Table I, page 13, with formula 5.55).
%Note: The numerical values are very close to
%\begin{equation}
%F^r_{\pm}=-\frac{3\mu Ef_0}{(1-3/r_0)r_0^2}\left[\pm r_0^2+2r_0-4\right]
%\end{equation}
%The fractional different is only $\sim 10^{-4}$ at worst, so wouldn't
%not make any practical difference if we used the analytic expression
%instead. Here, however, I use the exact numerical data.

%Useful formulas:
%\begin{eqnarray}
%Y^{lm}(\pi/2,\varphi)&=&\tilde A_{lm}\, e^{im\varphi}, \nonumber\\
%Y^{lm}_{\theta}(\pi/2,\varphi)&=&\tilde B_{lm}\, e^{im\varphi},
%\end{eqnarray}
%where
%\begin{equation}
%\tilde A_{lm}=\left\{\begin{array}{ll}
%(-1)^{(l+m)/2} \left[\frac{(2l+1)(l+m-1)!!(l-m-1)!!}{4\pi(l+m)!!(l-m)!!}\right]^{1/2},
%& l-m \  \text{even} \\
%0, & l-m \ \text{odd},
%\end{array}\right.
%\end{equation}
%\begin{equation}
%\tilde B_{lm}=\left\{\begin{array}{ll}
%0 & l-m \  \text{even} \\
%l\left[\frac{(l+1)^2-m^2}{(2l+3)(2l+1)}\right]^{1/2}\tilde A_{l+1,m}
%-(l+1)\left[\frac{l^2-m^2}{(2l+1)(2l-1)}\right]^{1/2}\tilde A_{l-1,m},
%& l-m \ \text{odd},
%\end{array}\right.
%\end{equation}

%%%%%%%%%%%%%%%%%%%%%%%%%%%%%%%%%%%%%%%%%%%%%%%%%%%%%%%%%%%%%%%%%%%
%                            REFERENCES
%%%%%%%%%%%%%%%%%%%%%%%%%%%%%%%%%%%%%%%%%%%%%%%%%%%%%%%%%%%%%%%%%%%

\end{document}